\documentclass[prd,twocolumn,superscriptaddress,showpacs,aps,amsmath,amssymb,nofootinbib]{revtex4-2}

\usepackage{natbib}
\usepackage{color}

\usepackage{graphicx}
\usepackage{amsmath}
\usepackage{amssymb}
\usepackage{verbatim}
\usepackage{float}
\usepackage{wasysym}
\usepackage{epsfig}
\usepackage{psfrag}
\usepackage{dsfont}
\usepackage{amsfonts}
\usepackage{mathrsfs}
\usepackage{multirow}
\usepackage{times}
\usepackage{bm}
\usepackage{hyperref}
\usepackage{xspace}
\usepackage{tikz}
\usepackage{afterpage}
\usetikzlibrary{matrix}
\usepackage[acronym]{glossaries} 

\usepackage[normalem]{ulem}
\usepackage{mwe}
\usepackage{aas_macros}
\hypersetup{
  colorlinks=true,        
  linkcolor=blue,         
  citecolor=cyan          
}
\usepackage{pifont}

\usepackage{ragged2e}

\allowdisplaybreaks                                                   

\newcommand{\eg}{e.g.,~}

\newacronym{bcs}{BCs}{boundary conditions}
\newacronym{bc}{BC}{boundary condition}
\newacronym{rh}{RH}{Rankine-Hugoniot}
\newacronym{cfc}{CFC}{conformally flat condition}
\newacronym{gr}{GR}{General Relativity}
\newacronym{rhs}{rhs}{right-hand side}
\newacronym{eos}{EoS}{Equation of State}
\newacronym{pdes}{PDEs}{partial differential equations}
\newacronym{gw}{GW}{gravitational wave}
\newacronym{gws}{GWs}{gravitational waves}
\newacronym{ns}{NS}{neutron star}
\newacronym{nss}{NSs}{neutron stars}
\newacronym{bh}{BH}{black hole}
\newacronym{bhs}{BHs}{black holes}
\newacronym{ccsn}{CCSN}{core-collapse supernova}
\newacronym{ccsne}{CCSNe}{core-collapse supernovae}
\newacronym{pns}{PNS}{proto-neutron star}
\newacronym{sasi}{SASI}{standing accretion shock instability}
\newacronym{sn}{SN}{supernova}

\begin{document}

\title{Proto-neutron star oscillations including accretion flows}

\author{Dimitra Tseneklidou}
\affiliation{Departamento de Astronom\'{\i}a y Astrof\'{\i}sica, Universitat de Val\`encia,
  Dr. Moliner 50, 46100, Burjassot (Valencia), Spain}
\author{Raimon Luna}
\affiliation{Departamento de Matem\'{a}tica da Universidade de Aveiro and Centre for Research and Development in Mathematics and Applications (CIDMA), Campus de Santiago, 3810-193 Aveiro, Portugal}
\author{Pablo Cerd\'a-Dur\'an}
\affiliation{Departamento de Astronom\'{\i}a y Astrof\'{\i}sica, Universitat de Val\`encia,
  Dr. Moliner 50, 46100, Burjassot (Valencia), Spain}
\affiliation{Observatori Astron\`omic, Universitat de Val\`encia,
C/ Catedr\'atico Jos\'e Beltr\'an 2, 46980, Paterna (Valencia), Spain}
\author{Alejandro Torres-Forn\'e}
\affiliation{Departamento de Astronom\'{\i}a y Astrof\'{\i}sica, Universitat de Val\`encia,
  Dr. Moliner 50, 46100, Burjassot (Valencia), Spain}
\affiliation{Observatori Astron\`omic, Universitat de Val\`encia,
C/ Catedr\'atico Jos\'e Beltr\'an 2, 46980, Paterna (Valencia), Spain}                              


\begin{abstract}

The gravitational wave signature from core-collapse supernovae (CCSNe) is dominated by quadrupolar oscillation modes of the newly born proto-neutron star (PNS), and could be detectable at galactic distances. We have developed a framework for computing the normal oscillation modes of a PNS in general relativity, including, for the first time, the presence of an accretion flow and a surrounding stalled accretion shock. These new ingredients are key to understand PNS oscillation modes, in particular those related to the standing-accretion-shock instability (SASI). Their incorporation is an important step towards accurate PNS asteroseismology. For this purpose, we perform linear and adiabatic perturbations of a spherically symmetric background, in the relativistic Cowling approximation, and cast the resulting equations as an eigenvalue problem. We discretize the eigenvalue problem using collocation Chebyshev spectral methods, which is then solved by means of standard and efficient linear algebra methods. We impose boundary conditions at the accretion shock compatible with the Rankine-Hugoniot conditions. We present several numerical examples to assess the accuracy and convergence of the numerical code, as well as to understand the effect of an accretion flow on the oscillation modes, as a stepping stone towards a complete analysis of the CCSNe case.

\end{abstract}

\maketitle


\section{Introduction}
\label{sec:Introduction}

The last decade has been extremely fruitful for \acrfull{gw} astronomy. Since 2015, the Advanced LIGO \cite{LIGOScientific:2014pky} and Advanced Virgo \cite{VIRGO:2014yos} detectors have reported more than 90 GW events, including  mergers of \acrfull{bhs}, \acrfull{nss} and mixed \acrshort{bh}-\acrshort{ns} mergers. Additionally, during the ongoing fourth observing run, the LIGO-Virgo-KAGRA collaboration has issued more than 200 public alerts for significant detection candidates\footnote{gracedb.ligo.org}.
In 2017 \acrshort{gw} astronomy entered the field of multi-messenger astronomy with the detection of a binary neutron star merger in coincidence with a gamma-ray burst and a kilonova \cite{LIGOScientific:2017ync}.

Among the most promising yet-missing candidates are \acrfull{ccsne}. Their simultaneous detection through \acrfull{gws}, neutrinos and photons across the whole electromagnetic spectrum will be a milestone for future multi-messenger astronomy, as well as for other fields of physics, such as high energy physics. \acrshort{gws} from \acrshort{ccsne} could be observed at galactic distances \cite{Szczepanczyk2021} using current detectors, with an expected rate of $1-3$ per century (see \cite{Gossan:2016} and references therein). The associated \acrshort{gw} signal is expected to be
exceptionally rich, because of the complex dynamics of the astrophysical scenario; fluid dynamics, \acrfull{gr}, neutrino interactions and the properties of matter at extreme densities are expected to play a crucial role \cite{Janka:2006fh}.
These waveforms represent therefore a sizable jump in complexity in contrast to the \acrshort{bh} and \acrshort{ns} merger case. The evolution of these systems displays non-linear effects and instabilities that result in an essentially stochastic dynamics, which is translated to the waveforms. As a result, it is impossible to use the template-matching techniques currently employed for parameter estimation in the case of mergers.
For this reason, data analysis methods for detecting \acrshort{gw} from \acrshort{ccsne} primarily focus on reconstructing the \acrshort{gw} strain amplitude using minimally-modeled burst search pipelines, such as coherent WaveBurst (cWB)\cite{Klimenko2008, Drago2021cWB}, which identify short-duration transients through excess power analyses in time–frequency representations. These searches typically examine data within an on-source window defined by electromagnetic or neutrino observations \cite{Abbott2020CCSN}.

The multidimensional nature of the explosion mechanism makes their study, through numerical simulations, a challenging task.   
Most massive stars are expected to have slowly rotating cores and undergo neutrino-driven explosions. We focus this work in this scenario, whose stages are outlined below.
For a detailed description of the mechanism the interested reader is referred to \cite{2017handbook_SN.1095Janka, 2024_Mezzacappa_et_al_review, 2020_Mueller_review, 2021Nature_CCSN}. 

Main sequence stars with masses above $8-10\, M_{\odot}$ (depending on metallicity) form shells of progressively heavier elements starting from the surface and going inwards leading to an iron core. As the core grows, it becomes unstable and collapses under its own gravity. The innermost part of the collapsing core bounces back, creating a shock wave that propagates outwards. Nuclear dissociation and heating of the still in-falling material halt the shock shortly afterwards.
The core bounce leaves behind a newborn hot \acrshort{ns}, the so-called \acrfull{pns}. At this moment the picture is the following: the \acrshort{pns} is located in the center and it is surrounded by the stalled shock, which is about $100 - 200$ km away from the star. The shock defines a sonic point
at which the flow transitions from the supersonic (outside) to the subsonic (inside) regimen. In the interior (subsonic area) 
neutrinos streaming out of the cooling \acrshort{pns} may lead to convection both inside the \acrshort{pns} and in the region between the \acrshort{pns} and the accretion shock. Furthermore, the interaction of the accreting flow with sound waves in the 
subsonic region above the \acrshort{pns} may lead to a global instability of the shock, the so-called \acrfull{sasi} \cite{Foglizzo2000,Blondin2003}. If the neutrinos are able to deposit sufficient energy behind the shock, then it will be revived, leading to a \acrfull{sn} explosion. Otherwise, the accumulation of mass at the \acrshort{pns} will eventually lead to its collapse to \acrshort{bh}. Our study focuses on the first $0.2-1$~s after bounce, before this bifurcation point is reached. At that stage, the system consists of a \acrshort{pns} and a stalled accretion shock, and the dynamics is sufficiently violent to emit copious amounts of \acrshort{gws}.

Multidimensional numerical simulations have shown that the \acrshort{gw} emission in \acrshort{ccsne} is dominated by the excitation of the normal quadrupolar modes of the \acrshort{pns}. These modes are continuously excited by the presence of convection and \acrshort{sasi} and produce a \acrshort{gw} signal with a strong stochastic component\cite{Murphy:2009dx, Mueller:2012sv, CerdaDuran2013, Kuroda_2016, Andresen:2016pdt}. The frequency of these oscillation modes, visible in the time-frequency representation of the \acrshort{gw} signal (spectrogram), traces the evolution of the properties of the \acrshort{pns} during its first second of life. This opens the door to performing \acrshort{pns} asteroseismology with the aim of inferring \acrshort{pns} properties (such as mass and radius) or properties of matter at high densities, e.g. the \acrfull{eos} of nuclear matter. 

The calculation of \acrshort{pns} oscillation modes was first performed for idealized setups, considering linear perturbations of non-rotating stars in hydrostatic equilibrium \cite{McDermott1983,Reisenegger1992, Ferrari2003, Passamonti2005,Dimmelmeier2006, Kruger:2014pva, Camelio:2017nka}.
Those studies focused on the cooling phase of the \acrshort{pns} between the \acrshort{sn} explosion and the formation of the solid crust. Possible oscillation modes during this phase include pressure-driven p-modes, gravity-driven g-modes and the fundamental f-mode.
The work of \cite{Sotani:2016uwn} was the first to consider the calculation of oscillation modes in the early phases of the evolution of the \acrshort{pns}, before the onset of the explosion. Using multidimensional simulations as background for the perturbations, it has been shown that the features present in the corresponding \acrshort{gw} spectrograms correlate directly to some of the $f$, $p$ and g-modes in the \acrshort{pns} \cite{Torres_et_al_I,Morozova:2018glm, Torres_et_al_II,Sotani2019,Westernacher-Schneider2019,Westernacher-Schneider2020,Sotani2020}. There have been proposals for \acrshort{eos}-independent universal relations linking observable \acrshort{gw} mode frequencies with \acrshort{pns} properties (combinations of \acrshort{pns} mass and radius) \cite{Sotani2017, Torres_et_al_letter,Sotani2019}. Using these kinds of relations it would be possible to infer the \acrshort{pns} properties for a nearby galactic \acrshort{sn} \cite{Bizouard2021,Bruel2023}. The core g-modes of \acrshort{pns} have also been suggested to be related to the \acrshort{eos} properties \cite{Jakobus2023,Wolfe2023,Jakobus2024} and may allow the inference of the \acrshort{eos} parameters.

State-of-the-art calculations of oscillation modes include the use of a background based on numerical simulations, the incorporation of \acrshort{gr}, the inclusion of metric perturbations (at least the dominant effects) and the incorporation of \acrfull{bcs} that reflect that the \acrshort{pns} is not isolated in vacuum, but surrounded by an accretion shock that effectively acts as boundary \cite{Torres_et_al_II}. Some calculations have relaxed some of these assumptions by using the relativistic Cowling approximation (no metric perturbations) \cite{Torres_et_al_I,Sotani2020} or by considering simplified \acrshort{bcs} at the \acrshort{pns} surface \cite{Morozova:2018glm, Sotani2020}. The treatment of \acrshort{bcs} has been shown to be relatively unimportant for the computation of internal modes of the \acrshort{pns}, such as g-modes \cite{Sotani2019}. However, a proper treatment of the shock has been shown to be crucial for the p-modes interacting with the shock \cite{Torres_et_al_I,Torres_et_al_II}. A consistent treatment of the \acrshort{bcs} at the shock is still missing in all previous work, and so far only crude approximations have been used.

So far, the presence of a subsonic accretion flow from the shock to the surface of the \acrshort{pns} has been neglected in all the asteroseismology effects. Mode computations without considering the accretion flow show that $f$ and p-modes related to the shock have important differences in frequencies with respect to the modes observed in simulations \cite{Torres_et_al_II}. It was suggested that the difference may be caused by the absence of an accretion flow in the analysis. Furthermore, the interaction of advection and sound waves in the region between the \acrshort{pns} and the shock has been shown to play an important role in the development of the \acrshort{sasi} \cite{Foglizzo2000,Foglizzo2002,Foglizzo2007}. Therefore, a consistent treatment of the accretion flow is crucial to understand shock instabilities and to perform asteroseismology with \acrshort{sasi} modes. Given that these modes are tightly linked to the shock, their study should necessarily include proper treatment of the \acrshort{bcs} at the shock. The main goal of this work is to develop a framework to compute \acrshort{pns} modes in the presence of an accretion flow with proper \acrshort{bcs} at the shock. Additionally, we present a new numerical implementation to solve the oscillation modes.

The calculation of oscillation modes is based on the linearization of the perturbations of the hydrodynamics equations around an equilibrium solution. These equations, together with an appropriate set of \acrshort{bcs}, result in an eigenvalue problem whose eigenvalues are the oscillation frequencies and eigenfunctions the oscillation patterns. For spherically symmetric backgrounds, oscillations can be decomposed in spherical harmonics. Oscillations with different $l$ and $m$ decouple and the eigenvalue problem becomes a set of one-dimensional eigenvalue problems for each value of $l$ and $m$.
Previous approaches to the problem were based on the shooting algorithm. Those require the radial integration of the differential equations, with different values of a trial mode frequency, aiming to identify which ones are compatible with the \acrshort{bcs}.
When the mode frequencies are real numbers, representing stable and undamped modes, the search is restricted to the real line and it is equivalent to numerically finding the roots of a function.
However, in the presence of an advection flow, the modes have a complex frequency in general, representing modes that are either damped or unstable. In that case, the shooting algorithm becomes significantly more complex since one would have to find roots of a two dimensional function.
Furthermore, as more physical realism is added to the system, the number of equations and \acrshort{bcs} grow making shooting algorithms more complex. For example, the relaxation of the relativistic Cowling approximation in previous work \cite{Torres_et_al_II} made it necessary to use nested shooting algorithms to be able to impose several \acrshort{bcs} at the same time. Finally, if the background is not spherically symmetric, the problem becomes multidimensional (e.g. two-dimensional for axisymmetric backgrounds) and it is not possible to use the shooting method anymore. This is the case if rotation or strong magnetic fields are considered.

In order to tackle these numerical difficulties, we use a different approach in this work, based on spectral collocation methods. Spectral methods are a way of discretizing functions into a computational grid, which uses interpolation to estimate derivatives \cite{BoydSpectral, ThefethenSpectral, VoigtSpectral}. Unlike the finite difference schemes, which use a Taylor expansion to interpolate functions locally, spectral methods decompose smooth functions globally in a basis of orthogonal functions. The power-law convergence rate of finite differences, given by the truncation order of their Taylor series, now gives way to an exponential convergence rate. The spectral discretization of the equations allows to write any system of differential equations as an algebraic system for the values at the collocation points cast as a matrix equation. This equation has the explicit form of a generalized eigenvalue problem, for which standard methods can be used. Any number of \acrshort{bcs} can be added to the matrix equations using a similar discretization procedure. This makes this approach extremely robust, flexible and easy to implement. Some examples of the use of spectral methods for eigenvalue problems can be found in \cite{ThefethenSpectral, Dias:2015nua, Jansen:2017oag}.

The paper is structured as follows: In Section \ref{sec:MathFr} we introduce the formalism that we have employed to describe the background hydrodynamics of the \acrshort{pns} in \acrshort{gr}. We then derive the linearized equations for the perturbations of the \acrshort{pns} in the Cowling approximation, including the advective effects. We devote Section \ref{sec:BoundaryConditions} to a discussion on the relativistic \acrfull{rh} \acrshort{bcs}. Section \ref{sec:SpectralMethods} introduces the basics of collocation spectral methods, which are the basis for the algorithms that have been used to solve the perturbation equations and extract the modes. The specific models that have been used to test the methods and the numerical results of each test case are explained in detail in Section \ref{sec:Results}, and we finally we explain the main conclusions of this work in Section \ref{sec:Conclusions}.

Greek indices run from $0$ to $3$, while the Latin ones run from $1$ to $3$. The prime symbol, $a^\prime$, denotes differentiation with respect to the radial coordinate $r$. We use geometrized units with $G=c=1$.

\section{Linear perturbations of a spherical background including accretion flows}
\label{sec:MathFr}

\subsection{General framework}
\label{subsec:GeneralFramework}

The line element of the spacetime in the $3+1$ decomposition of space-time is described by
\begin{equation}\label{eq:line_element}
    ds^2 = g_{\mu\nu}dx^{\mu}dx^{\nu}=\left(\beta^i \beta_i - \alpha^2 \right)dt^2 + 2\beta_i dtdx^i + \gamma_{ij} dx^idx^j,
\end{equation}
where $\alpha$ represents the lapse function, $\beta^i$ is the shift vector and $\gamma_{ij}$ the spatial $3-$metric. Considering now isotropic coordinates, for a static and spherically symmetric object the aforementioned line element will translate into
\begin{equation}\label{eq:line_el_isotropic}
    ds^2 = g_{\mu\nu}dx^{\mu}dx^{\nu}= - \alpha^2 dt^2 + \gamma_{ij} dx^idx^j,
\end{equation}
where the spatial $3-$metric is conformally flat and can be written as $\gamma_{ij} = \psi^4 f_{ij}$, with $\psi$ being the conformal factor and $f_{ij}$ the flat spatial $3-$metric. Unless explicitly noted we will use spherical coordinates $\{r, \theta, \varphi\}$ and a coordinate (non-orthonormal) basis to describe vector components.

We consider a perfect fluid, whose energy-momentum tensor is described by
\begin{equation}\label{eq:T_mn}
    {T}^{\mu\nu} = \rho h {u}^{\mu}{u}^{\nu} + p {g}^{\mu\nu},
\end{equation}
where $\rho$ stands for the rest-mass density, $p$ for the pressure, ${u}^{\mu}$ is the $4-$velocity, $h \equiv 1 + \epsilon +p/\rho$ is the specific enthalpy and $\epsilon$ the specific internal energy. The energy density can be described as $e\equiv \rho(1 + \epsilon)$. The rest-mass current density of the fluid reads $J^{\mu}=\rho u^{\mu}$. 

The conservation of rest mass (consequence of the conservation of the number of particles) can be written as $\nabla_\mu  J^\mu = 0$, whose explicit form reads 
\begin{equation}\label{eq:cons_Nmu_conservative}
    \partial_{\mu}\left( \sqrt{-g}\rho u^{\mu} \right) =0, 
\end{equation}
the so-called continuity equation. Here $g$ is the determinant of the metric.

As a consequence of the Bianchi identities, the conservation of energy and momentum is given by
\begin{equation}\label{eq:conservation_EnMom}
    \nabla_{\mu}T^{\mu\nu} = 0,
\end{equation}
which can be expressed explicitly as
\begin{equation}\label{eq:cons_EnMom_Christoffel}
    \frac{1}{\sqrt{-g}}\bigg[ \partial_t \left(\sqrt{-g}T^{\mu 0}\right) + \partial_j\left( \sqrt{-g}T^{\mu j} \right) \bigg] = -\Gamma^{\mu}_{\nu\delta}T^{\delta\nu}  \;.
\end{equation}

The next step is to express the above conservation equations in a conservative form in the 3+1 form. The number particle conservation (continuity equation) and momentum conservation in \acrshort{gr} \cite{Banyuls:1997zz} are given, respectively, by
\begin{equation}\label{eq:continuity}
    \frac{1}{\sqrt{\gamma}} \partial_t \left[ \sqrt{\gamma} D \right] + \frac{1}{\sqrt{\gamma}} \partial_i \left[ \sqrt{\gamma} D \upsilon^{*i} \right] = 0,
\end{equation}
\begin{equation}\label{eq:mom_cons}
    \frac{1}{\sqrt{\gamma}} \partial_t \left[ \sqrt{\gamma} S_j \right] + \frac{1}{\sqrt{\gamma}} \partial_i \left[ \sqrt{\gamma} S_j \upsilon^{*i} \right] + \alpha \partial_j p = \frac{\alpha\rho h}{2} {u}^{\mu} {u}^{\nu} \partial_j {g}_{\mu\nu} ,
\end{equation}
with $\gamma$ being the determinant of the $3-$metric, while $D = \rho W$ and $S_j = \rho h W^2 \upsilon_j$ are conserved quantities and $W = 1/\sqrt{1-\upsilon_i\upsilon^i}$ is the Lorentz factor. For a detailed description of the derivation of the relativistic hydrodynamic equations, the interested reader is referred to \cite{2008Font_review, Rezzolla_book_GR_hydro}.

Since different velocities are appearing in the equations we present them all here for clarity. These are the $4-$velocity,
\begin{equation}\label{eq:4_velocity}
    {u} ^{\mu} =  W \alpha^{-1} (1, \: \upsilon^{*i}),
\end{equation}
the advective (coordinate) velocity
\begin{equation}\label{eq:advective_vel}
    \upsilon^{*i} = \frac{u^i}{u^0} = \alpha \upsilon^i - \beta^i ,
\end{equation}
and the Eulerian velocity given by,
\begin{equation}\label{eq:eulerian_velocity}
    \upsilon^i = \frac{u^i}{W} + \frac{\beta^i}{\alpha}. 
\end{equation}
%


\subsection{Background configuration}
\label{subsec:BGconf}

The unperturbed configuration, hereafter the background configuration, consists of a stationary equilibrium solution $(\partial_t = 0)$ for a non-rotating spherically symmetric star $(\partial_\theta = \partial_\varphi = 0$. This configuration is in general \textit{not} static, but stationary, and may incorporate a radial velocity, $\upsilon^i \neq 0$, representing an accretion flow. 
As discussed in Section \ref{sec:Introduction}, our system consists of a \acrshort{pns} surrounded by a stalled accretion shock. The shock defines a sonic surface, separating an external region characterized by a supersonic flow from an internal region (between the \acrshort{pns} and the shock) dominated by a subsonic flow.
Next, we derive the conditions that such a background has to fulfill to be stationary.

In the spherical and stationary case, the hydrodynamics equations~\eqref{eq:continuity} and \eqref{eq:mom_cons} reduce to the next equations for the background, respectively, 
\begin{equation}\label{eq:continuityBG}
    \frac{2}{r} + \frac{D^{*'}}{D^*} + \frac{{\upsilon^r}^{\prime}}{\upsilon^r} = 0  ,
\end{equation}
\begin{align}\label{eq:momentumBG}
    \mathcal{G_P}  = \mathcal{G_\alpha} - {\upsilon}^2\left( \frac{2}{r} +\frac{\alpha^{\prime}}{\alpha} +4\frac{\psi^{\prime}}{\psi} \right)  ,
\end{align}
where $\upsilon^2=\psi^4{\upsilon^r}^2$, $D^* \equiv \alpha \psi^6 D$ and the ram pressure is
\begin{equation}\label{eq:Pram}
    p_{\rm ram} \equiv \rho h W^2 \upsilon^2.
\end{equation}
We also define
\begin{equation}\label{eq:P_caligr}
    \mathcal{G_P} = \frac{1}{W^2} \frac{p^{\prime} + p^{\prime}_{\rm ram}}{ \rho h}~,
\end{equation}
and the gravitational acceleration
\begin{equation}\label{eq:G_i}
    \mathcal{G_\alpha} \equiv -\frac{\alpha^{\prime}}{\alpha}  .
\end{equation}

Let us introduce two characteristic frequencies, the relativistic Brunt–V\"{a}is\"{a}l\"{a} $\mathcal{N}^2$ frequency, 
\begin{equation}\label{eq:Brunt_V_freq}
    \mathcal{N}^2 \equiv \frac{\alpha^2}{\psi^4}\mathcal{G}_{\alpha} \mathcal{B}   ,
\end{equation}
where 
\begin{equation}\label{eq:B_i}
    \mathcal{B} = \frac{e^{\prime}}{\rho h} - \frac{1}{\Gamma_1}\frac{p^{\prime}}{p}~,
\end{equation}
is the relativistic version of the Schwarzschild discriminant, 
and the relativistic Lamb $\mathcal{L}^2$ frequency
\begin{equation}\label{eq:Lamb_freq}
    \mathcal{L}^2 \equiv \frac{\alpha^2}{\psi^4}c_s^2 \frac{l(l+1)}{r^2} \; .
\end{equation}

For the case without an accretion flow, $\upsilon^r=0$, we obtain $W=1$ and $p_{\rm ram}=0$, and Eqs.~\eqref{eq:continuityBG} and \eqref{eq:momentumBG} reduce to $\mathcal{G_P} = \mathcal{G_{\alpha}}$\footnote{Eq. \eqref{eq:continuityBG} leads to ${\upsilon^r}^{\prime} = 0$, which is trivially true if $\upsilon^r=0$.}, recovering the hydrostatic equilibrium condition of e.g. the work of 
\cite{Torres_et_al_I} (see their Eq.(13) for $\mathcal{G}$). The variables $\mathcal{G_P}$ and $\mathcal{G_\alpha}$ appear in \cite{Torres_et_al_II}, where the authors compare the difference in the value of the Brunt–V\"{a}is\"{a}l\"{a} $\mathcal{N}^2$ frequency, calculating it separately for each definition of $\mathcal{G}$. Note that here $\mathcal{G_P}$ includes also the $p_{\rm ram}$ and is only equal to the quantity in \cite{Torres_et_al_II} if no accretion flow is present. In our derivations, 
Eq. (\ref{eq:continuityBG}) will be used frequently to simplify the fraction ${\upsilon^r}^{\prime}/\upsilon^r$.

The two main families of modes studied in this work are the p- (pressure-modes) and the g-modes (gravity-modes). For the first ones, pressure acts as the restoring force, while for the latter ones, buoyancy is the restoring force. g-modes appear when the Brunt–V\"{a}is\"{a}l\"{a} frequency is larger than zero, $\mathcal{N}^2>0$, while for $\mathcal{N}^2=0$ there will only be sound waves and thus p-modes. In classical asteroseismology, $\mathcal{N}^2$ and $\mathcal{L}^2$ map the regions of each type of modes in the so-called propagation diagrams, that show these characteristic frequencies as a function of $r$ (see e.g. \cite{1989nos..book.....U}).

Apart from these equations, we need to ensure that the outer boundary corresponds to a stationary accretion shock. This restricts the possible values for the variables at the shock and the velocity profile itself. We discuss in more detail these extra conditions in Section~\ref{sec:BoundaryConditions}. 

The background metric for stationary objects is such that the only non-zero metric functions are $\alpha$, $\psi$ and $\beta^r$. 
Hereafter, we consider the case in which $\beta^r=0$. This case simplifies the equations and, for the case of a \acrshort{pns}, can be justified. The typical compactness of a \acrshort{pns} is in the range $M/R \sim 0.05 - 0.2$\footnote{The typical \acrshort{pns} evolves from $\sim 0.7$~M$_\odot$ and $30$~km shortly after bounce, to $\sim 1.4$~M$_\odot$ and $10$~km, about $1$~s later.}. For such compactness, a post-Newtonian analysis is valid. In particular, the shift is of the order of $(M/R)^{3/2}$ while the velocity is of order $(M/R)^{1/2}$ \cite{Blanchet1990}. Therefore, the inclusion of the shift is expected to lead to a $5-20 \%$ correction of the velocity of the fluid. For failed supernova, in which the \acrshort{pns} keeps increasing its mass due to accretion, this approximation may have to be revisited.


\subsection{Perturbation Equations}
\label{subsec:PerturbationEquations}

Allow us now to introduce linear adiabatic perturbations of the hydrodynamic variables with respect to their background equilibrium state. The Eulerian perturbation, $\delta q$, of a generic fluid scalar quantity $q$ is of the form
\begin{equation}\label{eq:perturbation}
    q \xrightarrow{} q + \delta q,
\end{equation}
where $q$ refers to the background quantity, and
\begin{equation}\label{eq:spher_harm_expansion}
    \delta q = \delta \hat{q}(r) Y_{lm}(\theta, \phi)e^{-i\sigma t} \; .
\end{equation}
Here the variables with hat represent the part of the perturbation depending only on $r$ in the separation of variables.
For simplicity, here we consider the relativistic Cowling approximation and thus the spacetime perturbations are not taken into account. This assumption could be easily relaxed in the future but is sufficient for this work.

In addition to the Eulerian perturbations, the Lagrangian displacement, $\xi^{i}$, expresses the displacement of a fluid quantity with respect to its position at rest. Its components are given by,
\begin{align}
    \xi^r &= \eta_1 Y_{lm} e^{-i\sigma t} , \label{eq:xi_r} \\
    \xi^{\theta} &= \eta_2\frac{1}{r}\partial_{\theta}Y_{lm} e^{-i\sigma t} ,\label{eq:xi_theta} \\
    \xi^{\phi} &= \eta_2\frac{1}{r\sin^2{\theta}}\partial_{\phi}Y_{lm} e^{-i\sigma t} , \label{eq:xi_phi}
\end{align}
where $\eta_1$ and $\eta_2$ represent the radial and angular displacement amplitude.\footnote{Note that $\eta_1$ and $\eta_2$ relate to  $\eta_r$ and $\eta_{\perp}$ in \cite{Torres_et_al_I, Torres_et_al_II} as $\eta_1 = \eta_r$ and $\eta_2 = \eta_{\perp}/r$.}
The Lagrangian and Eulerian perturbations of scalar quantities are related through 
\begin{equation}\label{eq:Lagr_Euler_pert}
    \Delta q = \delta q + \xi^{i} \partial_i q  \; .
\end{equation}
Throughout the whole analysis, the capital letter $\Delta$ will be used to denote a Lagrangian perturbation, while the lowercase $\delta$ will denote an Eulerian one. 
For any vector quantity, e.g. the coordinate velocity, Lagrangian and Eulerian perturbations are related by
\begin{equation}
\Delta \upsilon^{*i} = \delta \upsilon^{*i} + \xi^j\cdot \nabla_j\upsilon^{*i}  .
\end{equation}

The Lagrangian displacement is linked to the perturbed advective velocity according to
\begin{equation}\label{eq:xi_du*}
    \Delta \upsilon^{*i}  = \frac{d\xi^i}{dt} \; ,
\end{equation}
where $d\xi^i/dt = \partial_t \xi^{i} + \upsilon^{*j}\cdot \nabla_j \xi^i$ is the Lagrangian derivative, following the fluid motion. In the absence of advective velocity, the above relation reduces to $\delta \upsilon^{*i} = \partial_t \xi^{i}$, which has been used in previous studies, \eg \cite{Torres_et_al_I, Torres_et_al_II}. 

We consider adiabatic perturbations that fulfill
\begin{equation}\label{eq:adiabaticity}
    \frac{\Delta p}{\Delta \rho} = \left . \frac{\partial p}{\partial \rho}\right |_{\rm adiabatic}=hc_s^2=\frac{p}{\rho}\Gamma_1,
\end{equation}
with $c_s$ being the relativistic speed of sound, and $\Gamma_1$ the adiabatic index. A direct consequence is that the perturbation of the product $\rho h$ can be written as
\begin{equation}\label{eq:drhoh}
    \delta\left( \rho h \right) = \left( 1 + \frac{1}{c_s^2} \right)\delta p - \rho h \xi^r \mathcal{B} \;.
\end{equation}
Taking into account the adiabaticity condition and the relation between the Eulerian and Lagrangian perturbations, we arrive at the next expression relating to the Eulerian perturbation of pressure and density,
\begin{equation}\label{eq:dp_drho}
    \delta \hat{p} = p \Gamma_1 \left( \frac{\delta\hat{\rho}}{\rho} + \mathcal{B_N} \eta_1   \right)  \; ,
\end{equation}
with $\mathcal{B_N} = - \frac{p^\prime}{p \Gamma_1} + \frac{\rho^\prime}{\rho} $ being the Newtonian analogue of $\mathcal{B}$.
The velocity perturbations read
\begin{equation}\label{eq:du^i}
    \delta u^i = \upsilon^{*i} \delta u^t + u^t\delta\upsilon^{*i}  
\end{equation}
and
\begin{equation}\label{eq:dupsilon^i}
    \delta \upsilon^i = \alpha^{-1}\delta\upsilon^{*i} \; ,
\end{equation}
where $\delta\upsilon^{*i}$ can be decomposed into its components, according to,
\begin{align}
    \delta\upsilon^{*r} &= \delta\upsilon^{*}_R Y_{lm} e^{-i\sigma t} , \label{eq:du*_r} \\
    \delta\upsilon^{*\theta} &= \delta\upsilon^{*}_{\perp}\frac{1}{r}\partial_{\theta}Y_{lm} e^{-i\sigma t} ,\label{eq:du*_theta} \\
    \delta\upsilon^{*\phi} &= \delta\upsilon^{*}_{\perp}\frac{1}{r\sin^2{\theta}}\partial_{\phi}Y_{lm} e^{-i\sigma t} . \label{eq:du*_phi}
\end{align}

The perturbed continuity equation and the momentum equations for the radial and the angular components have the following form, respectively,

\begin{widetext}
\begin{align}
    {A}_{_{00}}^{^{5\times 5}} \delta\hat{\rho} + {A}_{_{03}}^{^{5\times 5}}&\delta\upsilon^*_R + {A}_{_{04}}^{^{5\times 5}}\delta\upsilon^*_{\perp} = -i\sigma \left({B}_{_{00}}^{^{5\times 5}}\delta\hat{\rho} + {B}_{_{03}}^{^{5\times 5}}\delta\upsilon^*_R  \right)  \; ,\label{eq:continuityForm} \\
    {A}_{_{10}}^{^{5 \times 5}}\delta\hat{\rho} + {A}_{_{11}}^{^{5\times 5}}&\eta_1 + {A}_{_{13}}^{^{5 \times 5}}\delta\upsilon^*_R + {A}_{_{14}}^{^{5 \times 5}}\delta\upsilon^*_{\perp} = -i\sigma \left( {B}_{_{10}}^{^{5\times 5}}\delta\hat{\rho} + {B}_{_{11}}^{^{5\times 5}}\eta_1 + {B}_{_{13}}^{^{5\times 5}}\delta\upsilon^*_R \right) 
       \; , \label{eq:mom_r_form} \\
    {A}_{_{20}}^{^{5\times 5}}\delta\hat{\rho} + {A}_{_{21}}^{^{5\times 5}}&\eta_1 + {A}_{_{24}}^{^{5\times 5}}\delta\upsilon^*_{\perp} = -i\sigma {B}_{_{24}}^{^{5\times 5}}\delta\upsilon^*_{\perp}   \; , \label{eq:mom_theta_form} \\
    {A}_{_{31}}^{^{5\times 5}} &\eta_1 + {A}_{_{33}}^{^{5 \times 5}}\delta\upsilon^*_R = -i\sigma {B}_{_{31}}^{^{5\times 5}} \eta_1 \; , \label{eq:du*_R} \\
    {A}_{_{42}}^{^{5\times 5}} &\eta_2 + {A}_{_{44}}^{^{5 \times 5}}\delta\upsilon^*_{\perp} = -i\sigma {B}_{_{42}}^{^{5\times 5}} \eta_2 \; , \label{eq:du*_perp}
\end{align}
\end{widetext}
\enlargethispage{2\baselineskip}
\footnotetext[\value{footnote}]{Note that $\eta_1$ and $\eta_2$ relate to  $\eta_r$ and $\eta_{\perp}$ in \cite{Torres_et_al_I, Torres_et_al_II} as $\eta_1 = \eta_r$ and $\eta_2 = \eta_{\perp}/r$.}
where the nonzero coefficients are
\begin{align}
    {A}_{_{00}}^{^{5\times 5}} &= r\alpha \frac{\upsilon^r}{\rho}\left( \partial_r - \frac{\rho^{\prime}}{\rho} \right)  \; ,  \nonumber \\ 
    {A}_{_{03}}^{^{5\times 5}} &= rW^2\bigg( \partial_r +2\mathcal{G_P} -\mathcal{G_\alpha}  \nonumber \\
    &+\frac{D^{*\prime}}{D^*} +\frac{2}{r} -2\upsilon^2\frac{\rho^{\prime}}{\rho} \bigg)     \; ,  \nonumber \\
    {A}_{_{04}}^{^{5\times 5}} &= -l(l+1)     \; ,  \nonumber \\
    {B}_{_{00}}^{^{5\times 5}} &= -r\frac{1}{\rho} \; , \nonumber  \\
    {B}_{_{03}}^{^{5\times 5}} &= -r\upsilon^r W^2\psi^4 \alpha^{-1} \; ,  \label{eq:coef_contin_eq} 
\end{align}
\begin{align}
    {A}_{_{10}}^{^{5\times 5}} &= \frac{\alpha \psi^{-4}}{\rho}c_s^2\bigg\{\left( 1 + \frac{\upsilon^2}{c_s^2} \right)\partial_r + 
    \frac{(c_s^2)^{\prime}}{c_s^2} +W^{-2}\frac{h^\prime}{h}  \nonumber \\
    &+\left( 1 + \frac{1}{c_s^2} \right)\left[ \mathcal{G_P} -2\mathcal{G_\alpha} + \upsilon^2\left(  \frac{h^\prime}{h} -2\frac{\rho^\prime}{\rho} \right) \right] 
    \bigg\}  \; , \nonumber
       \\  
    {A}_{_{11}}^{^{5\times 5}} &= \frac{\alpha }{\psi^{4}}\Bigg\{ 
    c_s^2\left[ -\frac{\upsilon^2}{c_s^2} \mathcal{B} +  \mathcal{B_{N}}\left( 1+\frac{\upsilon^2}{c_s^2}\right) \right] \partial_r \nonumber \\
    &+\left( -\mathcal{G_P} +2\mathcal{G_\alpha} \right)\left[ \mathcal{B} - \mathcal{B_{N}} \left( 1+c_s^2 \right)\right]    \nonumber \\
    &-\frac{p^{\prime\prime}}{\rho h} +c_s^2\frac{\rho^{\prime\prime}}{\rho} 
    +c_s^2\frac{\rho^\prime}{\rho}\left( \frac{h^\prime}{h} + 2\frac{c_s^\prime}{c_s} \right)  \nonumber\\
    &+\upsilon^2\Bigg[ -\mathcal{B}^\prime +\mathcal{B_N}^\prime +\left(\mathcal{B}-\mathcal{B_N}\right)\left(\frac{\rho^\prime}{\rho} -\frac{h^\prime}{h} \right)  \nonumber \\
    &-2 c_s^2\frac{\rho^\prime}{\rho}\mathcal{B_{N}}
    \Bigg]  
    \Bigg\}  \; , \nonumber  \\ 
    {A}_{_{13}}^{^{5 \times 5}} &= W^2 \upsilon^r \bigg\{ 2\partial_r  +4\mathcal{G_P} -4\mathcal{G_\alpha}
    + 2\frac{W^\prime}{W} +4\frac{\psi^\prime}{\psi}  \nonumber \\
    &+2\frac{h^\prime}{h}   -4\upsilon^2 \frac{\rho^\prime}{\rho} 
    \bigg\}  \; , \nonumber  \\ 
    {A}_{_{14}}^{^{5 \times 5}} &= -\upsilon^r\frac{l(l+1)}{r} \; , \nonumber  \\  
    {B}_{_{10}}^{^{3 \times 3}} &= -\frac{\upsilon^r}{\rho}\left( 1+ c_s^2 \right)  \; , \nonumber  \\ 
    {B}_{_{11}}^{^{5 \times 5}} &= \upsilon^r \bigg[ \mathcal{B} -\mathcal{B_N}\left(1+c_s^2 \right) \bigg]  \; , \nonumber  \\ 
    {B}_{_{13}}^{^{5 \times 5}} &= -\frac{W^2}{\alpha}\left( 1 + \upsilon^2 \right)  \; , \label{eq:coef_mom_r} 
\end{align}
\begin{align}
    {A}_{_{20}}^{^{5\times 5}} &= \frac{\alpha}{W^2\psi^4}\frac{c_s^2}{\rho} \; , \nonumber \\ 
    {A}_{_{21}}^{^{5\times 5}} &=  \frac{\alpha c_s^2}{W^2\psi^4}\mathcal{B_N} \; , \nonumber \\ 
    {A}_{_{24}}^{^{5\times 5}} &= \upsilon^r\bigg[ r\partial_r +1
    +r\left(\mathcal{G_\alpha}  + \frac{W^\prime}{W}  +4\frac{\psi^\prime}{\psi} + \frac{h^\prime}{h} \right)
    \bigg] \; , \nonumber \\ 
    {B}_{_{24}}^{^{5\times 5}} &= -\alpha^{-1}r  \; , \label{eq:coef_mom_theta}
\end{align}
\begin{align}
    {A}_{_{31}}^{^{5\times 5}} &= \alpha\left(- \upsilon^{r}\partial_r  -\upsilon^r\mathcal{G}_\alpha +{\upsilon^{r}}^{\prime}\right)  \; , \nonumber \\ 
    {A}_{_{33}}^{^{5\times 5}} &= 1 \; , \nonumber \\ 
    {B}_{_{31}}^{^{5\times 5}} &= 1 \; , \label{eq:coef_du*_R}
\end{align}
\begin{align}
    {A}_{_{42}}^{^{5\times 5}} &= -\alpha\upsilon^{r}\left( \partial_r +2\frac{\psi^{\prime}}{\psi}\right)  \; , \nonumber \\ 
    {A}_{_{44}}^{^{5\times 5}} &= 1 \; , \nonumber \\ 
    {B}_{_{42}}^{^{5\times 5}} &= 1 \; , \label{eq:coef_du*_perp}
\end{align}
where all the coefficients are written in the form of $A_{ij}$, with index $i$ representing the equation ($0$: continuity, $1$: momentum $r$-component, $2$: momentum $\theta$-component, $3$: $\delta\upsilon^{*r}$, $4$: $\delta\upsilon^{*\theta}$), while $j$ stands for the perturbations in the following increasing order: $\left( \delta\hat{\rho}, \eta_1, \eta_2, \delta\upsilon^*_R, \delta\upsilon^*_{\perp} \right)$. The $5\times5$ label is used to indicate that these are the coefficients for a system of 5 equations with 5 unknowns and distinguish it from the coefficients of other systems of equations described below.
The $\phi-$component of the momentum equations leads to the same relation as Eq. (\ref{eq:mom_theta_form}) due to axisymmetry. Note that to simplify the above system of equations, we have used their background solutions (\ref{eq:continuityBG}), (\ref{eq:momentumBG}) and combinations of those, as well as the definitions, Eqs \eqref{eq:Lagr_Euler_pert}-\eqref{eq:dp_drho}.

The system of equations \eqref{eq:continuityForm}-\eqref{eq:du*_perp} can be written in the form of a generalized 
eigenvalue problem, as
\begin{equation}\label{eq:gen_eigenv_problem}
    \mathbf{A}{^{5\times 5}} \cdot \boldsymbol{\eta}^{(5)} = \sigma \mathbf{B}^{5\times 5} \cdot \boldsymbol{\eta}^{(5)} \; ,
\end{equation}
where $\boldsymbol{\eta}^{(5)}=(\delta\hat\rho, \eta_1, \eta_2, \delta\upsilon^*_R, \delta\upsilon^*_{\perp})^T$ is the matrix of the variables and $\mathbf{A}{^{5\times 5}}$, $\mathbf{B}^{5\times 5}$ are the coefficient matrices. 

In Section \ref{sec:SpectralMethods}, there is a thorough explanation of the numerical implementation of eigenvalue problems in the form
of Eq. (\ref{eq:gen_eigenv_problem}).

\subsection{Zero velocity limit}\label{subsec:zeroVel}

Next, it is interesting to consider the case of zero velocity and how it simplifies the system of equations (\ref{eq:continuityForm}) - (\ref{eq:du*_perp}). In this case, it is straightforward that the system can be reduced into a $2\times 2$ system, $\mathbf{A}{^{2\times 2}} \cdot \boldsymbol{\eta}^{(2)} = \sigma^2 \mathbf{B}^{2\times 2} \cdot \boldsymbol{\eta}^{(2)}$, where the only variables now will be $\boldsymbol{\eta}^{(2)}=(\eta_1, \eta_2)^T$. In this manner, we substitute Eq. (\ref{eq:mom_theta_form}) into the other two equations, obtaining the following coefficients for the matrices $\mathbf{A}^{2\times 2}$ and $\mathbf{B}^{2\times 2}$:
\begin{align}
    A_{_{00}}^{^{2 \times 2}} &= \frac{\alpha^2}{\psi^4} \mathcal{B} \mathcal{G_P},   \nonumber \\ 
    B_{_{00}}^{^{2 \times 2}} &= 1, \nonumber \\ 
    B_{_{01}}^{^{2 \times 2}} &= -r\partial_r -\left[ 1 + r\mathcal{G_P}\left( 1 - \frac{1}{c_s^2} \right) + r\left( \frac{\rho^\prime}{\rho} +\frac{h^\prime}{h} +4\frac{\psi^\prime}{\psi}\right) \right] , \nonumber \\ 
     A_{_{10}}^{^{2 \times 2}} &= -r\partial_r -2 -\frac{r}{c_s^2}\mathcal{G_P} -6r\frac{\psi^\prime}{\psi}, \nonumber \\ 
    A_{_{11}}^{^{2 \times 2}} &= l(l+1), \nonumber  \\ 
    B_{_{11}}^{^{2 \times 2}} &= \frac{r^2\psi^4}{\alpha^2 c_s^2} \; . \label{eq:coef_2x2} 
\end{align}

At this point, it is worth comparing our $2\times 2$ system of equations with the ones in \cite{Torres_et_al_I}, as the limit here is $\upsilon^r=0$. It is trivial to show after some calculations that the $2\times 2$ system is equivalent to Eqs (31) and (32) of \cite{Torres_et_al_I}. 


\section{Boundary Conditions}
\label{sec:BoundaryConditions}

The shock is defined as a sonic interface at which the flow is entering from a supersonic area (exterior) to a subsonic area (interior). 
At the shock location, some quantities are discontinuous (velocity, pressure, density, internal energy), however, these discontinuities have to fulfill the \acrshort{rh} conditions \cite{1948Courant}, to ensure the continuity of the mass flux and the energy-momentum conservation. 

The relativistic \acrshort{rh} conditions are given by the equations, 
\begin{equation}\label{eq:RH_rel_1}
    [[\rho u^{\mu}]] n_{\mu} = 0,
\end{equation}
\begin{equation}\label{eq:RH_rel_2}
    [[T^{\mu\nu}]] n_{\nu} = 0,
\end{equation}
where $n_{\mu}$ is a unit 4-vector normal to the surface of the shock. The double brackets are used to denote the subtraction between the two states, meaning the exterior and the interior, e.g. $[[F]] = F_{\mathrm{int}} - F_{\mathrm{ext}}$. 
Throughout the whole section, we will label with ${\mathrm{ext}}$ (exterior) the value of a variable at the shock location encountering it from the supersonic region and with ${\mathrm{int}}$ (interior) its value at the shock location right after the shock-discontinuity. As a direct consequence of Einstein's equations, all metric quantities remain continuous across the shock.

For a thorough derivation of Eqs. (\ref{eq:RH_rel_1}) and (\ref{eq:RH_rel_2}) the interested reader is referred to \cite{Rezzolla_book_GR_hydro}. To calculate the above conditions, we follow partially the methodology and notation as in \cite{Marti_et_al_1994}, where the authors present a comprehensive derivation of the \acrshort{rh} conditions for a Minkowski spacetime.

Previous attempts of imposing \acrshort{bcs} at the shock location \cite{Torres_et_al_I, Torres_et_al_II} faced the problem that, in the absence of an accretion flow, the \acrshort{rh} conditions cannot be applied. As a consequence, the authors in \cite{Torres_et_al_I, Torres_et_al_II} justified their choice of \acrshort{bcs} by appealing to the physical necessity that perturbations cannot propagate across the shock from the subsonic to the supersonic region. Here, we improve that work by deriving the \acrshort{bcs} directly from the \acrshort{rh} conditions.

The inner boundary at $r=0$ is not an actual physical boundary of the system since it is just the center of the domain. However, numerically one may need to impose \acrshort{bcs} to ensure regularity at this point \citep[see e.g][]{Torres_et_al_I,Torres_et_al_II}. In this work, the use of spectral methods with a spectral basis formed by Chebyshev polynomials of the first kind, ensures that any function is automatically regular at the origin. Therefore, there is no need to impose any \acrshort{bcs} there. In some of the tests described in the section \ref{sec:Results}, the domain does not extend down to $r=0$, but to a non-zero radius. The boundary conditions at the inner boundary in those cases are described in the corresponding sections.

\subsection{Normal vector to the shock}
Before continuing with the \acrshort{rh} conditions it is worth describing in detail the normal vector to the shock, which plays an important role in the \acrshort{bcs}.
For this purpose, we introduce two observers: the observer O, related to the coordinates $x^\mu$, for which the line element is Eq.~\eqref{eq:line_el_isotropic}, and the observer O$^\prime$, comoving with the shock, defined by the 4-velocity of the shock, $u^\mu_{s}$. 
We use primed and non-primed indices to refer to the components of vectors for the observer O$^\prime$ and O, respectively.
The components of the shock 4-velocity for the observer O$^\prime$ are trivially 
\begin{equation}\label{eq:u_mu'}
    u_s^{\mu'} = (1,0,0,0).
\end{equation}
and those of the normal vector
\begin{equation}\label{eq:n_mu'}
    n_{\mu'} = (0, n_{i'})  \;,
\end{equation}
where $n_{i'}$ is a unitary 3-vector normal to the hypersurface. In the frame of O$^\prime$ it is trivial to obtain that  $ u_s^{\mu'}n_{\mu'} = 0$ and  $ n^{\mu'}n_{\mu'} = 0$, therefore, in a general frame (e.g. O) we get
\begin{eqnarray}
    u_s^{\mu}u_{s\mu} &=& -1, \label{eq:u_square} \\
    u_s^{\mu}n_{\mu} &=& 0, \label{eq:u_n_mu}\\
     n^{\mu}n_{\mu} &=& 1 . \label{eq:n_square}
\end{eqnarray}
The 4-velocity of the shock for the observer O is given by
\begin{equation}\label{eq:4u_shock}
    u_s^{\mu} = \alpha^{-1} W_s (1, \upsilon_s^{*i}) ,
\end{equation}
where $\upsilon_s^{*i} = \alpha \upsilon_s^{i} - \beta^i$ is the advective (coordinate) velocity and $\upsilon_s^i$ is the 3-velocity of the shock, for an Eulerian observer. The associated Lorentz factor is
\begin{align}\label{eq:Ws}
    W_s &= \frac{1}{\sqrt{1-\upsilon_s^2}}  .
\end{align}

From Eq. (\ref{eq:u_n_mu}) for the Eulerian observer we have
\begin{align}\label{eq:n0}
    n_0 &= -\upsilon_s^{*i}n_i \;.
\end{align}
Thus, the normal vector to the shock surface for the observer O reads
\begin{equation}\label{eq:n_mu_Eulerian}
    n_{\mu} = \left(-\upsilon_s^{*j}n_j, n_i \right) \;.
\end{equation}
Lastly, using the normalisation condition (\ref{eq:n_square}) we find that
\begin{align}\label{eq:norm_n}
    n_i n_j\left(-\upsilon_s^i \upsilon_s^j + \gamma^{ij} \right) &= 1  ,
\end{align}
which for $\upsilon_s^i = 0$ is automatically fulfilled, as in the case of the observer comoving with the shock.

\subsection{Background \acrshort{rh} conditions}

For the background configuration, we consider a shock that satisfies the conditions typically found in accretion shocks around \acrshort{pns}, namely:
\begin{itemize}
    \item The shock is static, i.e. $\upsilon^{*i}_s = 0$.
    \item The exterior region is supersonic and thus the exterior velocity has to be always larger than the local sound speed. In contrast, the interior velocity has to be smaller than the local sound speed (subsonic region).
    \item The exterior velocity is larger than the interior velocity.
    \item The density and pressure in the interior should be larger than the exterior ones.
    \item The shock is spherically symmetric.
\end{itemize}
For a spherical shock with purely radial velocity, the normal vector, Eq.~\eqref{eq:n_mu_Eulerian}, reads 
\begin{equation}
n_\mu = \psi^2 W_s (-\upsilon^{*r}_s, 1,0,0).    
\end{equation}
If the shock is static ($\upsilon^{*r}_s=0$) then $W_s =1$ and $n_\mu = (0,\psi^{2},0,0)$. In that case equations (\ref{eq:RH_rel_1}) and (\ref{eq:RH_rel_2}), result in a system of three equations for the background quantities at the shock: 
\begin{align}
    \left( \rho u^r \right)_{\rm ext} &= \left( \rho u^r \right)_{\rm int} ,\label{eq:RH1_BG} \\
    \left( \rho h u^r u^t\right)_{\rm ext} &= \left( \rho h u^r u^t \right)_{\rm int} , \label{eq:RH2_BG} \\
    \left( \rho h {u^r}^2 + \psi^{-4}p \right)_{\rm ext} &= \left( \rho h {u^r}^2 + \psi^{-4}p \right)_{\rm int} \;. \label{eq:RH3_BG}
\end{align}
The $\theta$ and $\phi$ components of Eq. (\ref{eq:RH_rel_2}) are satisfied automatically.

Given the hydrodynamics state (values of density, velocity, pressure, enthalpy, internal energy) in one side of the shock, and the \acrshort{eos}, the above set of equations allows to compute the hydrodynamics state in the other side of the shock. It is important to note that not every hydrodynamics state will lead to a physical solution with a discontinuity or that fulfills all the conditions above. This restricts the set of possible hydrodynamics states. We provide particular examples in Section~\ref{sec:Results}.

\subsection{Perturbed \acrshort{rh} conditions}
Next, we consider a shock that has been perturbed by the action of the \acrshort{pns} oscillations that we are studying. In that case we make the next considerations:
\begin{itemize}
    \item The perturbed shock is not spherical in general and its coordinate location is given by $r_s \delta^i_r+ \zeta^i$, where $r_s$ is the coordinate radius of the unperturbed spherical shock, and $\zeta^i$ is the shock displacement.
    \item We decompose the shock displacement, expressed in the coordinate basis, as
\begin{equation}\label{eq:zeta_spher}
    \zeta^i = \left(Z_1, \frac{Z_2}{r}\partial_{\theta}, \frac{Z_2}{r\sin^2{\theta}}\partial_{\phi}\right)Y_{lm}e^{-i\sigma t},
\end{equation}
where $Z_1$ and $Z_2$ are two radial functions.
The decomposition of the shock displacement, $\zeta^i$, is analogous to the definition of the Lagrangian displacement, $\xi^i$. Note that, in the presence of an accreting flow, the shock displacement is not equal, in general, to the Lagrangian displacement of the fluid ($\zeta^i\ne \xi^i$).
\item We linearize the perturbed \acrshort{rh} conditions with respect to both $\xi^i$ and $\zeta^i$. Therefore, combinations of both will be considered of higher order and will be neglected.
    \item The perturbed shock moves with respect to its equilibrium and therefore it has a coordinate velocity given by
    \begin{equation}\label{eq:shock_velocity}
        \upsilon^{*i}_s = \partial_t \zeta^i \;.
    \end{equation}
    \item In the \textbf{\textit{interior}} there is a \textbf{\textit{subsonic}} flow. This means that perturbations (in particular sound waves) can propagate both upstream and downstream. Therefore, in this region (where the \acrshort{pns} is located) is where we solve our perturbation equations for $\xi^i$. The value of a quantity, $q$, at any point in the interior (in particular at the shock) can be decomposed as $q=q_0 + \delta q$, where here we use the subscript $0$ to denote the background.
    In this case, the value of $q$ at the perturbed shock can be computed considering a Taylor expansion around the unperturbed shock location, $r_s$, resulting in
    \begin{align}\label{eq:qTaylorINT}
        &q\left( r_s + \zeta \right) |_{\rm int} = q\left( r_s \right) + \zeta^i\partial_i q\left( r_s \right) |_{\rm int}  \nonumber \\
        &= q_0\left( r_s \right) + \delta q\left( r_s \right) + \zeta^i \partial_i q_0\left( r_s \right) |_{\rm int} \;,
    \end{align}
    Note that the sum $\delta q\left( r_s \right) + \zeta^i \partial_i q_0\left( r_s \right) $ resembles a Lagrangian perturbation, but with respect to the shock displacement $\zeta^{i}$.

    \item In the \textbf{\textit{exterior}}, the flow is \textbf{\textit{supersonic}}. The perturbations from the interior cannot propagate upstream. So, there will be no perturbations in the supersonic region. Contrary to the interior, in the exterior
    there are no Eulerian perturbations. As a consequence, any perturbed quantity in the exterior will be of the form
    \begin{align}\label{eq:qTaylorEXT}
        &q\left( r_s + \zeta \right) |_{\rm ext} = {q}\left( r_s \right) + \zeta^i\partial_i {q}\left( r_s \right) |_{\rm ext}  \nonumber \\
        &= q_0\left( r_s \right) + \zeta^i \partial_i q_0\left( r_s \right) |_{\rm ext}  \;.
    \end{align}
    The detailed calculation of the derivatives of the background quantities in the exterior can be found in Appendix \ref{App:RHCalculations}.
\end{itemize}

For the case of the perturbed shock, the normal vector is not radial anymore and has to be computed with care. Its detailed derivation is given in the Appendix \ref{App:NormalVector}, which results in
\begin{equation}\label{eq:n_mu_total}
    n_{\mu} = \psi^2 W_s\begin{pmatrix}
        i\sigma Z_r Y_{lm} e^{-i\sigma t} \\
            1 \\
         \left(-Z_r + Z_2 \right)\partial_{\theta} Y_{lm} e^{-i\sigma t}  \\
         \left(-Z_r + Z_2 \right)\partial_{\phi} Y_{lm} e^{-i\sigma t}
    \end{pmatrix}.
\end{equation}

The 4-velocity of the perturbed fluid can be written as
\begin{equation}\label{eq_4u}
    \left(u^t + \delta u^t, u^r + \delta u^r, \delta u^{\theta}, \delta u^{\phi}  \right),
\end{equation}
where 
\begin{eqnarray}\label{eq:du0}
    \delta u^t &=& 
    \alpha^{-2}\psi^4v^{*r} \delta u^r, \label{eq:du0} \\
    \delta u^r &=& \alpha^{-1}W^3 \delta\upsilon^{*r}, \label{eq:du^r} \\
    \delta u^{\theta} &=& u^t\delta\upsilon^{*\theta}\label{eq:du^theta}.
\end{eqnarray}

Finally, we derive the \acrshort{rh} conditions for a perturbed shock, by linearizing equations \eqref{eq:RH_rel_1} and \eqref{eq:RH_rel_2}. Given that we consider adiabatic perturbations, the energy flux does not result in an independent condition. Of the other four conditions, the conditions for the $\theta$ and $\phi$ momentum fluxes result in the same condition. The resulting three independent conditions read as follows

\begin{align}\label{eq:rh1_GR_expanded}
    &-i\sigma \alpha^{-1}\left[[ \rho W ]\right] Z_1 
     = \left( \rho\frac{W^3}{\alpha} \delta\upsilon^*_R + u^r\delta\hat{\rho} \right)_{\rm int},
\end{align}
\begin{align}\label{eq:rh2_r_GRexpanded}
    &-\bigg\{\mathcal{G_{\alpha}}\left[[ \rho h W^2]\right] +\left( \frac{2}{r_s} +4\frac{\psi^\prime}{\psi} -\mathcal{G_{\alpha}} \right)[[p]]
    \bigg\}Z_1  \nonumber \\
    &= \bigg\{ 2\psi^4 \frac{W^3}{\alpha} \rho h  u^r \delta\upsilon^*_R \nonumber \\
    & +\left[ -p^\prime +c_s^2 W^2 \rho h \left(\frac{\rho^\prime}{\rho} -\frac{\upsilon^2}{c_s^2}\frac{h^\prime}{h} \right) \right]\eta_r \nonumber \\
    &+c_s^2 W^2h\left( 1 + \frac{\upsilon^2}{c_s^2}\right)\delta\hat{\rho}
    \bigg\}_{\rm int},
\end{align}

\begin{equation}\label{eq:rh2_theta_GRexpanded}
    \left[[ p ]\right]\left( -Z_1 +Z_2\right) = - r_s  \psi^4 \alpha^{-1}\left( W\rho h u^r \delta\upsilon^*_{\perp}\right)_{int}.
\end{equation}
Note that it is possible to use Eq.~\eqref{eq:rh2_r_GRexpanded} to eliminate $Z_1$ from Eq.~\eqref{eq:rh1_GR_expanded}. The resulting equation does not depend on $Z_1$ or $Z_2$ and can be used to impose \acrshort{bcs} to the eigenvalue problem, without the need of using three equations. These two additional equations would only be needed to compute the shock displacement, a posteriori.


\section{Spectral Collocation Methods}
\label{sec:SpectralMethods}
 
All the numerical computations in this paper are produced by virtue of spectral collocation methods. These methods represent an extension and improvement to the codes developed in [26, 28], and are particularly useful for a number of reasons. Their exponential convergence rate makes it possible to produce extremely accurate results with a very limited number of grid points, typically reaching machine precision with as few as 32 points per grid function. Working with such small grids, together with the possibility of writing the discretized equations in matrix form, makes the method very efficient computationally. Therefore, it is straightforward to use optimized linear algebra libraries from high-level and interpreted programming languages, making its implementation much easier without a significant loss in efficiency. All the modes --- including complex ones --- are computed at once, just by solving a matrix eigenvalue problem, without the need for iterations or shooting algorithms.
Another significant advantage of spectral methods is that regularity of the solution is automatically imposed, by construction, due to the global interpolation of functions in terms of a linear combination of a basis of regular functions. This often eliminates the need to impose extra \acrshort{bcs} to enforce regularity, usually needed when using finite difference schemes. Standard references for spectral methods are \cite{BoydSpectral, ThefethenSpectral, VoigtSpectral}, while applications to \acrshort{gr} can be found in \cite{Dias:2015nua, Jansen:2017oag}.

\subsection{Discretization of Differential Eigenvalue Problems}
\label{subsec:discretization}

In spectral methods, we can express any function, $f$, as a truncated expansion in an appropriate basis, in our case Chebyshev polynomials. This is equivalent to giving the function in a discrete number of $N$ collocation grid points. In this representation derivatives of $f(x)$ can be approximated on the grid points as 
\begin{equation}
    {\bm f}' \approx {\bm D} \cdot {\bm f}
\end{equation}
where $(\cdot)$ denotes matrix multiplication, and ${\bm f}$ is a column vector with the values of $f(x)$ on the $N$ grid points. The differentiation matrix, $\bm D$, is an $N\times N$ matrix, corresponding to the derivative operator discretized in the spectral grid. A detailed derivation of the elements of ${\bm D}$ can be found in \cite{VoigtSpectral}. Using the differentiation matrix, it is then possible to write any differential equation in discrete form, replacing the derivative operators by ${\bm D}$. Multiplicative operators (multiplication by some grid function $f$) are just matrices with the grid function values along the diagonal, which we will denote with a bar on top, as $\bm{\bar f}$. In addition, \acrshort{bcs} can be imposed by replacing the equation, at the corresponding boundary point, by the equation describing such boundary condition (see example below). Systems of $S$ equations and $S$ functions are constructed by block-wise stacking of their discretized matrices, in order to form an $S\,N \times S\,N$ matrix. Then, the whole system of equations should be written in the form of a generalized eigenvalue problem
\begin{equation}
\begin{split}
    {\bm A} \cdot {\bm \eta} = \sigma {\bm B} \cdot {\bm \eta}\, ,
\label{eqn:generalized_eigenvalue_problem}
\end{split}
\end{equation}
where ${\bm \eta}$ is a column vector of length $S \, N$ with the values of the eigenfunctions on the grid, concatenated one after the other. $\sigma$ is the (possibly complex) eigenvalue, and ${\bm A, \bm B}$ are discretized differential operators. Such a system can be solved very efficiently with standard linear algebra libraries. 

As an example, we will now show the detailed construction for the discretized version of the $2\times 2$ system of equations (presented in Section \ref{subsec:zeroVel}) for the case with constant background quantities ($p'=0, \, \rho'=0, \, e'=0$), $\alpha=\psi=1$ and $c^2_s = 1$. Here, we impose the simple \acrshort{bc} $\eta_2(R)=0$, at the outer boundary $r=R$, as an example.
The equations are, in this case,
\begin{equation}
    \left[\begin{array}{cc}
    0 & 0 \\
    -2-r \partial_r & l(l+1)
    \end{array}\right]\, 
    \left[\begin{array}{cc}
    \eta_1 \\
    \eta_2
    \end{array}\right]\, = \sigma^2
    \left[\begin{array}{cc}
    1 & -1-r\partial_r \\
    0 & r^2
    \end{array}\right]\, 
    \left[\begin{array}{cc}
    \eta_1 \\
    \eta_2
    \end{array}\right]\, \nonumber
\end{equation}
We have two equations and two unknown functions. Therefore, we can write the problem in the form (\ref{eqn:generalized_eigenvalue_problem}), with ${\bm \eta}$ as the column vector of length $2\,N$
\begin{equation}
\begin{split}
    {\bm \eta} = \left[\begin{array}{c}
    \bm{\eta_1}\\ \\
    \bm{\eta_2} 
    \end{array}\right]\, .
\end{split}
\end{equation}
The discretized matrices ${\bm A}$ and ${\bm B}$ for the differential equations will have size $2\,N \times 2\,N$. They have the form
\begin{equation}
\begin{split}
    \bm{A} = \left[\begin{array}{cc}
    \bm{0} \quad & \bm{0} \\ \\
    -2 \bm{I} - \bm{ \bar r} \cdot \bm{D} \quad & l(l+1) \bm{I}
    \end{array}\right]\, ,
\end{split}
\end{equation}
\begin{equation}
\begin{split}
    \bm{B} = \left[\begin{array}{cc}
    \bm{I} & \quad - \bm{I} - \bm{\bar r} \cdot {\bm D} \\ \\
    \bm{0} & \quad \bm{\bar r^2}
    \end{array}\right]\, ,
\end{split}
\end{equation}
where $\bm{D}$ is the differentiation matrix defined above, ${\bm I}$ is the identity matrix and ${\bm 0}$ is the zero matrix. The \acrshort{bc} $\eta_2 (R) = 0$ is enforced
by replacing the last row with
\begin{equation}
    A_{2N, \, n} = \delta_{2N,\, n}, \, \qquad
    B_{2N, \, n} = 0,
\end{equation}
for $n = 0,\, \dots,\, 2N$. More complicated systems of equations or \acrshort{bcs} can be implemented following an analogous procedure. The only restriction is that the \acrshort{bcs} have to be written in a shape compatible with Eq.~\eqref{eqn:generalized_eigenvalue_problem}.

When a large number of points is needed, Chebyshev spectral grids can be extended to more than one domain. For numerical reasons, the use of more than 128 points per domain is generally not recommended,
so multi-domain grids are a good way to extend the number of points beyond this limit. Additionally, non-smooth functions are not handled well by spectral grids, so the boundaries between neighboring domains can be chosen to coincide with discontinuities of the interpolated function or its derivatives. A typical case is the presence of piecewise-defined functions. The procedure to compute the matrix $\bm D$ for the case of evenly distributed multi-domain spectral grids is described in many textbooks, e.g. \cite{ThefethenSpectral}.

The numerical solution of the discretized system will often generate spurious or unphysical modes, which have to be filtered out. This can be done by comparing the results under different number of grid points, as spurious modes tend to vary strongly with the resolution. Additionally, spurious eigenfunctions usually have a strong content of high frequencies, especially the Nyquist frequency, which is also a good diagnostic criterion.

In practice, we follow the procedure described below. First, we implement two different resolutions. In our standard setup, we compare results obtained using a grid with 2 domains of 128 points each to those from a grid with 3 domains of 64 points each.  
For resolution studies, we rescaled both resolutions using the same factor. When quoting the numerical resolution, we always refer to the highest of the two. We discard eigenvalues as spurious, if the eigenfrequency or the eigenfunction differ significantly for the two grids. First, we apply the criterion of the eigenfrequency discarding modes whose relative difference is larger than $10^{-2}$. In a second step, we compare eigenfunctions by computing the mean square difference and discard those with values larger than $10^{-1}$.
Furthermore, it has been observed that some of the nonphysical modes are very oscillatory. Therefore, as a third criterion, the modes that have a
number of nodes close to the Nyquist frequency are discarded (in practice if either the real part of $\eta_1$ or $\eta_2$ have more than $100$ nodes).


\section{Results}
\label{sec:Results}

In this section, we present the results of the numerical implementation of the system of equations describing the oscillation modes for three distinct cases. Two of them are test cases of the general equations presented in this work and one is a simple classical problem with analytical solution. The idea is to assess and test both the performance of the equations and the robustness of the code as we add progressively more complexity. Thus, first, we consider a test case to examine the two families of modes; p and g-modes, as well as their interaction in the absence of advective velocity. There, we recover also the case of $\mathcal{N}^2=0$, which has an analytical solution. That allows us to perform convergence tests for our code. Moving on, we consider the case of plane waves in an one-dimensional classical flow with \acrshort{rh} \acrshort{bcs}, where we allow for non-zero velocity. This simple case with analytical solution, allows to understand solutions with an accretion flow. In that way, we can "isolate" the effect of the velocity and the \acrshort{bcs} on the modes. Lastly, we examine the 
general case with an accretion flow, but with $\mathcal{N}^2=0$.
In all cases we focus in $l=2$ modes, since they are the most relevant for \acrshort{gw} observations. However, modes with other $l$ could be computed with the same code as well.

\subsection{Test case 1: $p$ and $g$-modes}
\label{subsec:TestCase1}
%
Our starting point is the extension of the test case presented in the Appendix of \cite{Torres_et_al_I}, where the authors consider a static sphere of radius unity ($r \in [0,1]$), density and sound speed. In that case, the solutions are the spherical Bessel functions of the first kind. Since the fluid variables are constant, that means their derivatives will be zero. From the definition of the Brunt–V\"{a}is\"{a}l\"{a} frequency $\mathcal{N}^2$, Eq. (\ref{eq:Brunt_V_freq}), it is easily deduced that, in that case, $\mathcal{N}^2 = 0$. Therefore, the only modes present were p-modes. Here, we generalize that case by adding a buoyant region ($\mathcal{N}^2>0$) for $ r \gtrsim r_{min}$. An additional difference is that we adopt a general relativistic framework, albeit in a simplified case. As a result, additionally to p-modes, the system also supports g-modes, and both families of modes interact with each other. 

We consider ideal gas \acrshort{eos} 
\begin{equation}
p =\rho\epsilon(\Gamma_1 -1). \label{eq:eos}
\end{equation}
The background is such that i) the sound speed, $c_s^2$, and the adiabatic index are constant, ii) the spatial three-metric is flat, $\psi=1$, but $\alpha$ is general, iii) the total gravitational mass of the system is $M$, iv) the outer boundary is located at $r=1$, and v) the Brunt-Väisälä frequency is
\begin{equation}\label{eq:test1_N2}
    \mathcal{N}^2 = \mathcal{N}^2_0 \frac{1}{4} \left\{ \tanh{\left[10(r-r_{\rm min})\right]} + 1\right\}^2 \; ,
\end{equation}
where $\mathcal{N}^2_0$ is a parameter of the model. The function corresponds to a smooth step between zero, at the center, and its maximum, $\mathcal{N}^2_0$, at $r=1$. We choose $r_{\rm min}=1/2$, which sets the radial scale at which this transition happens.

Since the velocity is zero
 then the hydrostatic equilibrium holds and according to Eq. (\ref{eq:momentumBG}), $\mathcal{G}_{\alpha}=\mathcal{G_P}$. Considering the adiabatic condition, Eq. (\ref{eq:adiabaticity}), and the definition of the Brunt–V\"{a}is\"{a}l\"{a} frequency, Eq. (\ref{eq:Brunt_V_freq}), one can find that
\begin{equation}\label{eq:p_g_modes_drho_rho}
    \alpha^\prime = \pm R_{\alpha}^{-1}  ,
\end{equation}
where
\begin{equation}\label{eq:test1_Rg}
    R_{\alpha}^{-1} = \sqrt{\frac{\mathcal{N}^2 c_s^2}{\Gamma_1-1-c_s^2}} \; .
\end{equation}

The lapse function can be calculated by integrating numerically Eq. (\ref{eq:p_g_modes_drho_rho}) for the buoyant region and choosing the integration constant so that at the outer boundary the metric matches the exterior Schwarzschild metric, which in isotropic coordinates is 
\begin{equation}
\alpha_R =\alpha(r=1)=\frac{1 - M/2}{1 + M/2},
\end{equation} 
Once the lapse function is known, then we integrate numerically the hydrostatic equilibrium equation, Eq. (\ref{eq:momentumBG}), which, combined with Eq.~\eqref{eq:adiabaticity} reads 
\begin{equation}
\frac{p'}{p} = -\frac{\Gamma_1}{c_s^2}\frac{\alpha'}{\alpha},
\end{equation}
to obtain the pressure, $p$, up to an integration constant $\mathcal{C}$, i.e. we can write our pressure as $p = \mathcal{C} \tilde p$, being $\tilde p$ the result of our numerical integration.
Due to Eq. (\ref{eq:adiabaticity}), the density will be
\begin{equation}\label{eq:test1_rhoC}
    \rho = \mathcal{C} \tilde p \frac{\Gamma_1}{c_s^2}\left( 1 - \frac{c_s^2}{\Gamma_1 - 1}  \right) ,
\end{equation}
where the enthalpy has been substituted by $h=\frac{\Gamma_1 - 1}{\Gamma_1 - 1 -c_s^2}$.
To fix the value of the integration constant, one needs to integrate over the density to calculate the total mass,
\begin{equation}\label{eq:total_M}
    M = \int 4\pi \rho r^2 dr,
\end{equation}
which is proportional to $\mathcal{C}$. 

We investigate three different test models, two resembling neutron stars (NS1 and NS2) and one resembling a white dwarf (WD). The parameters for the different models can be found in Table \ref{tab:test1_models}.
Note that the sound speed is limited by Eq. (\ref{eq:test1_Rg}) to be $c_s^2 < \Gamma_1 - 1$.\footnote{The same limit is obtained from the ideal gas \acrshort{eos} for the case $\epsilon >>1$. }
Their characteristics are presented in Table \ref{tab:test1_models}. Since in our units the outer radius is located at $r=R=1$, we use the radius in km in the table to rescale the units of all other quantities accordingly.

\begin{table}[h]
    \centering
    \begin{tabular}{cccccc}
    \hline\hline
        model & M $(M_\odot)$ & $R $(km) & $\Gamma_1$ & $c_s^2$ & $\alpha_R$\\  \hline
      NS1   & $1.4$   & 10  & 2 & 0.1 & 0.81\\ 
      NS2   &  2.0    & 10  & 2  & 0.1 & 0.74\\ 
      WD    &  1.2    & $5 \cdot 10^3 $ & 4/3 & 0.1 & 0.84 \\
      \hline\hline
    \end{tabular}
    \caption{\justifying The parameters for the three test models used in the test case 1. We also show the value of the lapse at the outer radius.}
    \label{tab:test1_models}
\end{table}

\begin{figure}[h]
\begin{center}
\includegraphics[width=0.5\textwidth]{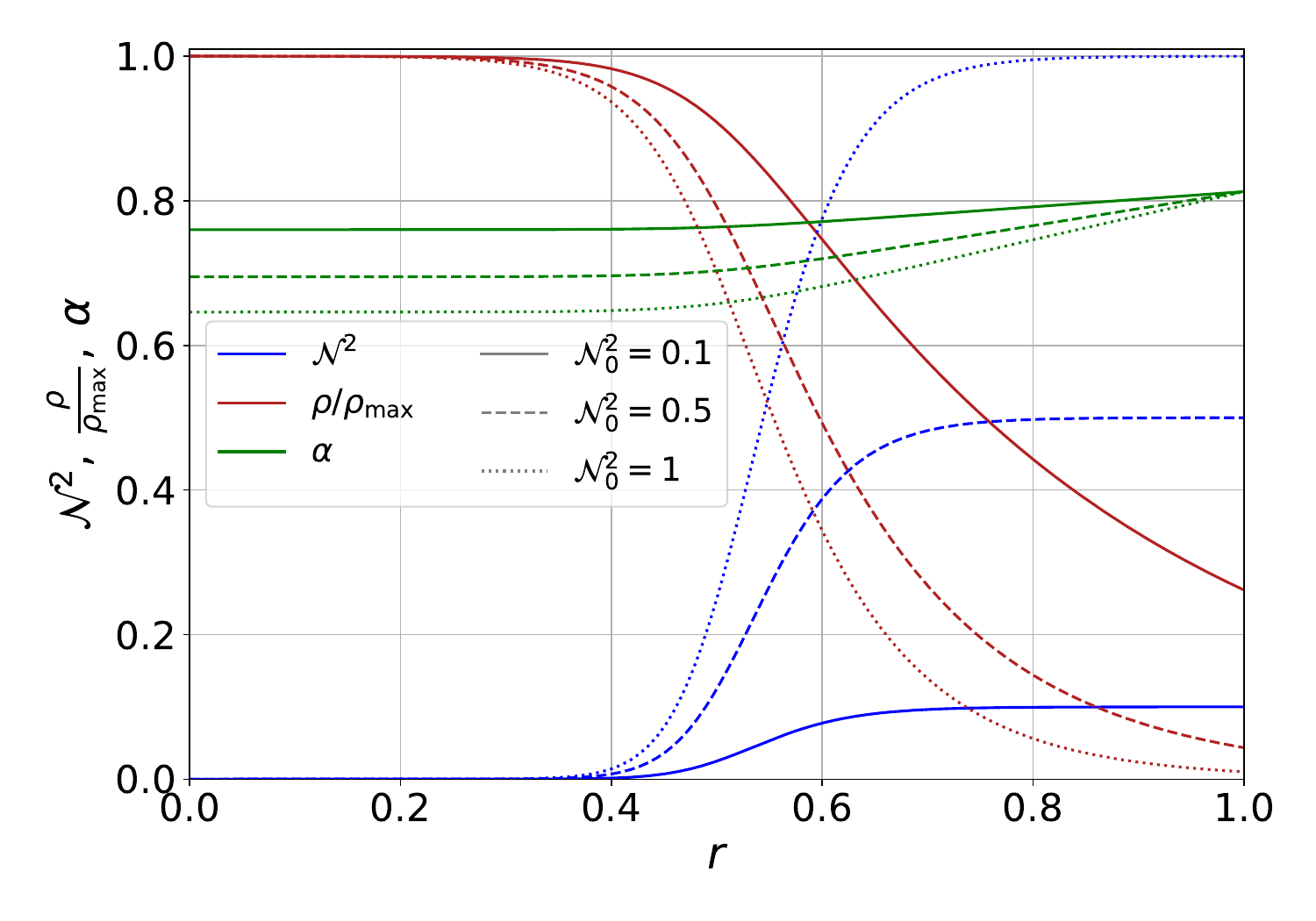}
\caption{\justifying The profiles of the lapse function, $\alpha$, density, $\rho$, and the Brunt–V\"{a}is\"{a}l\"{a} frequency, $\mathcal{N}^2$, with respect to $r$ for $NS1$. For larger values of $\mathcal{N}^2_0$, the density gradient becomes sharper. The density profile resembles the one of a neutron star, decreasing outwards.
}
\label{fig:N2_rho_profiles}
\end{center}
\end{figure}

The radial profiles of the lapse function, density and Brunt–V\"{a}is\"{a}l\"{a} frequency for the model NS1 are shown in Fig. (\ref{fig:N2_rho_profiles}).

Since there is no accretion flow in this case, we cannot apply the \acrshort{rh} conditions at the outer radius as \acrshort{bcs} for the perturbations. Instead, we impose the following \acrshort{bc}
\begin{equation}\label{eq:test1_BC}
    \eta_2(r=1)= 0  \; .
\end{equation}
For the particular case of $\mathcal{N}^2_0 = 0$, the resulting test is similar to that presented in the Appendix A of \cite{Torres_et_al_I}, but with different \acrshort{bcs}. In this case, it is possible to compute the analytical solution: 
\begin{eqnarray}\label{eq:test1_etas_Bessel}
    \eta_1 &=& \eta_0 \partial_r\left[ j_l \left( r\frac{\sigma_n}{\alpha_R c_s} \right) \right] , \\
    \eta_2 &=& \frac{\eta_0}{r} j_l \left( r\frac{\sigma_n}{\alpha_R c_s} \right), 
\end{eqnarray}
where $j_l$ is the spherical Bessel function of the first kind, $\eta_0$ a constant fixing the amplitude of the mode and $\sigma_n$ the eigenvalue of the $n$-th mode, that can be easily computed imposing Eq.~\eqref{eq:test1_BC}.

Moving on to the implementation of the test case, we would like to point out that since the velocity vanishes, then it is possible to use both systems of equations, namely the $5\times 5$ and the $2\times 2$ (presented in \ref{subsec:PerturbationEquations}). Both of them have been tested and it has been found that the relative difference between the numerical solutions (frequencies) obtained with the two systems for all three models and all values of $\mathcal{N}^2$ does not exceed $0.7 \%$, while the minimum value is of the order of $10^{-11} \%$.

\begin{figure}
\begin{center}
\includegraphics[width=0.5\textwidth]{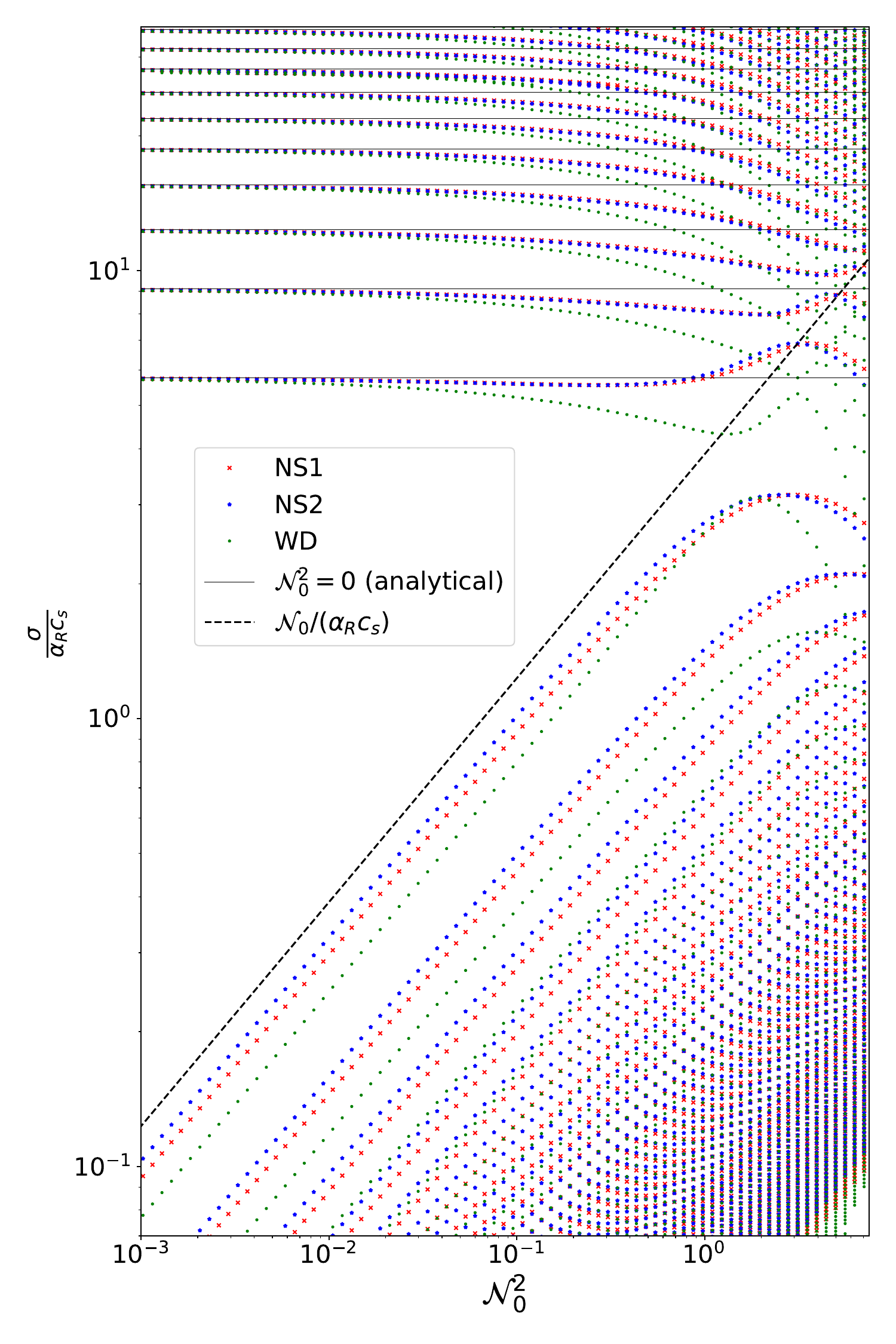}
\caption{\justifying Eigenfrequencies with respect to the value of $\mathcal{N}_0^2$ are plotted in logarithmic scale for three different models NS1, NS2 and WD. The first ten analytical solutions for $\mathcal{N}^2_0 = 0$ are shown as black solid lines. The two families of modes, p- and g-modes, seem to be naturally separated for small values of $\mathcal{N}^2_0$, while for higher values, there appear avoided crossings. The dashed black line represents the frequencies $\mathcal{N}_0/(\alpha c_s)$. The g-modes scale with the the Brunt–V\"{a}is\"{a}l\"{a} frequency, as expected theoretically.
}
\label{fig:N2_modes}
\end{center}
\end{figure}

\begin{figure*}
        \centering
            \includegraphics[width=0.47\textwidth]{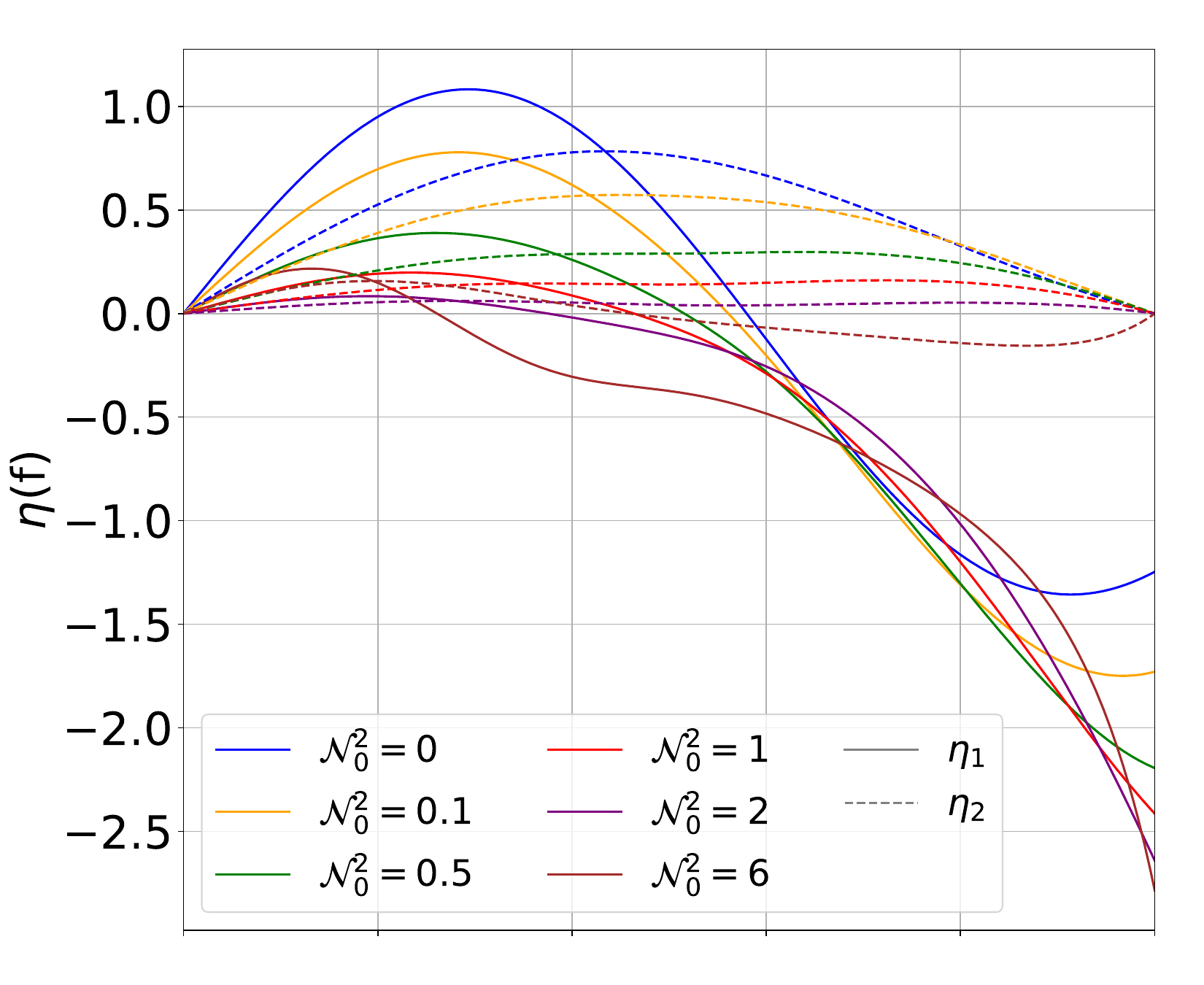}
            $\quad \;\; $
            \includegraphics[width=0.47\textwidth]{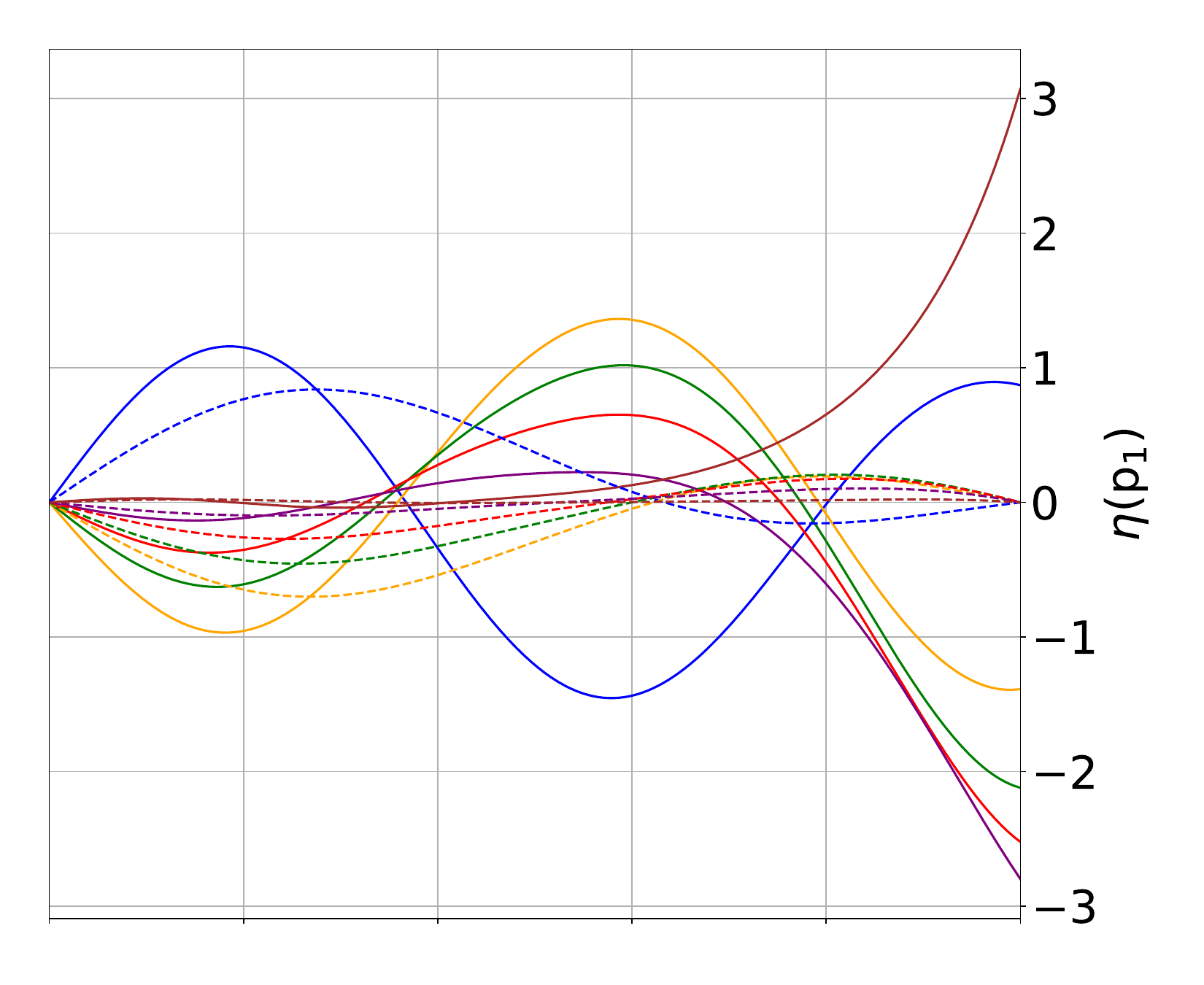}
            \\
            \includegraphics[width=0.49\textwidth]{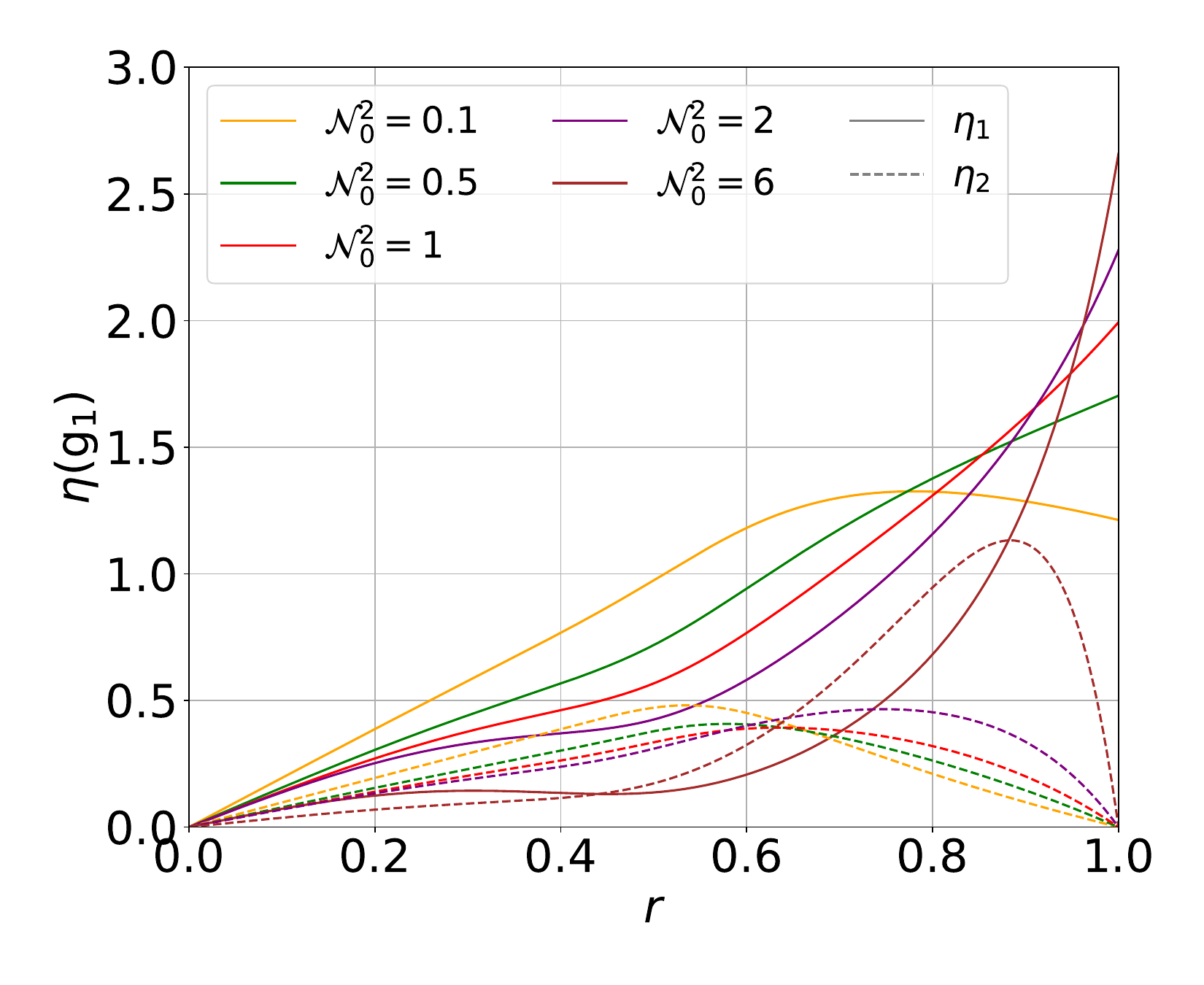}
            \includegraphics[width=0.48\textwidth]{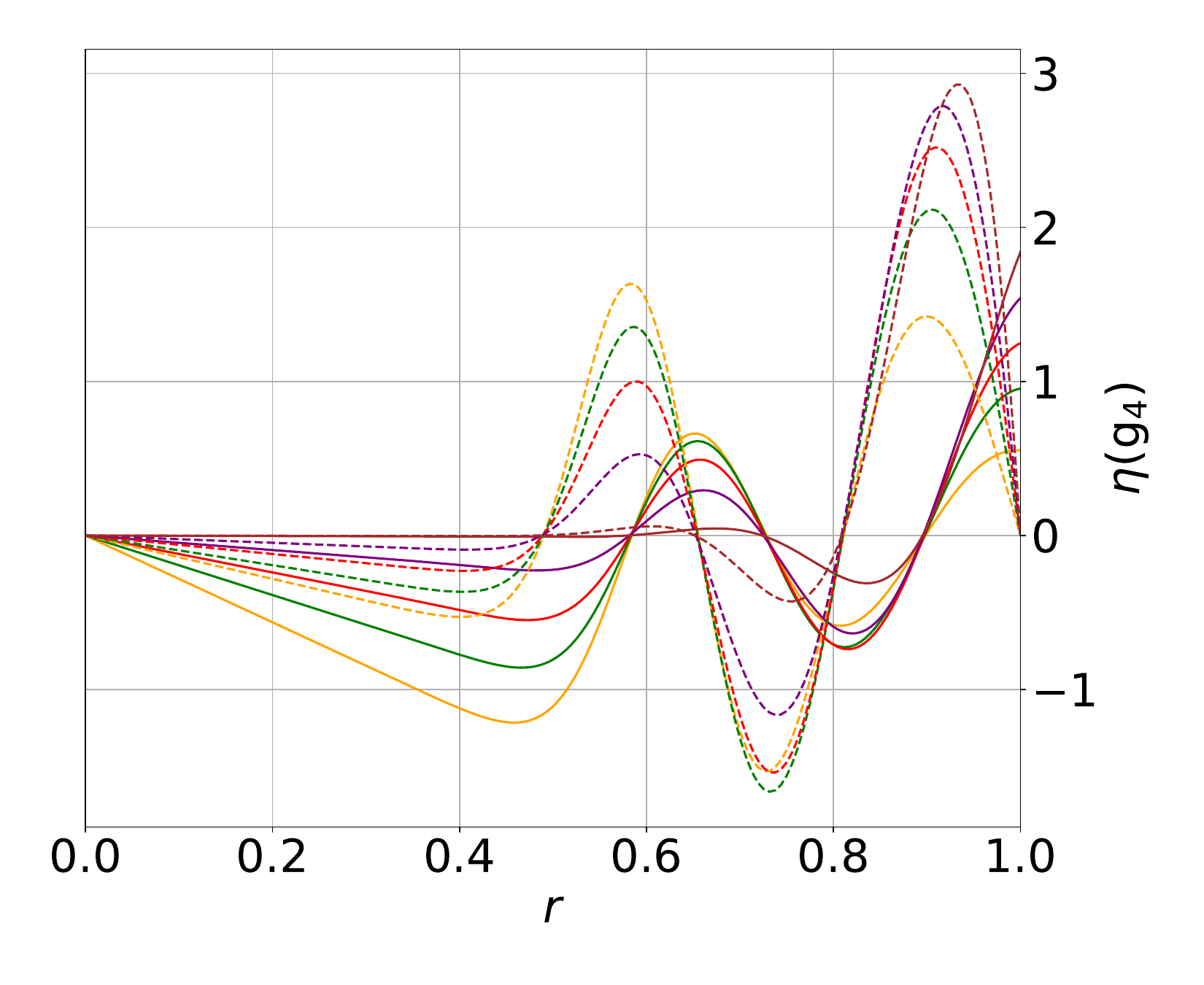}
        \caption{\justifying Radial profiles of the eigenfuctions of the two first p-modes (f and p$_1$) and two g-modes (g$_1$ and g$_4$) are plotted for different values of $\mathcal{N}^2_0$. As $\mathcal{N}^2_0$ increases the amplitude of the eigenfunctions is decreasing, while close to the avoided crossing the eigenfunctions are deformed. Although their shape changes, their number of nodes remains constant. This could be evidence that we are tracing the modes correctly after their interaction.} 
        \label{fig:test1_eigenfunctions}
    \end{figure*}

Figure \ref{fig:N2_modes} shows the frequencies for different values of $\mathcal{N}^2_0$, with constant $c_s^2=0.1$, for three different models. It is easily distinguishable that there are two families of modes for each type of star. When $\mathcal{N}^2=0$ we can compare the p-mode frequencies with the analytical solution (solid horizontal lines). For small values of $\mathcal{N}^2_0$, p-modes
follow closely the analytical solution,
while for larger values they start deviating. Regarding the eigenfunctions, the higher the frequency of the p-modes, the higher the number of nodes. The second family is recognised/classified as g- (gravity) modes as they scale with $\mathcal{N}^2_0$ (dashed black line). For higher values of $\mathcal{N}^2_0$ the interaction of the modes is evident as there are avoided crossings. Such abrupt changes in frequency have been known to happen when there is a continuous change of a background quantity, e.g. here the Brunt–V\"{a}is\"{a}l\"{a} frequency \cite{1983_Cox}. The behavior of the WD model is different than NS1 and NS2. While the WD model could be adequately described using Newtonian gravity, the NS1 and NS2 are sufficiently compact so that \acrshort{gr} effects are important, and this could be a source for the different behavior. Another possible explanation could be the different value of the adiabatic index.

In addition to the observed change in frequency, the variation of the Brunt–V\"{a}is\"{a}l\"{a} frequency has an impact on the shape
of the eigenfunctions $\eta_1$ and $\eta_2$. In Fig. \ref{fig:test1_eigenfunctions}, we present the eigenfunctions  $\eta_1$ and $\eta_2$ for two p- and two g-modes, normalized with the mean squared value. 
Among the values of $\mathcal{N}^2_0$ used for the plot, we show cases close to the avoided crossing.
In that manner, the effect of the interaction of the modes can be monitored through their eigenfunctions. For that reason the first modes of each family have been chosen, the f- and g$_1$- modes. We also display how modes with higher-order nodes are affected. As $\mathcal{N}_0^2$ increases, the maximum value of the eigenfunction decreases. However, as we approach the avoided crossing, the shape of the eigenfunction starts altering too. The number of nodes is conserved.

\begin{figure}[h]
\begin{center}
\includegraphics[width=0.49\textwidth]{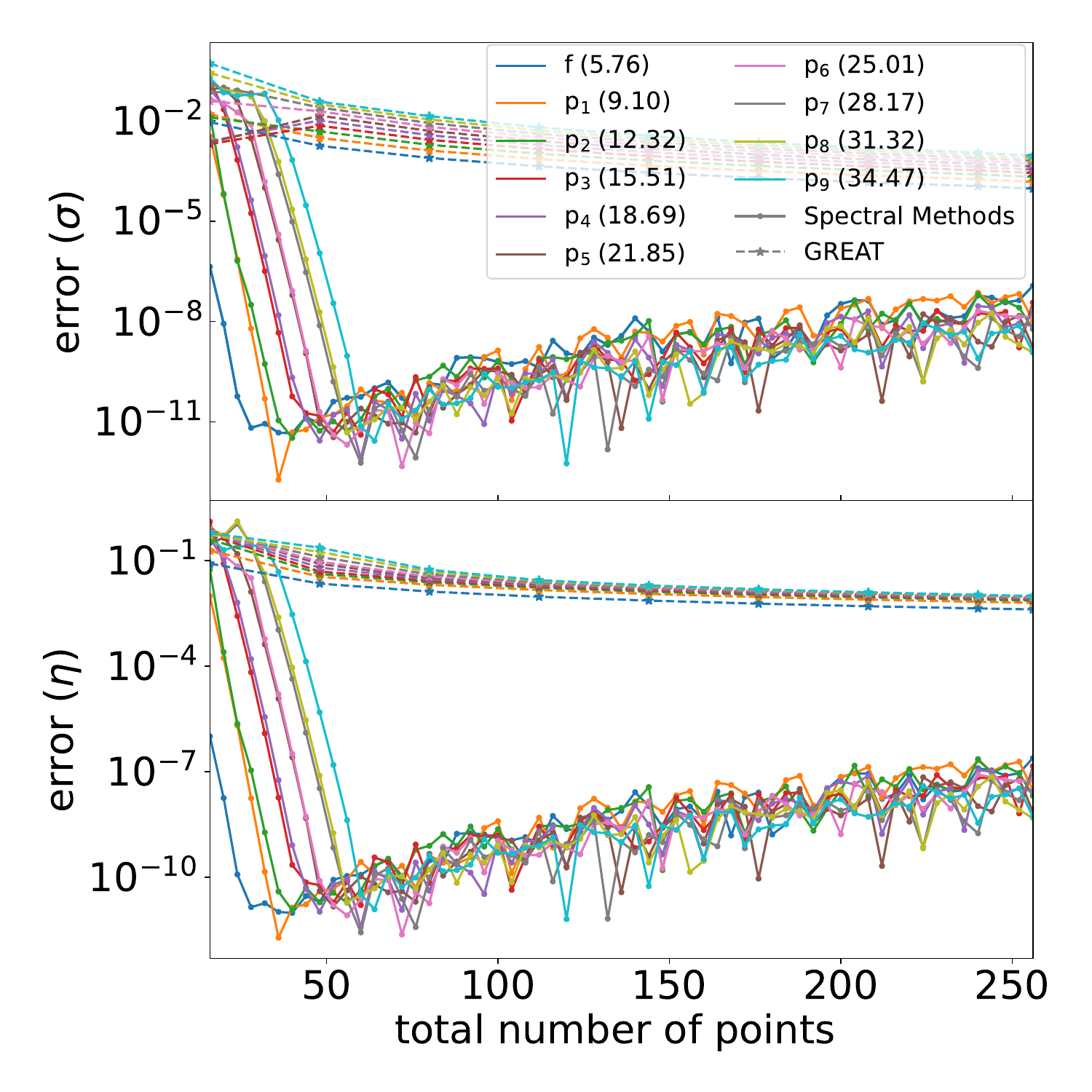}
\caption{\justifying Convergence test for the first ten p-modes in the absence of buoyancy, $\mathcal{N}^2=0$. In the upper figure, the absolute relative errors of the frequencies are plotted with respect to the total points used for all domains. In the lower one, the square root of the sums of the mean square errors of the eigenfunctions are plotted. In both graphs, the exponential convergence is evident.
In addition, the dashed lines in the plot correspond to the convergence of the code GREAT, which is of quadratic order. }
\label{fig:convergence}
\end{center}
\end{figure}

Before closing this section, we would like to present the convergence test we performed. We choose the case of $\mathcal{N}^2=0$ because we can compare directly with the analytical solution. In the upper figure of Fig. \ref{fig:convergence} the absolute relative errors of the frequencies of the first ten modes are plotted with respect to the total number of points (points per domain $\times$ domains). 
In the lower figure, the square root of the sum of the mean square errors of the two eigenfunctions is plotted with respect to the total points used. The numerical method converges exponentially until it reaches machine accuracy. It is easily noticed that the higher the mode, the more points it needs to converge. Nonetheless, all modes converge for total points more than 50. In our runs, we compare the results between 128 points per domain and 2 domains and 64 points and 3 domains. Both lie well within the convergence area. 

It is worth noticing that, beyond $30-60$ points (depending on the mode), the errors start increasing. In this regime the error is dominated by rounding error in machine arithmetic and depends on the condition number of the matrix defining the eigenvalue problem. The condition number is related to the ratio of the maximum and the minimum eigenvalues. As the number of points increases, it is possible to capture higher-order harmonics, with larger eigenvalues. As a result, the condition number, and hence the error in this regimen, increases with the number of points. Therefore, it is generally not recommended to use more than 128 points per domain.

\subsection{Plane Waves}
\label{subsec:PlaneWaves}

Before addressing the case of a spherical background with an advection flow in GR, which does not have analytical solutions, we consider the simpler case of a classical one-dimensional flow. In this case, there exists an analytical solution, that is presented in  Appendix~\ref{App:PlaneWaves}.
 Note that, as an exception to the rest of the paper, the analysis of this section is Newtonian and the coordinate system is Cartesian. 
 Here, our target is to investigate the effect of an accretion flow on the modes, in a controlled setup.

We consider the same ideal gas \acrshort{eos} as in the previous test (Eq.~\eqref{eq:eos}), and a domain $x\in[0,1]$. The background is such that i) the sound speed is constant and equal to $c_s^2=1$, ii) the adiabatic index is $\Gamma_1 = 4/3$, iii) the density is constant and has a value of $\rho=1$, iv) the interior velocity is negative, constant and subsonic, v) there is an accretion shock at $x=1$.
This solution is trivially consistent with the stationary conditions given by the momentum and continuity equations for the background. For this background it is possible to compute analytically the solution for linear perturbations moving along the fluid (see Appendix~\ref{App:PlaneWaves}).

$\upsilon_{\rm int}$ represents the value of the interior velocity at the shock and is specified by the \acrshort{rh} conditions at the location of the shock. Recall that, to have a shock, forming a supersonic area in the exterior and a subsonic one in the interior, the conditions described in \ref{sec:BoundaryConditions} must be met. Thus, the value of the velocity at the shock is not arbitrary, but is constrained to a range of possible values. For the case considered here, the range of possible values is $\upsilon_{\rm int} \in [-0.879731, -0.365076]$. In this range, the velocity profiles selected ensure that the velocity is negative and subsonic in the whole domain. A more detailed discussion can be found in section~\ref{subsec:TestCase3}.

Numerically, we solve the eigenvalue problem set by equations~\eqref{eq:pert_Euler_N_plane_1} and \eqref{eq:pert_Euler_N_plane_2} together with a set of \acrshort{bcs}. At $x=0$ we impose zero velocity perturbations, which results in 
\begin{equation}
\delta\upsilon_R^*(x=0) = 0.\label{eq:planeW_BCs_1}
\end{equation}
At $x=1$ the boundary conditions have to be consistent with the \acrshort{rh} conditions for the perturbations (see appendix \ref{App:PlaneWaves}) and result in the next condition 
\begin{equation}\label{eq:planeW_BCs_2}
  \left( c_s^2 + \upsilon_{\rm int}^2 \right) \delta \hat{\rho} + 2(\rho \upsilon )_{\rm int} \delta\upsilon^*_R = 0 \; .
\end{equation}

\begin{figure}[h]
\begin{center}
\includegraphics[width=0.49\textwidth]{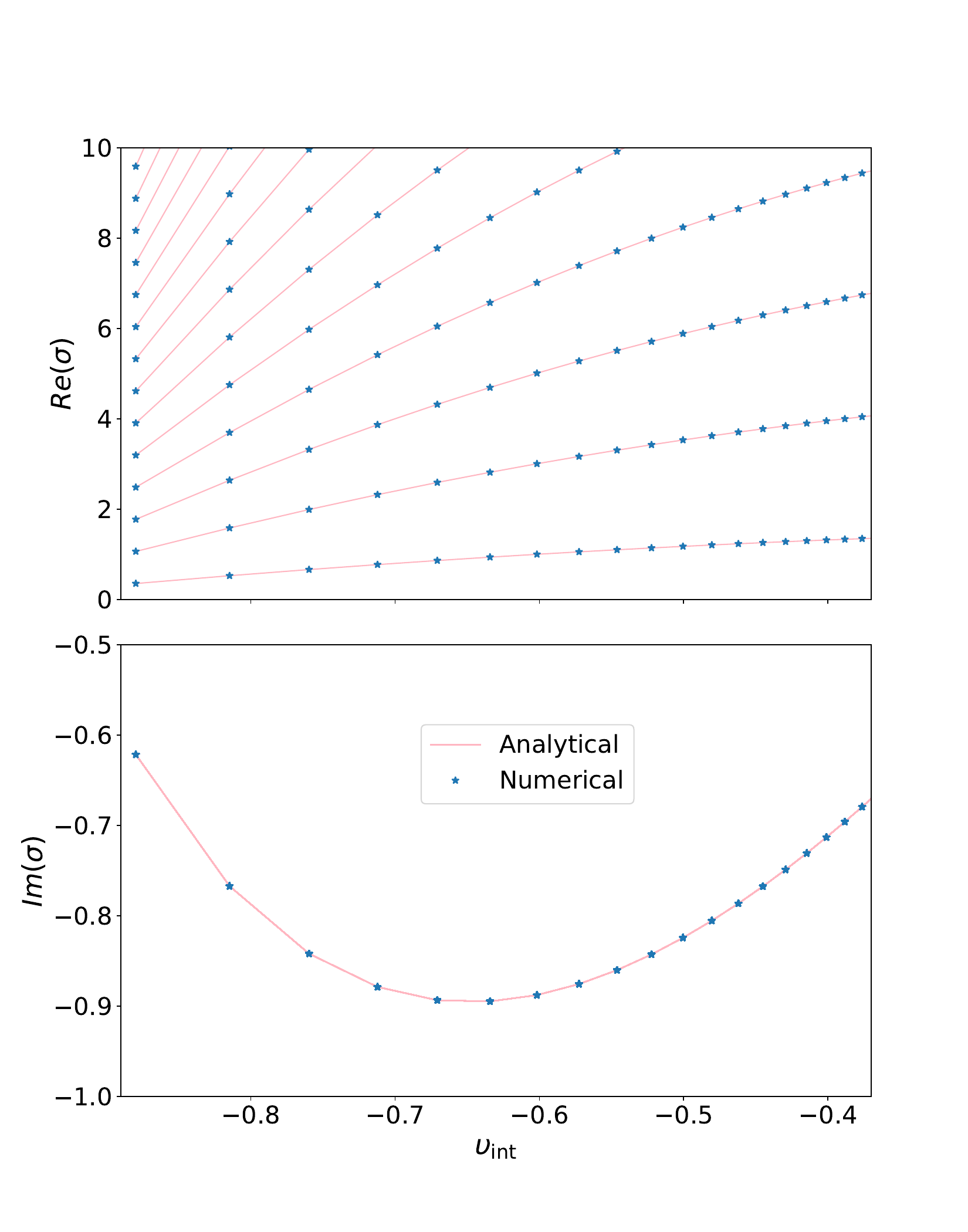}
\caption{\justifying Eigenvalues for the case of plane waves in an one-dimensional flow.
The solid lines represent the analytical solution for constant $\upsilon$, while the dots are the numerical ones. The upper/lower graph shows the real/imaginary part of the frequency as a function of the background velocity at the shock, $\upsilon_{\rm int}$. }
\label{fig:planeWaves_solutions}
\end{center}
\end{figure}

In Fig. \ref{fig:planeWaves_solutions}, we show the dependence of the numerically computed eigenvalues with the value of $\upsilon_{\rm int}$ used for the background. Additionally, we demonstrate that the analytical solution lies on top of the numerically calculated values.

For decreasing values of $|\upsilon_{\rm int}|$, the eigenvalues converge towards constant equidistant frequencies of p-modes. This is the regime in which the sound speed dominates over the velocity of the accretion flow, and the modes become progressively more similar to pure p-modes. Note that it is not possible to reach the $\upsilon_{\rm int} =0$ limit, because, for lower values than those shown in the figure, there are no accretion shock solutions (the accreting flow is too slow to form a shock).
For each value of $\upsilon_{\rm int}$, the different harmonics have the same value of the imaginary part, as it happens for the analytical case.
In addition, $Im(\sigma) < 0$ for all values of $\upsilon_{\rm int}$ and thus all modes are stable. The larger the $|Im(\sigma)|$ the faster the mode would damp. Note that an imaginary part of $\sigma$ results in an exponential time behavior of the mode. Depending on the sign, the modes will be damped (negative) or unstable (positive). The magnitude of the imaginary part is the damping or growth rate, in the respective cases. 

\subsection{Test case 2: Accretion flow}
\label{subsec:TestCase3}

Our last test case is an extension of the NS1 model of Section~\ref{subsec:TestCase1}, including an accretion flow, a standing accretion shock at the outer boundary and a modified density profile. For simplicity, we consider the case $\mathcal{N}^2=0$.
This is accomplished by using a relativistic polytropic fluid for the background with \acrshort{eos},
\begin{equation}\label{eq:polytrope}
    p = K e^{\Gamma} ,
\end{equation}
where $K$ and $\Gamma$, the polytropic index, are constants. This choice is equivalent to considering an ideal gas \acrshort{eos} and constant entropy (and composition). Integrating the first law of thermodynamics under these conditions it is possible to compute the specific internal energy resulting in
\begin{equation}
    \epsilon\equiv \left(1 -K\rho^{\Gamma - 1} \right)^{1/(1-\Gamma)} - 1.
\end{equation}
The integration constant has been fixed such that $\epsilon=0$ when $\rho=0$. In this case the adiabatic index is not constant but related to the polytropic index through $\Gamma_1 = \frac{\rho h}{e} \Gamma$.
The background is such that i) the spatial three-metric is flat, $\psi=1$, but $\alpha$ is general, ii) the total gravitational mass of the system is $M$, iii) the density profile is chosen to be steep and increasing inwards, and therefore the sound speed too, iv) the outer boundary is located at $r=1$, v) the velocity is negative and subsonic, and vi) there is an accretion shock at $x=1$. The parameters of the model can be found in Table~\ref{tab:test1_models}.

Due to the presence of an accretion shock, we impose the \acrshort{rh} conditions (at the shock location) in the background. The shock introduces naturally two regions, a supersonic (exterior) and a subsonic (interior) one, allowing for matter to flow from the exterior to the interior.
The value of $\upsilon_{\rm int}$ is constrained by the \acrshort{rh} conditions at the shock. The procedure to obtain the range of possible values consists in setting the interior thermodynamical quantities and the exterior velocity, $\upsilon_{\rm ext}$, and solving the \acrshort{rh} conditions in the background, Eqs (\ref{eq:RH1_BG}) - (\ref{eq:RH3_BG}), together with the \acrshort{eos}, to obtain the $\upsilon_{\rm int}$ and the thermodynamical quantities in the exterior. 
The exterior velocity cannot have arbitrary values as in order to form a shock, the conditions discussed in section \ref{subsec:PerturbationEquations} have to be fulfilled. The velocity will have a negative sign, denoting the in-falling matter in the direction $-r$. 
In both regions, at the shock location, $r=1$, the \acrshort{eos} is the one of a
perfect fluid with the same adiabatic index. The density value, $\rho_{int}$, is set to the same value as for model NS1 in section~\ref{subsec:TestCase1}. The value of the lapse function at $r=1$ is chosen to be the one of the Schwarzschild metric (see Table~\ref{tab:test1_models}). Finally, the sound speed at the shock location is set to be equal to $c_s^2(r=1)=0.1$.

\begin{figure}[h]
\begin{center}
\includegraphics[width=0.48\textwidth]{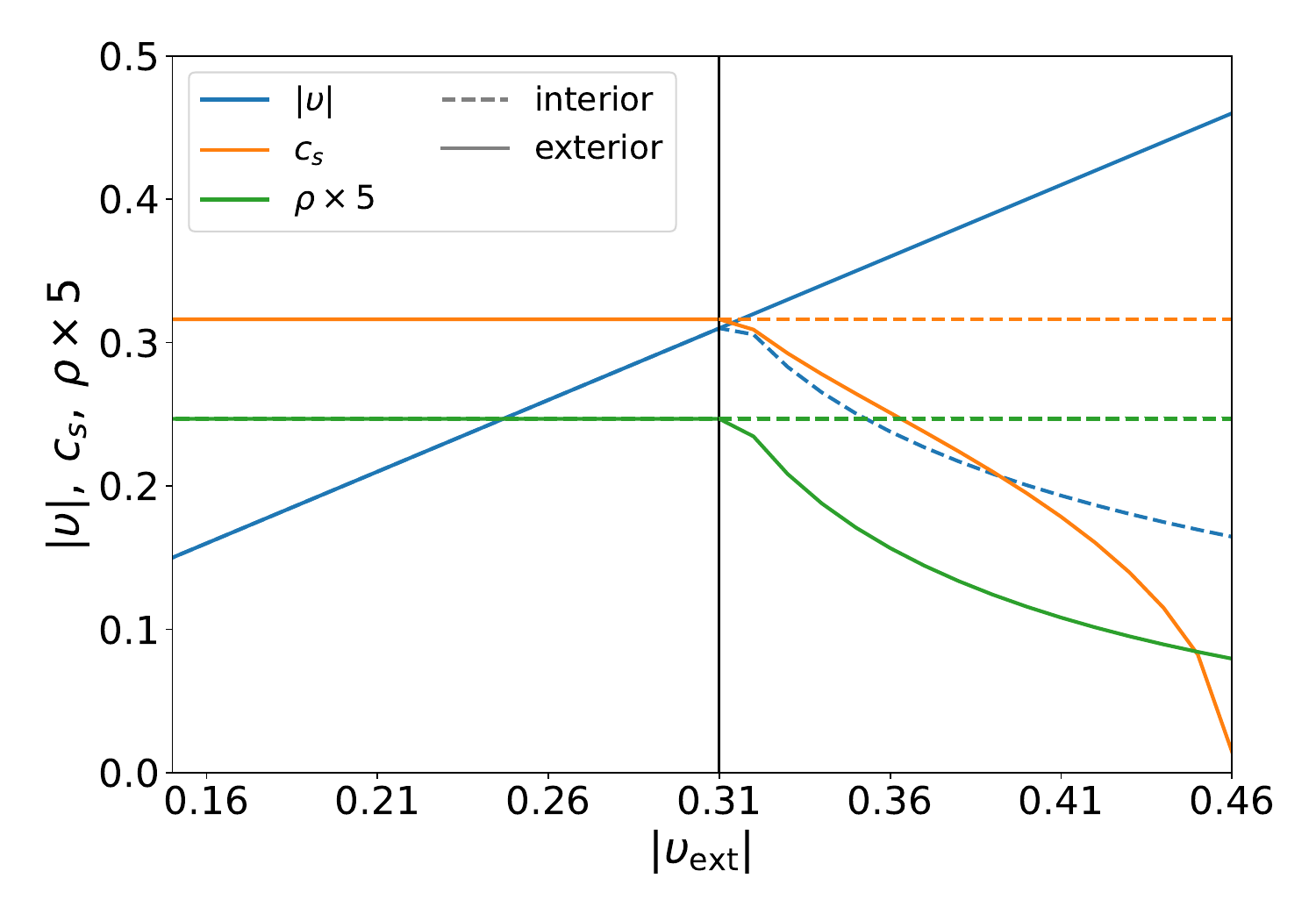}
\caption{\justifying The absolute value of the velocity, the sound speed and the density at the shock location for the interior (dashed line) and the exterior (solid line) are plotted with respect to the exterior velocity $|\upsilon_{\rm ext}|$. For the region $|\upsilon_{\rm ext}|\in [0.32,0.46]$, the conditions for 
an accretion shock
are met: $|\upsilon_{\rm ext}|>|\upsilon_{\rm int}|$, $\rho_{\rm int}>\rho_{\rm ext}$, $|\upsilon_{\rm int}|< c_{s_{\rm int}}$. For values outside this interval there is no shock formation, for which the flow follows the chosen direction (exterior$\xrightarrow{}$interior).}
\label{fig:advection_BG_quantities}
\end{center}
\end{figure}

In Fig. \ref{fig:advection_BG_quantities} the velocity, the sound speed and the density of both the interior and the exterior at the shock location are plotted with respect to the absolute exterior velocity. For values  $|\upsilon_{\rm ext}| < 0.31$, there is no shock formed and the only solution for Eqs (\ref{eq:RH1_BG})-(\ref{eq:RH3_BG}) is to have the same values for the interior and exterior. For larger values, in the range $|\upsilon_{\rm ext}| \in [0.32,0.46]$, a shock is formed and there are two distinct regions, a subsonic and a supersonic one. Note that the velocity in the exterior is always higher than that in the interior. Furthermore, the velocity in the interior is smaller than the sound speed, confirming that it is a subsonic flow. The density in the interior is higher than in the exterior. The density plot agrees qualitatively with the ones for pressure,
shown as $\mathcal{S}_{\xleftarrow{}}$ in Section 81 in \cite{1948Courant} for a classical 1D flow and with Fig. 2 in \cite{Marti_et_al_1994} for relativistic shock waves in Minkowski spacetime. 
For $|\upsilon_{\rm ext}| > 0.46$ there only two physical solutions, an inverted shock (the interior region now becomes supersonic and the exterior subsonic), or continuous flow, with no shock. Since any of both fulfill the requirements of our model, we will not consider those solutions.

For such a background, the hydrodynamics equations, \eqref{eq:continuityBG} and \eqref{eq:momentumBG}, after some calculations will result, respectively, in
\begin{equation}\label{eq:advection_continuity_BG}
    r^2 \alpha \rho W\upsilon^r = \mathfrak{C},
\end{equation}
where $ \mathfrak{C}$ is a constant, which can be calculated for the shock location, and 
\begin{equation}\label{eq:advection_momentum_BG}
    \partial_r \ln {\alpha} = -\frac{1}{1+\upsilon^2}\left( \mathcal{G_P} + \upsilon^2\frac{2}{r} \right) .
\end{equation}
We prescribe the following density profile
\begin{equation}\label{eq:advection_rho_profile}
    \rho = \frac{\rho_{int}}{r^{10}}.
\end{equation}
The steep increase in density when moving radially inwards, mimics the conditions at the \acrshort{pns} surface onto which matter is accreting. The increase of sound speed permits the reflection of sound waves traveling down. This allows the formation of standing waves, and hence oscillation modes, trapped between the shock and the density gradient. In order for this effect to appear in the presence of an accretion flow, the gradient has to be sufficiently steep. If a sufficiently small value of the exponent of $r$ is chosen, then all modes disappear, because any perturbation is just advected downwards through the inner boundary.

We solve the coupled system of Eqs. \eqref{eq:advection_continuity_BG} and \eqref{eq:advection_momentum_BG} by means of a fixed point iteration method. As initial guess we use constant lapse equal to the boundary value, $\alpha = \alpha(r=1)$. At each step of the iteration we compute the velocity from Eq. \eqref{eq:advection_continuity_BG} (remember that $\rho$ is prescribed) keeping $\alpha$ fixed. Then we integrate numerically the momentum equation \eqref{eq:advection_momentum_BG} to update the value of the lapse function, $\alpha$. We use a relaxation factor of $0.1$ in the update, and terminate the iteration when the relative change in $\alpha$ is smaller than $10^{-10}$.

At this point, allow us to comment on the choice of the radial domain for which the aforementioned analysis is taking place. The density profile prohibits the radial coordinate to start from zero, because the velocity would diverge. Therefore, on the one hand, the density profile limits the domain in use and on the other hand, there is the physical necessity that the interior velocity has to be subsonic. In other words, we need to ensure that the interior velocity is smaller than the sound speed at all times. Taking in account all the above requirements, we solve the system of equations, both the background and the perturbed ones, in the region $r\in[0.8, 1]$. The mass enclosed in this region is $1.13 M_\odot$, which is smaller than the total mass of the system, $1.4 M_\odot$, that we use to impose boundary conditions. This means that the mass difference is in the region $r<0.8$, and only contributes to the curvature but not to the eigenvalue problem itself.

The system of equations describing the perturbations of the interior is the $5\times 5$ system derived in \ref{subsec:PerturbationEquations}.
As \acrshort{bcs} we test two possibilities: i) the \acrshort{rh} conditions at $r=1$, given by Eqs (\ref{eq:rh1_GR_expanded}) and (\ref{eq:rh2_r_GRexpanded}), and ii) the condition $\eta_1(r=1)=0$.
Although the \acrshort{rh} conditions are the appropriate ones when there is an accretion shock present, they only allow for a limited set of velocity values. In contrast, for the condition of a vanishing perturbation at the outer boundary, we can also include the zero velocity limit. As inner \acrshort{bc}, we choose $\delta\upsilon^*_R (r=0.8) = 0$.

\begin{figure*}
        \centering
            \includegraphics[width=0.49\textwidth]{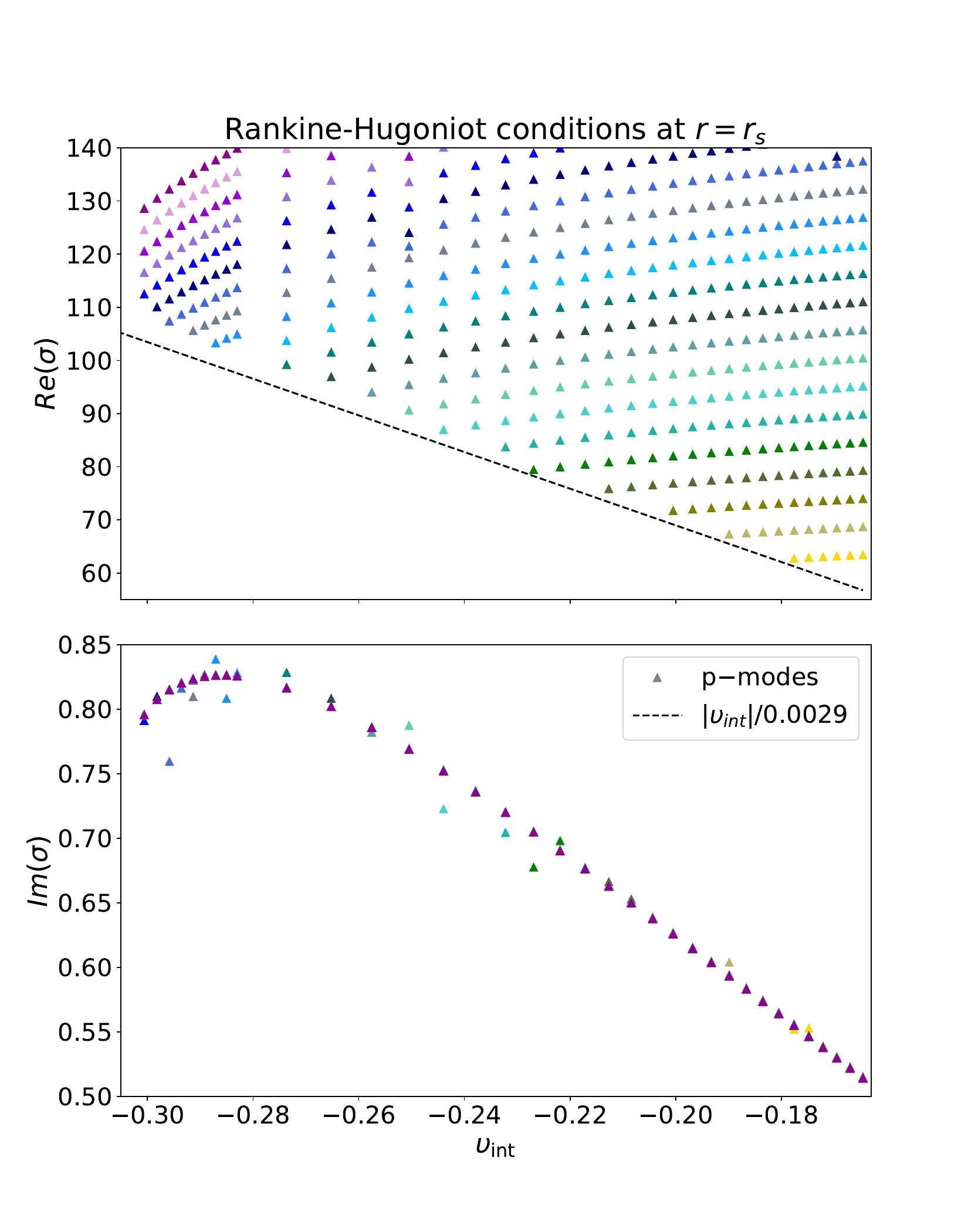}
            \includegraphics[width=0.49\textwidth]{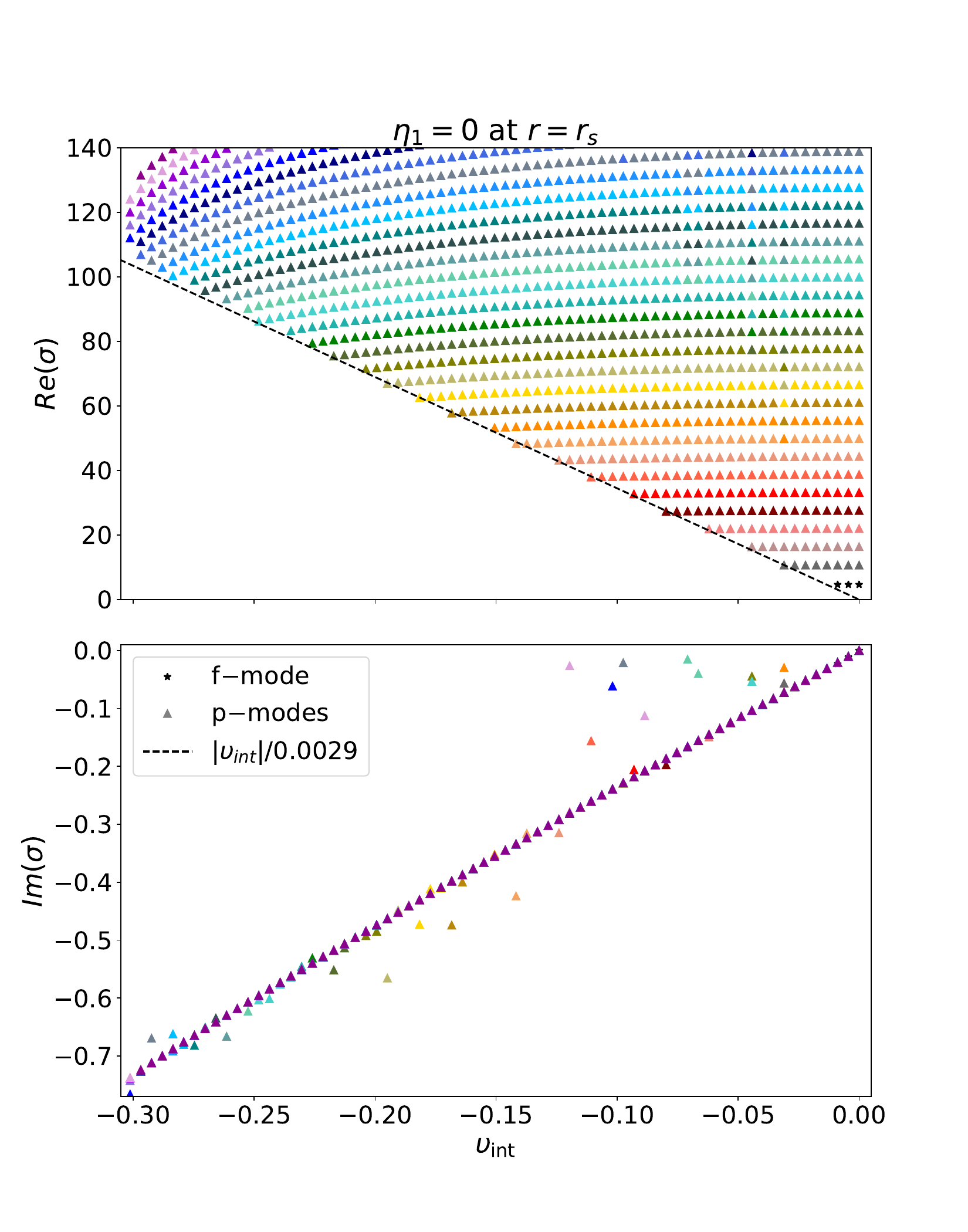}
        \caption[ The eigenfunction ]
        {\justifying The real (upper) and the imaginary (lower) parts of the frequencies are plotted with respect to the interior velocity at the shock location for two different boundary conditions (at the shock location); the \acrshort{rh} conditions (left) and $\eta_1(r=1)=0$ (right). Each colour corresponds to a mode and the scale is the same for both sets of \acrshort{bcs}. The $f-$mode is depicted with black stars, while the p-modes with triangles. The dashed black line scales with $|\upsilon_{int}|$.} 
        \label{fig:advection_profiles}
    \end{figure*}

In Fig. \ref{fig:advection_profiles} we present the computed eigenfrequencies using the \acrshort{rh} conditions at the shock location and alternatively the condition $\eta_1(r=1)=0$. The behavior of the eigenfrequencies with respect to $\upsilon_{\rm int}$, resembles the case of the plane waves of section~\ref{subsec:PlaneWaves} for both sets of \acrshort{bcs}. Specifically, we observe modes with decreasing real part of frequencies as $|\upsilon_{int}|$ increases. In addition, for each value of the velocity all the modes have the same imaginary part. However, we notice that for the \acrshort{rh} conditions the imaginary part of the frequencies is positive (unstable modes) and growing with higher velocities, while for $\eta_1(r=1)=0$, the imaginary part is always negative (stable damped modes) and decreasing as $|\upsilon_{int}|$ increases.

Even more striking in Fig. \ref{fig:advection_profiles} is the evanescent region, where there are no modes present. For each value of the interior velocity there is a cut-off frequency, below which all frequencies are extinct. We observe that the cut-off frequency scales with the absolute value of the interior velocity. In practice this means that only harmonics above certain value exist, while low order harmonics and the fundamental mode do not exist for most of the cases. It is thus evident that the cut-off frequency depends on the wavelength of the mode.

Our interpretation of this cut-off frequency is based on the concept of acoustic impedance. The transmission and reflection coefficients of waves propagating in a medium with increasing impedance (as it is the case of increasing density and sound speed) can be computed in the small wavelength approximation. However, if the wavelength becomes larger and comparable to the density scale height, the reflection coefficient decreases.
As a consequence, the modes with small wavelengths will be reflected and thus create standing waves in the region between the gradient and the shock. On the contrary, when the wavelengths are comparable to the scale of the gradient, then the modes escape and they are not reflected. As the velocity increases, perturbations advected down with the flow, have an increasing difficulty in being reflected back upwards, thus increasing the cut-off wavelength, and consequently the cut-off frequency. When the velocity reaches the maximum allowed by the \acrshort{rh} conditions value then all modes escape and there are no standing waves. For the case of the \acrshort{rh} conditions we can observe the $p_{11}$-mode as the lowest mode for the smallest value of $|\upsilon_{int}|$. For the condition of $\eta_1(r=1)=0$, we start from zero velocity, where all modes are present, including the fundamental, and disappear progressively as the magnitude of the velocity increases. Note that the scale of the colours for both cases are the same and thus the yellow points represent the 11th mode in both plots. In Fig. \ref{fig:advection_etas} we show the eigenfunctions of $\eta_1$ and $\eta_2$ for each set of \acrshort{bcs}.

Before moving on to the description of the eigenfunctions, we would like to point out the existence of spurious modes in both graphs of Fig. \ref{fig:advection_profiles}. The spurious modes can be easily detected as the ones not following the general trend. In the case of $\eta_1(r=1)=0$, they are more clear. We can identify them as spurious because we plotted their radial profiles and they are very oscillatory or located only in one part of the domain. In addition, if we decrease the threshold we use for the eigenfrequencies and/or the eigenfunctions then these spurious modes disappear. However, we avoid decreasing the threshold systematically as then the higher order modes for higher values of $|\upsilon_{int}|$ disappear (the upper-left corner of the plot). In addition, some of the modes close to the cut-off frequency might become extinct.
The reason for that is, as we show below, that the eigenfunctions of the mode of the cut-off frequency become very oscillatory at the last value of the velocity at which they appear. The last value of $|\upsilon_{int}|$ for which a mode appears is also the maximum velocity that this mode is being reflected, as we loose modes going to higher magnitude of the velocity. A careful reader might wonder at this point if by increasing the thresholds of the frequencies and the eigenfunctions, the lower order modes will be present and thus the cut-off is artificial. The answer is that this is not the case. the cut-off frequencies will always exist and the slope will be almost the same, but we might miss (or gain) some of the points closer to the dashed line.

\begin{figure}[h]
\begin{center}
\includegraphics[width=0.49\textwidth]{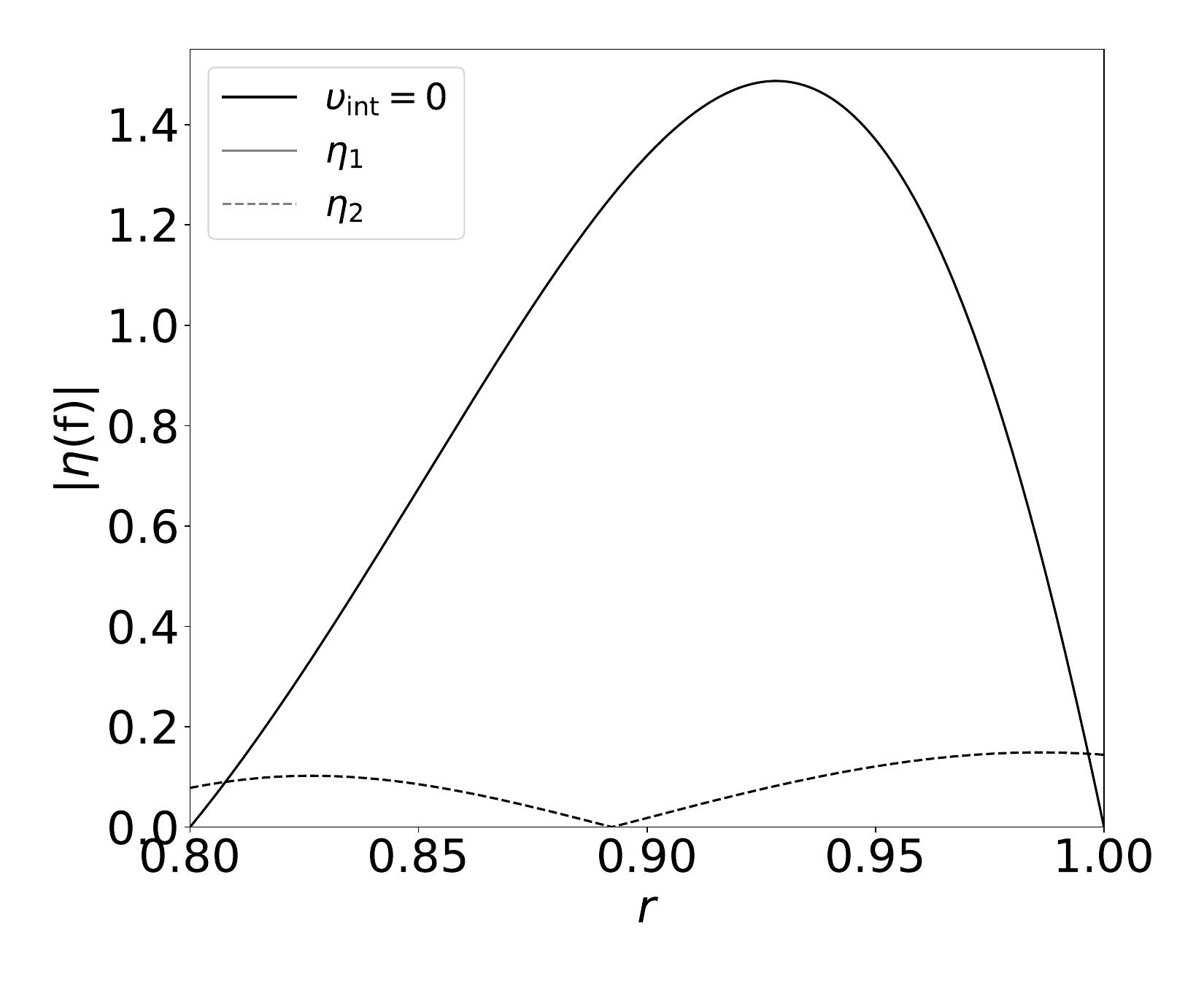}
\caption{\justifying The radial profile of the f-mode is plotted for the case of zero velocity and boundary condition at the shock: $\eta_1(r=1)=0$. The eigenfunctions $\eta_1$ (solid line) and $\eta_2$ (dashed line) are shown.}
\label{fig:advection_f_mode}
\end{center}
\end{figure}

\begin{figure*}
        \centering
            \includegraphics[width=0.49\textwidth]{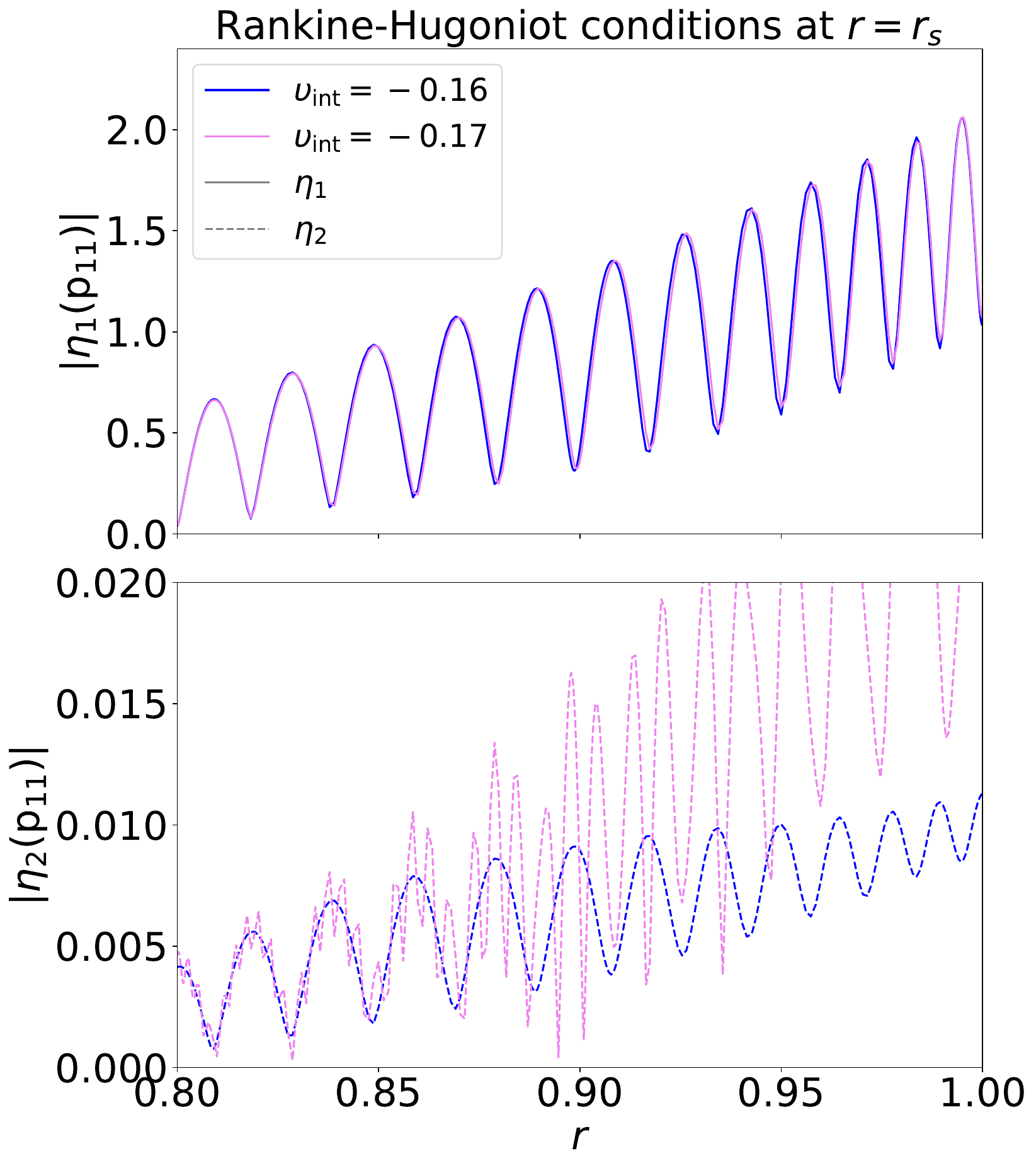}
            \includegraphics[width=0.49\textwidth]{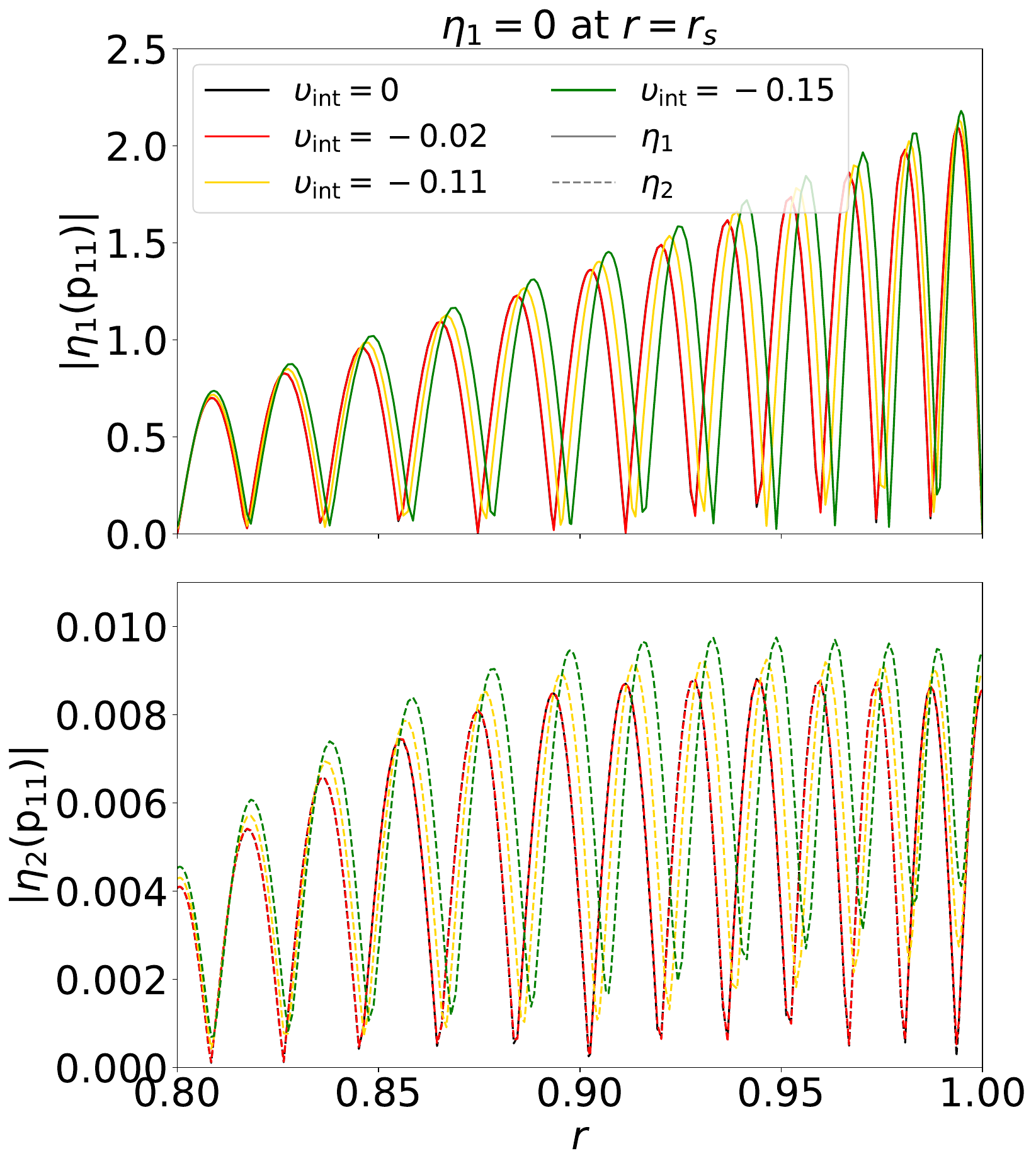}
        \caption[ The eigenfunction ]
        {\justifying The radial profile of the $p_{11}-$mode for the \acrshort{bc} at the shock location $\eta_1(r=1)=0$ (right) and the \acrshort{rh} conditions (left). The eigenfunctions $\eta_1$ (upper; solid line) and $\eta_2$ (lower; dashed line) are shown for a range of values of the interior velocity at the shock location. In black, the eigenfunctions of the zero velocity case are also depicted.} 
        \label{fig:advection_etas}
    \end{figure*}

Closing this section, we would like to show the eigenfunctions of $\eta_1$ and $\eta_2$ of the $p_{11}-$mode for both sets of \acrshort{bcs}, while for $\eta_1(r=1)=0$ we present also the ones of the $f-$mode. In Fig. \ref{fig:advection_f_mode} the eigenfunctions of $\eta_1$ and $\eta_2$ are plotted for zero interior velocity. $\eta_1$ has zero nodes for the fundamental mode, as expected analytically. In Fig. \ref{fig:advection_etas}, the eigenfunctions of the $p_{11}$-mode are plotted for different values of the interior velocity for both cases of \acrshort{bcs}. In the case of \acrshort{rh} \acrshort{bcs}, we observe that for the last value of the velocity, for which the mode exists, the eigenfunction becomes oscillatory/noisy. The closest the value of the frequency is to the cut-off frequency, the more oscillatory it becomes, probably an indication that the existence of the mode is barely possible (for that value of the velocity). Similar behavior is also exhibited for $\eta_1(r=1)=0$. Note also that the eigenfunctions in both figures are normalized using the square root of the mean of the sum of $|\eta_1| ^2$ and $|\eta_2|^2$.

\section{Conclusions and Discussion}
\label{sec:Conclusions}

In this work we have developed, for the first time, a formalism to compute the normal oscillation modes of a \acrshort{pns} with the presence of an accretion flow and surrounded by a standing accretion shock. This has applications to the \acrshort{gw} emission during \acrshort{ccsn} explosions and will have an impact on the advancement of asteroseismology in this context.
For this purpose, we have derived the linear perturbation equations in \acrshort{gr}, using the relativistic Cowling approximation and considering adiabatic perturbations. Those result in an eigenvalue problem, in the form of a set of linear differential equations, that can be solved numerically. \acrshort{bcs} are applied at the accretion shock location, consistent with the \acrshort{rh} conditions, which ensure that perturbations conserve mass, momentum and energy across the shock.
We use spectral collocation methods for the numerical discretization of the differential equations appearing in the eigenvalue problem. By doing so, the system is transformed in an algebraic eigenvalue problem for the values of the perturbation variables at the collocation points, that can be solved by means of standard and efficient linear algebra methods.
It is the first time that such methods are adopted in the context of asteroseismology of \acrshort{ccsne}.

We present three numerical experiments in order to test the numerical code and to explore the impact of all three features mentioned above; the inclusion of the accretion flow, the \acrshort{rh} conditions at the shock and the implementation of spectral methods. In the first two cases cases, we consider an ideal gas \acrshort{eos}, and constant values for the sound speed and adiabatic index, across the domain, while for the last one we study a sphere of fluid with varying sound speed and no stratification, modeled by a relativistic polytropic \acrshort{eos}. The first experiment is a numerical test of the code, similar to the test case presented in \cite{Torres_et_al_I}, but including buoyancy.
We consider a spherically symmetric fluid sphere, resembling a star, 
and no accretion flow. In addition, we allow for a buoyant region $(\mathcal{N}^2 > 0)$. In that fashion, we are able to study the two families of modes, p-modes and g-modes, as well as their interaction with respect to the value of the Brunt–V\"{a}is\"{a}l\"{a} frequency. It is worth mentioning, that the interaction of the modes is revealed monitoring their frequencies, but also their eigenfunctions. In the limit of $\mathcal{N}^2=0$, we recover a case with analytical solutions. In doing so, we are able to perform convergence tests for our code and compare with the code GREAT, previously used in the context of asteroseismology of \acrshort{ccsne}. We show that spectral methods converge exponentially, while GREAT converges quadratically.

Next, we consider the case of a classical one-dimensional perturbed flow with an accretion shock as a boundary for the incoming matter. This scenario allows us to study the effect of the accretion flow and the use of \acrshort{rh} at the shock, in a simplified scenario setup. We use constant velocity profile, which has an analytical solution.
Our numerical results are compatible with damping p-modes and are in perfect agreement with the analytical ones.

Our last test case combines the two previous ones by considering a spherically symmetric fluid sphere in \acrshort{gr} with an accretion flow and an accretion shock at the outer boundary. For simplicity, we consider the case for $\mathcal{N}^2=0$.
If the whole sphere down to $r=0$ were to be considered in the test, since the approximation of the hydrostatic equilibrium has been lifted, the background equilibrium would be non-stationary. An accreting object cannot be in stationary equilibrium, because the matter that is being accreted will increase the mass of the object continuously, changing it over time. This is indeed what happens in the \acrshort{ccsn} scenario; during the first $\sim 0.5-1$~s after bounce, the \acrshort{pns} accretes about $1$~M$_\odot$, growing in mass and becoming more compact. All previous asteroseismology studies of the scenario have assumed that the background changes in a timescale ($\sim 100$~ms) that is longer than the typical period of the modes studied ($f\sim 50 - 3000$~Hz, i.e. $T\sim0.3 - 20$~ms). This allows to consider the background as being effectively frozen for the mode computations, albeit not strictly speaking stationary.
We construct a non-stratified background configuration consistent with the continuity and the momentum equations. In order to avoid diverging values of the density or the velocity at the center, we consider a section of the radial domain by introducing an inner boundary, allowing the matter to flow out of the domain \citep[see e.g.][for a similar set up]{Guilet2010}.

The presence of the shock requires the use of the \acrshort{rh} conditions at the outer boundary. We also consider the case of a vanishing perturbation as a \acrshort{bc} at the shock location, allowing us to study the zero velocity limit. The numerically computed frequencies for both sets of outer \acrshort{bcs} exhibit some similarities with the case of one-dimensional flow. We recover the $p-$modes with decreasing (real part of the) frequencies, while the imaginary part is the same for all modes for each value of the interior velocity. For the case of $\eta_1(r=1)=0$ the modes are damped (negative imaginary part), while for the \acrshort{rh} conditions the modes are unstable, as the imaginary part is positive.

We believe that the ingredients present in the test are sufficient as a toy model for \acrshort{sasi}. In order to have \acrshort{sasi}, it is not enough to have advection. We also need to have a realistic model of a \acrshort{ns}, such that pressure waves are reflected by its surface, initiating an advective-acoustic cycle. Here, the steep gradient of the density and the velocity can act as the reflecting surface. In that fashion, the down-going advection perturbation can be reflected into an up-going sound wave, creating standing waves, when interacting with the shock. The results of this toy model suggest that \acrshort{sasi} modes are just the same family of p-modes that exist in the non-accreting limit that become unstable in the presence of an accretion flow and a standing shock. In particular, the \acrshort{bcs} at the shock seem to play an important role in the stability of the modes, since we only find unstable modes for the case in which we use consistent \acrshort{rh} conditions. However, we cannot be certain that in a more realistic scenario these are the only modes present and responsible for the \acrshort{sasi} observed in multidimensional \acrshort{ccsne} simulations. There could be, for example, a completely new family of unstable modes, with positive imaginary part. We plan to study \acrshort{sasi} extensively in the future with the code presented here, using the results of numerical simulations of \acrshort{ccsne} as a background, following a similar procedure as in \cite{Torres_et_al_I,Morozova:2018glm,Torres_et_al_II}. However, this is clearly out of the scope of this paper.

An intriguing aspect, that arises from the case of the spherical accretion flow with a standing accretion shock, is the presence of an evanescent region below a cut-off frequency. The gradient of the density forms a "wall" for the waves. When the wavelength of an ingoing acoustic wave is shorter than the scale of the gradient of the density, then it will be reflected and travel upwards. Since the exterior region is supersonic the perturbations cannot escape and they are bound to travel to the area defined by the inner and outer boundaries. In that way, we will have the formation of standing waves. In contrast, if the wavelength is larger than the aforementioned "wall", then the modes will encounter a "transparent" gradient and pass through. As the interior velocity increases, its gradient becomes steeper, and more modes disappear. The existence of an acoustic refraction cut-off frequency in the presence of an accretion flow has been studied e.g. by \cite{Foglizzo2000,Foglizzo2001}. 
The phenomenon of cut-off frequencies is well known in systems in which waves propagate in media with varying properties. Additionally to variations in the sound speed, a cut-off frequency can also appear in the presence of rotation \cite[see][for examples in accreting disks]{Nowak1991,Kato2008}, shear layers \cite{CAMPOS_KOBAYASHI_2000} or, in the case of pipes with varying geometry \citep[see e.g.][]{Campos1986, Kinsler2000, 2019_Pierce}. The analogous effect can be found for electromagnetic waves \citep[see][for examples in the case of waveguides]{2006_Okamoto, 2023_Balani} and quantum mechanics \citep[see][ for an example in the context of quantum tunneling]{1978_Bender_Orszag}.

Another topic for discussion emerging from our work is the classification of the modes. Classical asteroseismology work (see e.g. \cite{1983_Cox}) classifies the modes according to the number of radial nodes of the eigenfunctions, into two classes, p and g-modes, depending on their frequency with respect to the node-less f-mode. The work of \cite{Torres_et_al_II} has shown that this classification procedure can be problematic for complicated scenarios such as \acrshort{ccsn}, in which p and g-modes can coexist and interact at similar frequencies.
The presence of complex eigenvalues (with non-zero imaginary part) adds yet another layer of complication to the problem. As long as the spectrum of the eigenvalues is real, the eigenfunctions are also real\footnote{Note that the numerically obtained eigenfunctions 
have an imaginary part, in general, but it can be scaled away by multiplying the eigenfunction by the appropriate complex amplitude, since
it only represents a phase. In that case, the number of nodes is a well defined quantity, that
can provide some useful information about the nature of the mode.}.
However, in the presence of an accretion flow, the eigenvalues and the eigenfunctions become complex. In this case, the question of how we should count the number of nodes arises, since the real and imaginary parts of the eigenfunction could have different number of nodes. Furthermore, it is possible to multiply the eigenfunction by any complex constant, and it will still be an eigenfunction, but the number of nodes may have changed (e.g. if the real and imaginary part have different number of nodes and one multiplies by the complex number $i$, the number of nodes will swap between the real and the imaginary part).
In this work, we have labeled the modes according to their ordering in frequency and by analogy with the zero-velocity limit. This is possible because no g-modes are considered in those cases, simplifying considerably the analysis.

Spectral methods have proven to be a powerful tool for solving eigenvalues problems, here in an astrophysical scenario. Their implementation is straightforward through the use of standard libraries and the convergence of the method is exponential. These methods are also much more flexible and easier to implement, in particular regarding \acrshort{bcs}.
Although we present here a particular application, \acrshort{pns} with accretion flows, we think that these methods may open a pathway to more complex scenarios including extra physics (e.g. elasticity of the crust, non-adiabatic perturbations, core superfluidity) or dimensionality (e.g. rotation or magnetic fields). 

For the sake of simplicity, 
we have taken into account a few approximations in this work, as a first approach and in an effort of "isolating" the effect of the accretion flow. Although our analysis is performed in \acrshort{gr}, the relativistic Cowling approximation is applied. This approximation could be removed in the future by following an approach similar to \cite{Morozova:2018glm,Torres_et_al_II}. The effect of the radial shift, $\beta^r$ in the background could also be included without much difficulty.

We believe that this work offers numerous extensions in the field of asteroseismology. Spherical symmetry provides a good approximation for \acrshort{ccsne} in the scenario of a neutrino-driven explosion. If rotation and/or magnetic field are accounted for, then except for \acrshort{ccsne} in the case of magneto-rotation explosion, remnants of binary neutron star mergers can also be explored. In that event, asteroseismology could be an efficient tool to infer the properties of the star and its \acrshort{eos}, as well as to study its stability against convection.


\section{Acknowledgements}

We would like to thank Prof. J. M. Marti for fruitful discussions and suggestions. This research was initiated by DT, PCD and AT during their stay/participation at the Institute for Pure \& Applied Mathematics (IPAM), University of California Los Angeles (UCLA) at the occasion of the Mathematical and Computational Challenges in the Era of Gravitational-Wave Astronomy workshop. IPAM is partially supported by the National Science Foundation through award DMS-1925919. These authors would like to thank IPAM and UCLA for their warm hospitality. DT, PCD and AT acknowledge the support from the grants Prometeo CIPROM/2022/49 from the Generalitat Valenciana, and  PID2021-125485NB-C21 from the Spanish Agencia Estatal de Investigación funded by MCIN/AEI/10.13039/501100011033 and ERDF A way of making Europe. RL acknowledges financial support provided by Generalitat Valenciana / Fons Social Europeu ``L'FSE inverteix en el teu futur'' through APOSTD 2022 post-doctoral grant CIAPOS/2021/150. RL also acknowledges financial support provided by the Individual CEEC program 2023.06381.CEECIND/CP2840/CT0002 (DOI identifier \url{https://doi.org/10.54499/2023.06381.CEECIND/CP2840/CT0002}), funded by the Portuguese Foundation for Science and Technology (Funda\c{c}\~{a}o para a Ci\^{e}ncia e a Tecnologia). This work is supported by the Center for Research and Development in Mathematics and Applications (CIDMA) through the Portuguese Foundation for Science and Technology (FCT – Funda\c{c}\~{a}o para a Ci\^{e}ncia e a Tecnologia) under the Multi-Annual Financing Program for R\&D Units, 2022.04560.PTDC (DOI identifier \url{https://doi.org/10.54499/2022.04560.PTDC}) and 2024.05617.CERN (DOI identifier \url{https://doi.org/10.54499/2024.05617.CERN}). This work has further been supported by the European Horizon Europe staff exchange (SE) programme HORIZON-MSCA-2021-SE-01 Grant No. NewFunFiCO-101086251. 

\appendix  

\section{Normal vector}\label{App:NormalVector}
Our starting point is the assumption of the shock as a spherical surface, given by
\begin{equation}\label{eq:f_surface}
    \vec{r}_s = r\cos{\phi}\sin{\theta} \hat{x}+ r\sin{\phi}\sin{\theta}\hat{y}+ r\cos{\theta}\hat{z}
\end{equation}
The displaced shock of the surface for an Eulerian observer will be
\begin{equation}\label{eq:f'}
    \vec{r}_s = \vec{r}_{s0} + \vec \zeta.
\end{equation}

In the aforementioned relation, the displacement has been decomposed into its radial and angular components using spherical harmonics, according to Eq.~\eqref{eq:zeta_spher}. First, the above vector will be written with respect to the Cartesian coordinates,
\begin{eqnarray}\label{eq:zeta_cart}
 \vec{\zeta} &=& (\cos{\phi} \sin{\theta}\zeta^r + r\cos{\theta}\cos{\phi}\zeta^{\theta} - r\sin{\theta}\sin{\phi}\zeta^{\phi}) \hat x \nonumber \\
 &       +&(\sin{\theta}\sin{\phi}\zeta^r + r\cos{\theta}\sin{\phi}\zeta^{\theta} + r\cos{\phi}\sin{\theta}\zeta^{\phi} ) \hat y \nonumber \\
   &     +&(\cos{\theta}\zeta^r -r\sin{\theta}\zeta^{\theta}) \hat z \; .  
\end{eqnarray}

The normal vector of the new surface for a static background shock is such that, at the linear level $\vec n \cdot \vec n=1$, so we can compute it as
\begin{equation}\label{eq:n'}
    \vec{n} = \left (\frac{\partial \vec{r}_s}{\partial \theta} \times \frac{\partial \vec{r}_s}{\partial \phi} \right )
    \left |\left| \frac{\partial \vec{r}_s}{\partial \theta} \times \frac{\partial \vec{r}_s}{\partial \phi} \right |\right |^{-1} \; . \end{equation}
Lastly, we need to transform the aforementioned normal vector to spherical coordinates.
For this purpose, first we transform to spherical coordinates the cross product in the first term, which reads 
\begin{equation}\label{eq:n_spher}
    \begin{pmatrix}
        1 + \left[ \frac{2}{r}Z_1 - \frac{l(l+1)}{r}Z_2 \right]Y_{lm} e^{-i\sigma t} \\
        \frac{1}{r^2}\left(-Z_1 + Z_2 \right)\partial_{\theta} Y_{lm} e^{-i\sigma t} \\
        \frac{1}{r^2\sin{\theta}^2}\left(-Z_1 + Z_2 \right)\partial_{\phi} Y_{lm} e^{-i\sigma t}
    \end{pmatrix}\psi^6 r^2\sin{\theta}.
\end{equation}
Remember that the cross product $\times$ and the norm of vectors in the hypersurface are defined with respect to the spatial metric $\gamma_{ij}$.
Next, the norm of the cross product is 
\begin{equation}\label{eq:norm_condition}
    \psi^8 r^2\sin{\theta} \sqrt{ 1+ 2\left(\frac{2}{r}Z_1 - \frac{l(l+1)}{r}Z_2 \right)Y_{lm} e^{-i\sigma t} }
\end{equation}
Therefore, the normalized normal vector in terms of the coordinate basis will be
\begin{equation}\label{eq_n_spher_norm}
    n^i = \psi^{-2}\begin{pmatrix}
        1   \\
        \frac{1}{r^2}\left(-Z_1 + Z_2 \right)\partial_{\theta} Y_{lm} e^{-i\sigma t} \\
        \frac{1}{r^2\sin{\theta}^2}\left(-Z_1 + Z_2 \right)\partial_{\phi} Y_{lm} e^{-i\sigma t}
    \end{pmatrix}
\end{equation}
or
\begin{equation}\label{eq:n_i}
    n_i = \gamma_{ij} n^i = \psi^2 \begin{pmatrix}
        1   \\
        \left(-Z_1 + Z_2 \right)\partial_{\theta} Y_{lm} e^{-i\sigma t} \\
        \left(-Z_1 + Z_2 \right)\partial_{\phi} Y_{lm} e^{-i\sigma t}
    \end{pmatrix}
\end{equation}
The only missing part is the calculation of the time component of the normal vector. According to Eq.~\eqref{eq:n_mu_Eulerian},
\begin{align}\label{eq:n_t}
    n_t &= -\psi^2 W_s \partial_t \zeta^r = i\sigma \psi^2 W_s Z_1 e^{-i\sigma t}  ,
\end{align}
where we have taken into account that the shock velocity is given by Eq. (\ref{eq:shock_velocity}), while in the background it is zero as the shock is static. For a vanishing shock velocity in the static solution the shock Lorentz factor is equal to unity.

\section{Derivatives of background quantities in the exterior
}\label{App:RHCalculations}
In this section, we show some details of the calculation of the exterior terms of the \acrshort{rh} conditions, in particular, the computation of the derivatives of the background quantities, $\partial_i q_0\left( r_s \right) |_{ext}$, appearing in Eq.~\eqref{eq:qTaylorEXT}.
We use the conservation of the rest-mass, Eq. (\ref{eq:cons_Nmu_conservative}) and the momentum, Eq. (\ref{eq:cons_EnMom_Christoffel}). 
From the first one, we arrive to the following relation
\begin{equation}\label{eq:derivative_massFlux}
    \partial_r\left( \rho u^r \right) = -\left( \frac{2}{r_s} + 6\frac{\psi^\prime}{\psi} \right)\left( \rho u^r \right).
\end{equation}
Eq. (\ref{eq:cons_EnMom_Christoffel}) is used to calculate the derivative $\partial_r T^{rr}$,
\begin{align}\label{eq:derivative_fluxMom}
    &\partial_r T^{rr} = 
    -\left( \frac{2}{r_s} +8\frac{\psi^\prime}{\psi} -G \right)T^{rr}  \nonumber \\
    &+\psi^{-4}\bigg[ \rho h W^2G +\left( \frac{2}{r_s} +4\frac{\psi^\prime}{\psi} -G \right)p
    \bigg]
\end{align}
Note that $T^{r\theta}$ and $T^{r\phi}$ are zero, so their derivatives are trivially zero. Therefore, these two derivatives are sufficient to linearize the \acrshort{rh} conditions.

\section{Plane waves - Analytical solution}\label{App:PlaneWaves}
In this section, we derive the analytical solution of plane wave perturbations in a classical one-dimensional flow and the corresponding \acrshort{rh} conditions, for the setup described in section~\ref{subsec:PlaneWaves}.
We consider the case in which the flow and the perturbations propagate in the x-direction with no transversal velocity. The Euler equations in the classical limit, with no gravity, take the form
\begin{align}\label{eq:Euler_N_plane}
    \partial_t \rho +  \partial_x \left(\rho\upsilon \right)  &= 0, \\
    \partial_t \left( \rho\upsilon \right) + \partial_x \left( \rho\upsilon^2 + p \right) &=0,
\end{align}
where $\upsilon$ refers to the velocity in the x direction. A simple solution of the stationary problem is the case in which $\rho$, $p$ and $\upsilon$ are constant. We use that case as a background condition for our perturbations. We will consider the case of an accretion flow with $\upsilon<0$. We search for the solutions of the eigenvalue problem in the domain $x\in[0,1]$.
At $x=0$ we impose zero velocity perturbation, $\delta \upsilon(x=0)=0$. At $x=1$ we consider a stationary accretion shock for the background. The perturbations at the shock have to be consistent with the \acrshort{rh} conditions, which sets the \acrshort{bc} at that point.

\subsection{Perturbations}
The general perturbed Euler equations read
\begin{align}\label{eq:pert_Euler_N_plane_1}
    \upsilon \partial_x \delta \hat{\rho} +\rho \partial_x \delta\upsilon^*_R &= i\sigma \delta \hat{\rho}  \;,\\ \label{eq:pert_Euler_N_plane_2}
    \left( \upsilon^2 + c_s^2 \right) \partial_x \delta \hat{\rho} +2\rho\upsilon \partial_x \delta\upsilon_R^*&=  i\sigma\left( \upsilon\delta \hat{\rho}  + \rho \delta \upsilon^*_R \right) \; ,
\end{align}
where the only fluid quantities appearing are from the interior of the domain, and perturbed quantities only depend on $x$. These equations are used for the numerical calculations described in Section~\ref{subsec:PlaneWaves}.

In order to derive an analytic solution, we assume plane wave solutions of the form
\begin{equation}\label{eq:plane_W_perturbation}
    \delta \rho = \delta \tilde{\rho} \, e^{i\left( kx - \sigma t\right)},
\end{equation}
i.e. we are considering the case $\delta \hat \rho (x) = \delta \tilde \rho \, e^{ikx}$, being $\delta \tilde \rho$ a constant (and similarly for all perturbed quantities).
Since the background is independent of $x$, Eulerian and Lagrangian perturbations have the same value (e.g. $\Delta \rho = \delta \rho$). We consider adiabatic perturbations, which in the classical case read
\begin{equation}
\frac{\delta p}{\delta \rho} = c_s^2 = \frac{p}{\rho} \Gamma_1 \; .
\end{equation}

Taking into account all the above, the perturbed Euler equations will be
\begin{align}
    k\upsilon \delta \tilde{\rho} + k\rho \delta\tilde \upsilon &= \sigma \delta\tilde{\rho} \; ,\label{eq:plane_anal_pert_cont} \\
    k \left( \upsilon^2 + c_s^2 \right) \delta \tilde{\rho} + 2k\rho \upsilon\delta \tilde \upsilon &= \sigma \left( \upsilon \delta \tilde{\rho} + \rho \delta \tilde\upsilon \right) \; .\label{eq:plane_anal_pert_mom}
\end{align}
By substituting the continuity equation (\ref{eq:plane_anal_pert_cont}) into the momentum (\ref{eq:plane_anal_pert_mom}) we arrive at
\begin{equation}\label{eq:plane_analyt}
    \left[ k^2\left( \upsilon^2 - c_s^2\right)  -2k\upsilon\sigma + \sigma^2 \right] \delta \tilde{\rho} = 0 \; .
\end{equation}
Since $ \delta \tilde{\rho} \neq 0$, the above relation leads to the following solutions
\begin{align}
    \sigma_{\pm} &= k ( \upsilon\pm c_s )  \; , \label{eq:plane_sol} 
\end{align}
which correspond to upstream ($+$) and downstream ($-$) waves.
Let us consider a general wave of the form $e^{i \phi(x,t)}$, being $\phi(x,t)$ the wave phase.
The differential of the wave phase is 
\begin{equation}\label{eq:plane_wave_phase}
    d\phi(x,t) = \partial_t \phi(x,t)dt + \partial_x \phi(x,t) dx  \; .
\end{equation}
The phase velocity is defined as the velocity of a point with constant phase. At such point, $d\phi = 0$, and the phase velocity will be
\begin{equation}\label{eq:plane_phase_velocity}
    \upsilon_{\rm phase} = \left . \frac{dx}{dt}\right|_{\textrm{constant phase}} = -\frac{\partial_t \phi(x,t)}{\partial_x \phi(x,t)} \; .
\end{equation}
For the plane waves considered here, the wave phase is $\phi(x,t) = kx - \sigma t$, therefore, the phase velocity for the downstream and upstream waves,  is
\begin{align}
     \upsilon_{\rm phase}^{\pm} &= \upsilon \pm c_s  \; .
\end{align}
In the case that the fluid has a velocity higher than the speed of sound, $\upsilon > c_s$, then both modes have positive $v_{\rm phase}$, meaning that it is not possible to have upstream waves in a supersonic flow. 
From Eq.~\eqref{eq:plane_anal_pert_cont} the perturbation amplitudes for the upstream and downstream waves for $k \neq 0$ fulfill,
\begin{align}
\delta\tilde{\rho}_{\pm} &= \pm \frac{\rho}{c_s}\delta \tilde \upsilon_{\pm} . \label{eq:plane_ampl}  
\end{align}
The eigenfunctions will be linear combinations of the upstream and downstream solutions for a given eigenfrequency $\sigma$.
The wavenumbers of the downstream and upstream waves are
\begin{align}
    k_{\pm} &= \frac{\sigma}{\upsilon\pm c_s}.\label{eq:plane_k} 
\end{align}
A general density and velocity perturbation will be a linear combination of both waves
\begin{eqnarray}
     \delta \rho &= \delta \tilde{\rho}_{+}\, e^{i\left(k_{+}x - \sigma t \right)} + \delta \tilde{\rho}_{-}\,e^{i\left(k_{-}x - \sigma t + \varphi \right)}, \nonumber 
\\
     \delta \upsilon &= \delta \tilde{\upsilon}_{+}\, e^{i\left(k_{+}x - \sigma t \right)} + \delta \tilde{\upsilon}_{-}\,e^{i\left(k_{-}x - \sigma t + \varphi \right)},\label{eq:plane_perturbation_up_down}
\end{eqnarray}
respectively,
where $\varphi$ is the phase difference between both waves. The ratio between $\delta \tilde \rho_+$ and $\delta \tilde \rho_-$, as well as the value of $\varphi$, can be computed imposing the \acrshort{bcs}.

\subsection{Classical \acrshort{rh} conditions}
At this point, let us calculate the classical \acrshort{rh} conditions, as they are needed to deal with the shock at $x=1$.
In the classical framework they are given by
\begin{equation}\label{eq:RH_Newt}
    \left[[U ]\right]\Vec{\upsilon}_{shock}\cdot\Vec{n} = \left[[F ]\right]\Vec{n} ,
\end{equation}
where $U = \left( \rho, \; \Vec{j}, \; E^t \right)^T$ is the matrix of the conserved quantities, $F = \left( \Vec{j}, \; \frac{1}{\rho}\Vec{j}\times \Vec{j} + p I,\; 
\left( E^t + p \right)\frac{1}{\rho}\Vec{j} \right)^T$ of the fluxes, 
 $\Vec{j} = \rho \Vec{v}$ is the momentum density, $E^t = \rho \epsilon +\frac{1}{2}\rho u^2$ the total energy density, 
 and $I$ is the identity matrix. For both the background and the perturbations considered here, the normal vector is constant and given by $\Vec{n} = \left( 1, 0, 0\right).$

In the case of plane waves, for a static shock, the above conditions for the background translate into the following 
relations,
\begin{align}
    \left( \rho \upsilon \right)_{\rm ext} &= \left(\rho \upsilon \right)_{\rm int} , \label{eq:RH_Newt_bg_1} \\
    \left( \rho {\upsilon}^2 + p\right)_{\rm ext} &= \left( \rho {\upsilon}^2 + p\right)_{\rm int} , \label{eq:RH_Newt_bg_2} \\
    \left[ \rho \upsilon \left( \epsilon +\frac{1}{2} {\upsilon}^2 + \frac{p}{\rho }\right)\right]_{\rm ext} &= \left[ \rho \upsilon\left( \epsilon +\frac{1}{2} {\upsilon}^2 + \frac{p}{\rho } \right)\right]_{\rm int} \;,  \label{eq:RH_Newt_bg_3}
\end{align}
where $\rm int$ and $\rm ext$ refer to the values at $x=1$ in the side of the interior of the domain and of the exterior, respectively. Since the background solution has constant $\rho$, $p$ and $\upsilon$, the interior value is the same in the whole domain. Given these three values, one can compute the background values at the exterior using Eqs.~\eqref{eq:RH_Newt_bg_1}-\eqref{eq:RH_Newt_bg_3} and the \acrshort{eos}. Note that not every combination of $\rho$, $p$ and $\upsilon$ produces a solution with an accretion shock, in which the region $x\in [0,1]$ is subsonic. This restricts the possible values that can be used as a background.

Let us consider now the shock in the presence of perturbations, such that the shock displacement velocity is $\upsilon_s = \tilde \upsilon_s e^{-i\sigma t}$, being $\tilde \upsilon_s$ the amplitude of the shock perturbation.
In this case, the three \acrshort{rh} conditions 
result in
\begin{align}
 &\frac{\upsilon_{\rm ext}}{c_s + \upsilon_{\rm int}} e^{i\left( \frac{x}{\upsilon_{\rm int}-c_s }\sigma + \varphi \right)} \delta  \tilde{\upsilon}_{-}
 +
 \frac{\upsilon_{\rm ext} }{c_s - \upsilon_{\rm int}}e^{i\frac{x}{\upsilon_{\rm int}+c_s}\sigma}\delta \tilde{\upsilon}_{+} \nonumber \\&= \frac{c_s (\upsilon_{\rm ext}-\upsilon_{\rm int}) }{c_s^2 - \upsilon_{\rm int}^2}\tilde \upsilon_{s} , \label{eq:plane_RH1} \\
    &- e^{i\left( \frac{x}{\upsilon_{\rm int} -c_s}\sigma + \varphi \right)} \delta \tilde{\upsilon}_{-} 
    + e^{i\frac{x}{\upsilon_{\rm int}+c_s}\sigma}\left( \frac{\upsilon_{\rm int}+c_s}{\upsilon_{\rm int}-c_s} \right)^2 \delta \tilde{\upsilon}_{+} = 0 , \label{eq:plane_RH2}
\end{align}
The third \acrshort{rh} condition does not provide any additional information to the system because the perturbations are adiabatic. 

\subsection{Boundary conditions}

Finally, we impose our \acrshort{bcs}. At $x=0$ the perturbation of the fluid velocity (Eq.~\eqref{eq:plane_perturbation_up_down}) is zero at all times,
\begin{equation}\label{eq:plane_inner_BC}
    \delta \upsilon (x=0, t) = 
    \left (\delta \tilde \upsilon_+ 
    + \delta \tilde \upsilon_+ e^{i\varphi} \right )
    e^{-i\sigma t} =
    0.
\end{equation}
We have freedom of choice for the phase difference $\varphi$, which corresponds just to the initial phase of the oscillation mode. For simplicity, we choose $\varphi=0$, which leads to
\begin{equation}\label{eq:plane_dvu_dvd}
    \delta \tilde{\upsilon}_{+} = - \delta \tilde{\upsilon}_{-}  \;.
\end{equation}

Next, the perturbed \acrshort{rh} conditions will be evaluated at the location of the shock, $x=1$, resulting in
\begin{align}
    & 
    \left( \frac{ e^{\frac{i\sigma}{ \upsilon_{\rm int}+c_s} }}{\upsilon_{\rm int}-c_s} 
    + \frac{ e^{ \frac{i\sigma}{\upsilon_{\rm int}-c_s} }}{ \upsilon_{\rm int}+c_s}
    \right)\upsilon_{\rm ext} \delta  \tilde{\upsilon}_{-}   
    = \frac{c_s (\upsilon_{\rm ext}-\upsilon_{\rm int}) }{c_s^2 - \upsilon_{\rm int}^2} \tilde \upsilon_{s}  \label{eq:plane_RH1_IC} \\
    &
    \bigg[ e^{\frac{i\sigma}{\upsilon_{\rm int}-c_s}} +e^{\frac{i\sigma}{\upsilon_{\rm int}+c_s }}\left(\frac{\upsilon_{\rm int}+c_s}{\upsilon_{\rm int}-c_s}\right)^2\bigg]\delta \tilde{\upsilon}_{-} = 0 \label{eq:plane_RH2_IC}
\end{align}
The 1st \acrshort{rh} condition provides a relation for the amplitude of the shock velocity, $\tilde \upsilon_{s}$. The wave amplitude, $\delta \tilde{\upsilon}_{-}$, is arbitrary and can be eliminated from the two \acrshort{rh} conditions leading to,
\begin{equation}\label{eq:planeWaves_final_RH}
    \bigg[ \left(\upsilon_{\rm int}-c_s\right)^2
     +\left( \upsilon_{\rm int}+c_s\right)^2e^{\frac{2ic_s\sigma}{c_s^2 - \upsilon_{\rm int}^2}}
    \bigg] \tilde \upsilon_{s} = 0
\end{equation}
Note that we can arrive at the last equation using the 2nd and the 3rd \acrshort{rh} conditions too, confirming that the 3rd \acrshort{rh} is unnecessary. The nontrivial solutions, i.e. $\tilde \upsilon_{s} \neq 0$, allow us to compute the eigenfrequencies
\begin{equation}\label{eq:plane_eigenvalues}
\sigma_n = \frac{c_s^2-\upsilon_{\rm int}^2}{2c_s}\bigg\{ 2\,\pi \,n -i\log{\bigg[-\left( \frac{\upsilon_{\rm int}-c_s}{\upsilon_{\rm int}+c_s} \right)^2\bigg]}\bigg\},
\end{equation}
where $n \in \mathbb{Z}$. Note that, for a given value of the velocity, $\upsilon_{\rm int}$, the real part of the eigenvalues, $Re(\sigma_n)$, increases with increasing $n$ and is an integer multiple of $Re(\sigma_1)$,
while the imaginary part, $Im(\sigma_n)$, is constant.

\bibliographystyle{apsrev4-1}
\bibliography{references}

\begin{thebibliography}{68}%
\makeatletter
\providecommand \@ifxundefined [1]{%
 \@ifx{#1\undefined}
}%
\providecommand \@ifnum [1]{%
 \ifnum #1\expandafter \@firstoftwo
 \else \expandafter \@secondoftwo
 \fi
}%
\providecommand \@ifx [1]{%
 \ifx #1\expandafter \@firstoftwo
 \else \expandafter \@secondoftwo
 \fi
}%
\providecommand \natexlab [1]{#1}%
\providecommand \enquote  [1]{``#1''}%
\providecommand \bibnamefont  [1]{#1}%
\providecommand \bibfnamefont [1]{#1}%
\providecommand \citenamefont [1]{#1}%
\providecommand \href@noop [0]{\@secondoftwo}%
\providecommand \href [0]{\begingroup \@sanitize@url \@href}%
\providecommand \@href[1]{\@@startlink{#1}\@@href}%
\providecommand \@@href[1]{\endgroup#1\@@endlink}%
\providecommand \@sanitize@url [0]{\catcode `\\12\catcode `\$12\catcode `\&12\catcode `\#12\catcode `\^12\catcode `\_12\catcode `\%12\relax}%
\providecommand \@@startlink[1]{}%
\providecommand \@@endlink[0]{}%
\providecommand \url  [0]{\begingroup\@sanitize@url \@url }%
\providecommand \@url [1]{\endgroup\@href {#1}{\urlprefix }}%
\providecommand \urlprefix  [0]{URL }%
\providecommand \Eprint [0]{\href }%
\providecommand \doibase [0]{http://dx.doi.org/}%
\providecommand \selectlanguage [0]{\@gobble}%
\providecommand \bibinfo  [0]{\@secondoftwo}%
\providecommand \bibfield  [0]{\@secondoftwo}%
\providecommand \translation [1]{[#1]}%
\providecommand \BibitemOpen [0]{}%
\providecommand \bibitemStop [0]{}%
\providecommand \bibitemNoStop [0]{.\EOS\space}%
\providecommand \EOS [0]{\spacefactor3000\relax}%
\providecommand \BibitemShut  [1]{\csname bibitem#1\endcsname}%
\let\auto@bib@innerbib\@empty
\bibitem [{\citenamefont {Aasi}\ \emph {et~al.}(2015)\citenamefont {Aasi} \emph {et~al.}}]{LIGOScientific:2014pky}%
  \BibitemOpen
  \bibfield  {author} {\bibinfo {author} {\bibfnamefont {J.}~\bibnamefont {Aasi}} \emph {et~al.} (\bibinfo {collaboration} {LIGO Scientific}),\ }\href {\doibase 10.1088/0264-9381/32/7/074001} {\bibfield  {journal} {\bibinfo  {journal} {Classical and Quantum Gravity}\ }\textbf {\bibinfo {volume} {32}},\ \bibinfo {eid} {074001} (\bibinfo {year} {2015})},\ \Eprint {http://arxiv.org/abs/1411.4547} {arXiv:1411.4547 [gr-qc]} \BibitemShut {NoStop}%
\bibitem [{\citenamefont {Acernese}\ \emph {et~al.}(2015)\citenamefont {Acernese} \emph {et~al.}}]{VIRGO:2014yos}%
  \BibitemOpen
  \bibfield  {author} {\bibinfo {author} {\bibfnamefont {F.}~\bibnamefont {Acernese}} \emph {et~al.} (\bibinfo {collaboration} {VIRGO}),\ }\href {\doibase 10.1088/0264-9381/32/2/024001} {\bibfield  {journal} {\bibinfo  {journal} {Class. Quant. Grav.}\ }\textbf {\bibinfo {volume} {32}},\ \bibinfo {pages} {024001} (\bibinfo {year} {2015})},\ \Eprint {http://arxiv.org/abs/1408.3978} {arXiv:1408.3978 [gr-qc]} \BibitemShut {NoStop}%
\bibitem [{\citenamefont {Abbott}\ \emph {et~al.}(2017)\citenamefont {Abbott} \emph {et~al.}}]{LIGOScientific:2017ync}%
  \BibitemOpen
  \bibfield  {author} {\bibinfo {author} {\bibfnamefont {B.~P.}\ \bibnamefont {Abbott}} \emph {et~al.} (\bibinfo {collaboration} {LIGO Scientific, Virgo}),\ }\href {\doibase 10.3847/2041-8213/aa91c9} {\bibfield  {journal} {\bibinfo  {journal} {Astrophys. J. Lett.}\ }\textbf {\bibinfo {volume} {848}},\ \bibinfo {pages} {L12} (\bibinfo {year} {2017})},\ \Eprint {http://arxiv.org/abs/1710.05833} {arXiv:1710.05833 [astro-ph.HE]} \BibitemShut {NoStop}%
\bibitem [{\citenamefont {{Szczepa{\'n}czyk}}\ \emph {et~al.}(2021)\citenamefont {{Szczepa{\'n}czyk}}, \citenamefont {{Antelis}}, \citenamefont {{Benjamin}}, \citenamefont {{Cavagli{\`a}}}, \citenamefont {{Gondek-Rosi{\'n}ska}}, \citenamefont {{Hansen}}, \citenamefont {{Klimenko}}, \citenamefont {{Morales}}, \citenamefont {{Moreno}}, \citenamefont {{Mukherjee}}, \citenamefont {{Nurbek}}, \citenamefont {{Powell}}, \citenamefont {{Singh}}, \citenamefont {{Sitmukhambetov}}, \citenamefont {{Szewczyk}}, \citenamefont {{Valdez}}, \citenamefont {{Vedovato}}, \citenamefont {{Westhouse}}, \citenamefont {{Zanolin}},\ and\ \citenamefont {{Zheng}}}]{Szczepanczyk2021}%
  \BibitemOpen
  \bibfield  {author} {\bibinfo {author} {\bibfnamefont {M.~J.}\ \bibnamefont {{Szczepa{\'n}czyk}}}, \bibinfo {author} {\bibfnamefont {J.~M.}\ \bibnamefont {{Antelis}}}, \bibinfo {author} {\bibfnamefont {M.}~\bibnamefont {{Benjamin}}}, \bibinfo {author} {\bibfnamefont {M.}~\bibnamefont {{Cavagli{\`a}}}}, \bibinfo {author} {\bibfnamefont {D.}~\bibnamefont {{Gondek-Rosi{\'n}ska}}}, \bibinfo {author} {\bibfnamefont {T.}~\bibnamefont {{Hansen}}}, \bibinfo {author} {\bibfnamefont {S.}~\bibnamefont {{Klimenko}}}, \bibinfo {author} {\bibfnamefont {M.~D.}\ \bibnamefont {{Morales}}}, \bibinfo {author} {\bibfnamefont {C.}~\bibnamefont {{Moreno}}}, \bibinfo {author} {\bibfnamefont {S.}~\bibnamefont {{Mukherjee}}}, \bibinfo {author} {\bibfnamefont {G.}~\bibnamefont {{Nurbek}}}, \bibinfo {author} {\bibfnamefont {J.}~\bibnamefont {{Powell}}}, \bibinfo {author} {\bibfnamefont {N.}~\bibnamefont {{Singh}}}, \bibinfo {author} {\bibfnamefont {S.}~\bibnamefont {{Sitmukhambetov}}}, \bibinfo {author} {\bibfnamefont
  {P.}~\bibnamefont {{Szewczyk}}}, \bibinfo {author} {\bibfnamefont {O.}~\bibnamefont {{Valdez}}}, \bibinfo {author} {\bibfnamefont {G.}~\bibnamefont {{Vedovato}}}, \bibinfo {author} {\bibfnamefont {J.}~\bibnamefont {{Westhouse}}}, \bibinfo {author} {\bibfnamefont {M.}~\bibnamefont {{Zanolin}}}, \ and\ \bibinfo {author} {\bibfnamefont {Y.}~\bibnamefont {{Zheng}}},\ }\href {\doibase 10.1103/PhysRevD.104.102002} {\bibfield  {journal} {\bibinfo  {journal} {\prd}\ }\textbf {\bibinfo {volume} {104}},\ \bibinfo {eid} {102002} (\bibinfo {year} {2021})},\ \Eprint {http://arxiv.org/abs/2104.06462} {arXiv:2104.06462 [astro-ph.HE]} \BibitemShut {NoStop}%
\bibitem [{\citenamefont {{Gossan}}\ \emph {et~al.}(2016)\citenamefont {{Gossan}}, \citenamefont {{Sutton}}, \citenamefont {{Stuver}}, \citenamefont {{Zanolin}}, \citenamefont {{Gill}},\ and\ \citenamefont {{Ott}}}]{Gossan:2016}%
  \BibitemOpen
  \bibfield  {author} {\bibinfo {author} {\bibfnamefont {S.~E.}\ \bibnamefont {{Gossan}}}, \bibinfo {author} {\bibfnamefont {P.}~\bibnamefont {{Sutton}}}, \bibinfo {author} {\bibfnamefont {A.}~\bibnamefont {{Stuver}}}, \bibinfo {author} {\bibfnamefont {M.}~\bibnamefont {{Zanolin}}}, \bibinfo {author} {\bibfnamefont {K.}~\bibnamefont {{Gill}}}, \ and\ \bibinfo {author} {\bibfnamefont {C.~D.}\ \bibnamefont {{Ott}}},\ }\href {\doibase 10.1103/PhysRevD.93.042002} {\bibfield  {journal} {\bibinfo  {journal} {\prd}\ }\textbf {\bibinfo {volume} {93}},\ \bibinfo {eid} {042002} (\bibinfo {year} {2016})},\ \Eprint {http://arxiv.org/abs/1511.02836} {arXiv:1511.02836 [astro-ph.HE]} \BibitemShut {NoStop}%
\bibitem [{\citenamefont {Janka}\ \emph {et~al.}(2007)\citenamefont {Janka}, \citenamefont {Langanke}, \citenamefont {Marek}, \citenamefont {Martinez-Pinedo},\ and\ \citenamefont {Mueller}}]{Janka:2006fh}%
  \BibitemOpen
  \bibfield  {author} {\bibinfo {author} {\bibfnamefont {H.-T.}\ \bibnamefont {Janka}}, \bibinfo {author} {\bibfnamefont {K.}~\bibnamefont {Langanke}}, \bibinfo {author} {\bibfnamefont {A.}~\bibnamefont {Marek}}, \bibinfo {author} {\bibfnamefont {G.}~\bibnamefont {Martinez-Pinedo}}, \ and\ \bibinfo {author} {\bibfnamefont {B.}~\bibnamefont {Mueller}},\ }\href {\doibase 10.1016/j.physrep.2007.02.002} {\bibfield  {journal} {\bibinfo  {journal} {Phys. Rept.}\ }\textbf {\bibinfo {volume} {442}},\ \bibinfo {pages} {38} (\bibinfo {year} {2007})},\ \Eprint {http://arxiv.org/abs/astro-ph/0612072} {arXiv:astro-ph/0612072} \BibitemShut {NoStop}%
\bibitem [{\citenamefont {Klimenko}\ \emph {et~al.}(2008)\citenamefont {Klimenko}, \citenamefont {Yakushin}, \citenamefont {Mercer}, \citenamefont {Drago}, \citenamefont {Necula}, \citenamefont {Mitselmakher},\ and\ \citenamefont {Mohanty}}]{Klimenko2008}%
  \BibitemOpen
  \bibfield  {author} {\bibinfo {author} {\bibfnamefont {S.}~\bibnamefont {Klimenko}}, \bibinfo {author} {\bibfnamefont {I.}~\bibnamefont {Yakushin}}, \bibinfo {author} {\bibfnamefont {A.}~\bibnamefont {Mercer}}, \bibinfo {author} {\bibfnamefont {M.}~\bibnamefont {Drago}}, \bibinfo {author} {\bibfnamefont {V.}~\bibnamefont {Necula}}, \bibinfo {author} {\bibfnamefont {G.}~\bibnamefont {Mitselmakher}}, \ and\ \bibinfo {author} {\bibfnamefont {S.}~\bibnamefont {Mohanty}},\ }\href {\doibase 10.1088/0264-9381/25/11/114029} {\bibfield  {journal} {\bibinfo  {journal} {Classical and Quantum Gravity}\ }\textbf {\bibinfo {volume} {25}},\ \bibinfo {pages} {114029} (\bibinfo {year} {2008})}\BibitemShut {NoStop}%
\bibitem [{\citenamefont {Drago}\ \emph {et~al.}(2021)\citenamefont {Drago}, \citenamefont {Klimenko}, \citenamefont {Lazzaro}, \citenamefont {Milotti}, \citenamefont {Mitselmakher}, \citenamefont {Necula}, \citenamefont {O'Brien}, \citenamefont {Prodi}, \citenamefont {Salemi},\ and\ \citenamefont {Szczepanczyk}}]{Drago2021cWB}%
  \BibitemOpen
  \bibfield  {author} {\bibinfo {author} {\bibfnamefont {M.}~\bibnamefont {Drago}}, \bibinfo {author} {\bibfnamefont {S.}~\bibnamefont {Klimenko}}, \bibinfo {author} {\bibfnamefont {C.}~\bibnamefont {Lazzaro}}, \bibinfo {author} {\bibfnamefont {E.}~\bibnamefont {Milotti}}, \bibinfo {author} {\bibfnamefont {G.}~\bibnamefont {Mitselmakher}}, \bibinfo {author} {\bibfnamefont {V.}~\bibnamefont {Necula}}, \bibinfo {author} {\bibfnamefont {B.}~\bibnamefont {O'Brien}}, \bibinfo {author} {\bibfnamefont {G.~A.}\ \bibnamefont {Prodi}}, \bibinfo {author} {\bibfnamefont {F.}~\bibnamefont {Salemi}}, \ and\ \bibinfo {author} {\bibfnamefont {M.}~\bibnamefont {Szczepanczyk}},\ }\href {\doibase 10.1016/j.softx.2021.100678} {\bibfield  {journal} {\bibinfo  {journal} {SoftwareX}\ }\textbf {\bibinfo {volume} {14}},\ \bibinfo {pages} {100678} (\bibinfo {year} {2021})}\BibitemShut {NoStop}%
\bibitem [{\citenamefont {Abbott}\ \emph {et~al.}(2020)\citenamefont {Abbott}, \citenamefont {others (LIGO Scientific~Collaboration},\ and\ \citenamefont {Collaboration)}}]{Abbott2020CCSN}%
  \BibitemOpen
  \bibfield  {author} {\bibinfo {author} {\bibfnamefont {B.~P.}\ \bibnamefont {Abbott}}, \bibinfo {author} {\bibnamefont {others (LIGO Scientific~Collaboration}}, \ and\ \bibinfo {author} {\bibfnamefont {V.}~\bibnamefont {Collaboration)}},\ }\href {https://journals.aps.org/prd/abstract/10.1103/PhysRevD.101.084002} {\bibfield  {journal} {\bibinfo  {journal} {Physical Review D}\ }\textbf {\bibinfo {volume} {101}},\ \bibinfo {pages} {084002} (\bibinfo {year} {2020})}\BibitemShut {NoStop}%
\bibitem [{\citenamefont {{Alsabti}}\ and\ \citenamefont {{Murdin}}(2017)}]{2017handbook_SN.1095Janka}%
  \BibitemOpen
  \bibfield  {author} {\bibinfo {author} {\bibfnamefont {A.~W.}\ \bibnamefont {{Alsabti}}}\ and\ \bibinfo {author} {\bibfnamefont {P.}~\bibnamefont {{Murdin}}},\ }\href {\doibase 10.1007/978-3-319-21846-5} {\emph {\bibinfo {title} {{Handbook of Supernovae}}}}\ (\bibinfo  {publisher} {Springer Cham},\ \bibinfo {year} {2017})\BibitemShut {NoStop}%
\bibitem [{\citenamefont {{Mezzacappa}}\ and\ \citenamefont {{Zanolin}}(2024)}]{2024_Mezzacappa_et_al_review}%
  \BibitemOpen
  \bibfield  {author} {\bibinfo {author} {\bibfnamefont {A.}~\bibnamefont {{Mezzacappa}}}\ and\ \bibinfo {author} {\bibfnamefont {M.}~\bibnamefont {{Zanolin}}},\ }\href {\doibase 10.48550/arXiv.2401.11635} {\bibfield  {journal} {\bibinfo  {journal} {arXiv e-prints}\ ,\ \bibinfo {eid} {arXiv:2401.11635}} (\bibinfo {year} {2024})},\ \Eprint {http://arxiv.org/abs/2401.11635} {arXiv:2401.11635 [astro-ph.HE]} \BibitemShut {NoStop}%
\bibitem [{\citenamefont {{M{\"u}ller}}(2020)}]{2020_Mueller_review}%
  \BibitemOpen
  \bibfield  {author} {\bibinfo {author} {\bibfnamefont {B.}~\bibnamefont {{M{\"u}ller}}},\ }\href {\doibase 10.1007/s41115-020-0008-5} {\bibfield  {journal} {\bibinfo  {journal} {Living Reviews in Computational Astrophysics}\ }\textbf {\bibinfo {volume} {6}},\ \bibinfo {eid} {3} (\bibinfo {year} {2020})},\ \Eprint {http://arxiv.org/abs/2006.05083} {arXiv:2006.05083 [astro-ph.SR]} \BibitemShut {NoStop}%
\bibitem [{\citenamefont {{Burrows}}\ and\ \citenamefont {{Vartanyan}}(2021)}]{2021Nature_CCSN}%
  \BibitemOpen
  \bibfield  {author} {\bibinfo {author} {\bibfnamefont {A.}~\bibnamefont {{Burrows}}}\ and\ \bibinfo {author} {\bibfnamefont {D.}~\bibnamefont {{Vartanyan}}},\ }\href {\doibase 10.1038/s41586-020-03059-w} {\bibfield  {journal} {\bibinfo  {journal} {\nat}\ }\textbf {\bibinfo {volume} {589}},\ \bibinfo {pages} {29} (\bibinfo {year} {2021})},\ \Eprint {http://arxiv.org/abs/2009.14157} {arXiv:2009.14157 [astro-ph.SR]} \BibitemShut {NoStop}%
\bibitem [{\citenamefont {{Foglizzo}}\ and\ \citenamefont {{Tagger}}(2000)}]{Foglizzo2000}%
  \BibitemOpen
  \bibfield  {author} {\bibinfo {author} {\bibfnamefont {T.}~\bibnamefont {{Foglizzo}}}\ and\ \bibinfo {author} {\bibfnamefont {M.}~\bibnamefont {{Tagger}}},\ }\href {\doibase 10.48550/arXiv.astro-ph/0009193} {\bibfield  {journal} {\bibinfo  {journal} {\aap}\ }\textbf {\bibinfo {volume} {363}},\ \bibinfo {pages} {174} (\bibinfo {year} {2000})},\ \Eprint {http://arxiv.org/abs/astro-ph/0009193} {arXiv:astro-ph/0009193 [astro-ph]} \BibitemShut {NoStop}%
\bibitem [{\citenamefont {{Blondin}}\ \emph {et~al.}(2003)\citenamefont {{Blondin}}, \citenamefont {{Mezzacappa}},\ and\ \citenamefont {{DeMarino}}}]{Blondin2003}%
  \BibitemOpen
  \bibfield  {author} {\bibinfo {author} {\bibfnamefont {J.~M.}\ \bibnamefont {{Blondin}}}, \bibinfo {author} {\bibfnamefont {A.}~\bibnamefont {{Mezzacappa}}}, \ and\ \bibinfo {author} {\bibfnamefont {C.}~\bibnamefont {{DeMarino}}},\ }\href {\doibase 10.1086/345812} {\bibfield  {journal} {\bibinfo  {journal} {\apj}\ }\textbf {\bibinfo {volume} {584}},\ \bibinfo {pages} {971} (\bibinfo {year} {2003})},\ \Eprint {http://arxiv.org/abs/astro-ph/0210634} {arXiv:astro-ph/0210634 [astro-ph]} \BibitemShut {NoStop}%
\bibitem [{\citenamefont {Murphy}\ \emph {et~al.}(2009)\citenamefont {Murphy}, \citenamefont {Ott},\ and\ \citenamefont {Burrows}}]{Murphy:2009dx}%
  \BibitemOpen
  \bibfield  {author} {\bibinfo {author} {\bibfnamefont {J.~W.}\ \bibnamefont {Murphy}}, \bibinfo {author} {\bibfnamefont {C.~D.}\ \bibnamefont {Ott}}, \ and\ \bibinfo {author} {\bibfnamefont {A.}~\bibnamefont {Burrows}},\ }\href {\doibase 10.1088/0004-637X/707/2/1173} {\bibfield  {journal} {\bibinfo  {journal} {Astrophys. J.}\ }\textbf {\bibinfo {volume} {707}},\ \bibinfo {pages} {1173} (\bibinfo {year} {2009})},\ \Eprint {http://arxiv.org/abs/0907.4762} {arXiv:0907.4762 [astro-ph.SR]} \BibitemShut {NoStop}%
\bibitem [{\citenamefont {Mueller}\ \emph {et~al.}(2013)\citenamefont {Mueller}, \citenamefont {Janka},\ and\ \citenamefont {Marek}}]{Mueller:2012sv}%
  \BibitemOpen
  \bibfield  {author} {\bibinfo {author} {\bibfnamefont {B.}~\bibnamefont {Mueller}}, \bibinfo {author} {\bibfnamefont {H.-T.}\ \bibnamefont {Janka}}, \ and\ \bibinfo {author} {\bibfnamefont {A.}~\bibnamefont {Marek}},\ }\href {\doibase 10.1088/0004-637X/766/1/43} {\bibfield  {journal} {\bibinfo  {journal} {Astrophys. J.}\ }\textbf {\bibinfo {volume} {766}},\ \bibinfo {pages} {43} (\bibinfo {year} {2013})},\ \Eprint {http://arxiv.org/abs/1210.6984} {arXiv:1210.6984 [astro-ph.SR]} \BibitemShut {NoStop}%
\bibitem [{\citenamefont {{Cerd{\'a}-Dur{\'a}n}}\ \emph {et~al.}(2013)\citenamefont {{Cerd{\'a}-Dur{\'a}n}}, \citenamefont {{DeBrye}}, \citenamefont {{Aloy}}, \citenamefont {{Font}},\ and\ \citenamefont {{Obergaulinger}}}]{CerdaDuran2013}%
  \BibitemOpen
  \bibfield  {author} {\bibinfo {author} {\bibfnamefont {P.}~\bibnamefont {{Cerd{\'a}-Dur{\'a}n}}}, \bibinfo {author} {\bibfnamefont {N.}~\bibnamefont {{DeBrye}}}, \bibinfo {author} {\bibfnamefont {M.~A.}\ \bibnamefont {{Aloy}}}, \bibinfo {author} {\bibfnamefont {J.~A.}\ \bibnamefont {{Font}}}, \ and\ \bibinfo {author} {\bibfnamefont {M.}~\bibnamefont {{Obergaulinger}}},\ }\href {\doibase 10.1088/2041-8205/779/2/L18} {\bibfield  {journal} {\bibinfo  {journal} {\apjl}\ }\textbf {\bibinfo {volume} {779}},\ \bibinfo {eid} {L18} (\bibinfo {year} {2013})},\ \Eprint {http://arxiv.org/abs/1310.8290} {arXiv:1310.8290 [astro-ph.SR]} \BibitemShut {NoStop}%
\bibitem [{\citenamefont {Kuroda}\ \emph {et~al.}(2016)\citenamefont {Kuroda}, \citenamefont {Kotake},\ and\ \citenamefont {Takiwaki}}]{Kuroda_2016}%
  \BibitemOpen
  \bibfield  {author} {\bibinfo {author} {\bibfnamefont {T.}~\bibnamefont {Kuroda}}, \bibinfo {author} {\bibfnamefont {K.}~\bibnamefont {Kotake}}, \ and\ \bibinfo {author} {\bibfnamefont {T.}~\bibnamefont {Takiwaki}},\ }\href {\doibase 10.3847/2041-8205/829/1/L14} {\bibfield  {journal} {\bibinfo  {journal} {The Astrophysical Journal Letters}\ }\textbf {\bibinfo {volume} {829}},\ \bibinfo {pages} {L14} (\bibinfo {year} {2016})}\BibitemShut {NoStop}%
\bibitem [{\citenamefont {Andresen}\ \emph {et~al.}(2017)\citenamefont {Andresen}, \citenamefont {M\"uller}, \citenamefont {M\"uller},\ and\ \citenamefont {Janka}}]{Andresen:2016pdt}%
  \BibitemOpen
  \bibfield  {author} {\bibinfo {author} {\bibfnamefont {H.}~\bibnamefont {Andresen}}, \bibinfo {author} {\bibfnamefont {B.}~\bibnamefont {M\"uller}}, \bibinfo {author} {\bibfnamefont {E.}~\bibnamefont {M\"uller}}, \ and\ \bibinfo {author} {\bibfnamefont {H.-T.}\ \bibnamefont {Janka}},\ }\href {\doibase 10.1093/mnras/stx618} {\bibfield  {journal} {\bibinfo  {journal} {Mon. Not. Roy. Astron. Soc.}\ }\textbf {\bibinfo {volume} {468}},\ \bibinfo {pages} {2032} (\bibinfo {year} {2017})},\ \Eprint {http://arxiv.org/abs/1607.05199} {arXiv:1607.05199 [astro-ph.HE]} \BibitemShut {NoStop}%
\bibitem [{\citenamefont {{McDermott}}\ \emph {et~al.}(1983)\citenamefont {{McDermott}}, \citenamefont {{van Horn}},\ and\ \citenamefont {{Scholl}}}]{McDermott1983}%
  \BibitemOpen
  \bibfield  {author} {\bibinfo {author} {\bibfnamefont {P.~N.}\ \bibnamefont {{McDermott}}}, \bibinfo {author} {\bibfnamefont {H.~M.}\ \bibnamefont {{van Horn}}}, \ and\ \bibinfo {author} {\bibfnamefont {J.~F.}\ \bibnamefont {{Scholl}}},\ }\href {\doibase 10.1086/161006} {\bibfield  {journal} {\bibinfo  {journal} {\apj}\ }\textbf {\bibinfo {volume} {268}},\ \bibinfo {pages} {837} (\bibinfo {year} {1983})}\BibitemShut {NoStop}%
\bibitem [{\citenamefont {{Reisenegger}}\ and\ \citenamefont {{Goldreich}}(1992)}]{Reisenegger1992}%
  \BibitemOpen
  \bibfield  {author} {\bibinfo {author} {\bibfnamefont {A.}~\bibnamefont {{Reisenegger}}}\ and\ \bibinfo {author} {\bibfnamefont {P.}~\bibnamefont {{Goldreich}}},\ }\href {\doibase 10.1086/171645} {\bibfield  {journal} {\bibinfo  {journal} {\apj}\ }\textbf {\bibinfo {volume} {395}},\ \bibinfo {pages} {240} (\bibinfo {year} {1992})}\BibitemShut {NoStop}%
\bibitem [{\citenamefont {{Ferrari}}\ \emph {et~al.}(2003)\citenamefont {{Ferrari}}, \citenamefont {{Miniutti}},\ and\ \citenamefont {{Pons}}}]{Ferrari2003}%
  \BibitemOpen
  \bibfield  {author} {\bibinfo {author} {\bibfnamefont {V.}~\bibnamefont {{Ferrari}}}, \bibinfo {author} {\bibfnamefont {G.}~\bibnamefont {{Miniutti}}}, \ and\ \bibinfo {author} {\bibfnamefont {J.~A.}\ \bibnamefont {{Pons}}},\ }\href {\doibase 10.1046/j.1365-8711.2003.06580.x} {\bibfield  {journal} {\bibinfo  {journal} {\mnras}\ }\textbf {\bibinfo {volume} {342}},\ \bibinfo {pages} {629} (\bibinfo {year} {2003})},\ \Eprint {http://arxiv.org/abs/astro-ph/0210581} {arXiv:astro-ph/0210581 [astro-ph]} \BibitemShut {NoStop}%
\bibitem [{\citenamefont {{Passamonti}}\ \emph {et~al.}(2005)\citenamefont {{Passamonti}}, \citenamefont {{Bruni}}, \citenamefont {{Gualtieri}},\ and\ \citenamefont {{Sopuerta}}}]{Passamonti2005}%
  \BibitemOpen
  \bibfield  {author} {\bibinfo {author} {\bibfnamefont {A.}~\bibnamefont {{Passamonti}}}, \bibinfo {author} {\bibfnamefont {M.}~\bibnamefont {{Bruni}}}, \bibinfo {author} {\bibfnamefont {L.}~\bibnamefont {{Gualtieri}}}, \ and\ \bibinfo {author} {\bibfnamefont {C.~F.}\ \bibnamefont {{Sopuerta}}},\ }\href {\doibase 10.1103/PhysRevD.71.024022} {\bibfield  {journal} {\bibinfo  {journal} {\prd}\ }\textbf {\bibinfo {volume} {71}},\ \bibinfo {eid} {024022} (\bibinfo {year} {2005})},\ \Eprint {http://arxiv.org/abs/gr-qc/0407108} {arXiv:gr-qc/0407108 [gr-qc]} \BibitemShut {NoStop}%
\bibitem [{\citenamefont {{Dimmelmeier}}\ \emph {et~al.}(2006)\citenamefont {{Dimmelmeier}}, \citenamefont {{Stergioulas}},\ and\ \citenamefont {{Font}}}]{Dimmelmeier2006}%
  \BibitemOpen
  \bibfield  {author} {\bibinfo {author} {\bibfnamefont {H.}~\bibnamefont {{Dimmelmeier}}}, \bibinfo {author} {\bibfnamefont {N.}~\bibnamefont {{Stergioulas}}}, \ and\ \bibinfo {author} {\bibfnamefont {J.~A.}\ \bibnamefont {{Font}}},\ }\href {\doibase 10.1111/j.1365-2966.2006.10274.x} {\bibfield  {journal} {\bibinfo  {journal} {\mnras}\ }\textbf {\bibinfo {volume} {368}},\ \bibinfo {pages} {1609} (\bibinfo {year} {2006})},\ \Eprint {http://arxiv.org/abs/astro-ph/0511394} {arXiv:astro-ph/0511394 [astro-ph]} \BibitemShut {NoStop}%
\bibitem [{\citenamefont {Kr\"uger}\ \emph {et~al.}(2015)\citenamefont {Kr\"uger}, \citenamefont {Ho},\ and\ \citenamefont {Andersson}}]{Kruger:2014pva}%
  \BibitemOpen
  \bibfield  {author} {\bibinfo {author} {\bibfnamefont {C.~J.}\ \bibnamefont {Kr\"uger}}, \bibinfo {author} {\bibfnamefont {W.~C.~G.}\ \bibnamefont {Ho}}, \ and\ \bibinfo {author} {\bibfnamefont {N.}~\bibnamefont {Andersson}},\ }\href {\doibase 10.1103/PhysRevD.92.063009} {\bibfield  {journal} {\bibinfo  {journal} {Phys. Rev. D}\ }\textbf {\bibinfo {volume} {92}},\ \bibinfo {pages} {063009} (\bibinfo {year} {2015})},\ \Eprint {http://arxiv.org/abs/1402.5656} {arXiv:1402.5656 [gr-qc]} \BibitemShut {NoStop}%
\bibitem [{\citenamefont {Camelio}\ \emph {et~al.}(2017)\citenamefont {Camelio}, \citenamefont {Lovato}, \citenamefont {Gualtieri}, \citenamefont {Benhar}, \citenamefont {Pons},\ and\ \citenamefont {Ferrari}}]{Camelio:2017nka}%
  \BibitemOpen
  \bibfield  {author} {\bibinfo {author} {\bibfnamefont {G.}~\bibnamefont {Camelio}}, \bibinfo {author} {\bibfnamefont {A.}~\bibnamefont {Lovato}}, \bibinfo {author} {\bibfnamefont {L.}~\bibnamefont {Gualtieri}}, \bibinfo {author} {\bibfnamefont {O.}~\bibnamefont {Benhar}}, \bibinfo {author} {\bibfnamefont {J.~A.}\ \bibnamefont {Pons}}, \ and\ \bibinfo {author} {\bibfnamefont {V.}~\bibnamefont {Ferrari}},\ }\href {\doibase 10.1103/PhysRevD.96.043015} {\bibfield  {journal} {\bibinfo  {journal} {Phys. Rev. D}\ }\textbf {\bibinfo {volume} {96}},\ \bibinfo {pages} {043015} (\bibinfo {year} {2017})},\ \Eprint {http://arxiv.org/abs/1704.01923} {arXiv:1704.01923 [astro-ph.HE]} \BibitemShut {NoStop}%
\bibitem [{\citenamefont {Sotani}\ and\ \citenamefont {Takiwaki}(2016)}]{Sotani:2016uwn}%
  \BibitemOpen
  \bibfield  {author} {\bibinfo {author} {\bibfnamefont {H.}~\bibnamefont {Sotani}}\ and\ \bibinfo {author} {\bibfnamefont {T.}~\bibnamefont {Takiwaki}},\ }\href {\doibase 10.1103/PhysRevD.94.044043} {\bibfield  {journal} {\bibinfo  {journal} {Phys. Rev. D}\ }\textbf {\bibinfo {volume} {94}},\ \bibinfo {pages} {044043} (\bibinfo {year} {2016})},\ \Eprint {http://arxiv.org/abs/1608.01048} {arXiv:1608.01048 [astro-ph.HE]} \BibitemShut {NoStop}%
\bibitem [{\citenamefont {{Torres-Forn{\'e}}}\ \emph {et~al.}(2018)\citenamefont {{Torres-Forn{\'e}}}, \citenamefont {{Cerd{\'a}-Dur{\'a}n}}, \citenamefont {{Passamonti}},\ and\ \citenamefont {{Font}}}]{Torres_et_al_I}%
  \BibitemOpen
  \bibfield  {author} {\bibinfo {author} {\bibfnamefont {A.}~\bibnamefont {{Torres-Forn{\'e}}}}, \bibinfo {author} {\bibfnamefont {P.}~\bibnamefont {{Cerd{\'a}-Dur{\'a}n}}}, \bibinfo {author} {\bibfnamefont {A.}~\bibnamefont {{Passamonti}}}, \ and\ \bibinfo {author} {\bibfnamefont {J.~A.}\ \bibnamefont {{Font}}},\ }\href {\doibase 10.1093/mnras/stx3067} {\bibfield  {journal} {\bibinfo  {journal} {\mnras}\ }\textbf {\bibinfo {volume} {474}},\ \bibinfo {pages} {5272} (\bibinfo {year} {2018})},\ \Eprint {http://arxiv.org/abs/1708.01920} {arXiv:1708.01920 [astro-ph.SR]} \BibitemShut {NoStop}%
\bibitem [{\citenamefont {Morozova}\ \emph {et~al.}(2018)\citenamefont {Morozova}, \citenamefont {Radice}, \citenamefont {Burrows},\ and\ \citenamefont {Vartanyan}}]{Morozova:2018glm}%
  \BibitemOpen
  \bibfield  {author} {\bibinfo {author} {\bibfnamefont {V.}~\bibnamefont {Morozova}}, \bibinfo {author} {\bibfnamefont {D.}~\bibnamefont {Radice}}, \bibinfo {author} {\bibfnamefont {A.}~\bibnamefont {Burrows}}, \ and\ \bibinfo {author} {\bibfnamefont {D.}~\bibnamefont {Vartanyan}},\ }\href {\doibase 10.3847/1538-4357/aac5f1} {\bibfield  {journal} {\bibinfo  {journal} {Astrophys. J.}\ }\textbf {\bibinfo {volume} {861}},\ \bibinfo {pages} {10} (\bibinfo {year} {2018})},\ \Eprint {http://arxiv.org/abs/1801.01914} {arXiv:1801.01914 [astro-ph.HE]} \BibitemShut {NoStop}%
\bibitem [{\citenamefont {{Torres-Forn{\'e}}}\ \emph {et~al.}(2019{\natexlab{a}})\citenamefont {{Torres-Forn{\'e}}}, \citenamefont {{Cerd{\'a}-Dur{\'a}n}}, \citenamefont {{Passamonti}}, \citenamefont {{Obergaulinger}},\ and\ \citenamefont {{Font}}}]{Torres_et_al_II}%
  \BibitemOpen
  \bibfield  {author} {\bibinfo {author} {\bibfnamefont {A.}~\bibnamefont {{Torres-Forn{\'e}}}}, \bibinfo {author} {\bibfnamefont {P.}~\bibnamefont {{Cerd{\'a}-Dur{\'a}n}}}, \bibinfo {author} {\bibfnamefont {A.}~\bibnamefont {{Passamonti}}}, \bibinfo {author} {\bibfnamefont {M.}~\bibnamefont {{Obergaulinger}}}, \ and\ \bibinfo {author} {\bibfnamefont {J.~A.}\ \bibnamefont {{Font}}},\ }\href {\doibase 10.1093/mnras/sty2854} {\bibfield  {journal} {\bibinfo  {journal} {\mnras}\ }\textbf {\bibinfo {volume} {482}},\ \bibinfo {pages} {3967} (\bibinfo {year} {2019}{\natexlab{a}})},\ \Eprint {http://arxiv.org/abs/1806.11366} {arXiv:1806.11366 [astro-ph.HE]} \BibitemShut {NoStop}%
\bibitem [{\citenamefont {{Sotani}}\ \emph {et~al.}(2019)\citenamefont {{Sotani}} \emph {et~al.}}]{Sotani2019}%
  \BibitemOpen
  \bibfield  {author} {\bibinfo {author} {\bibfnamefont {H.}~\bibnamefont {{Sotani}}} \emph {et~al.},\ }\href {\doibase 10.1103/PhysRevD.99.123024} {\bibfield  {journal} {\bibinfo  {journal} {\prd}\ }\textbf {\bibinfo {volume} {99}},\ \bibinfo {eid} {123024} (\bibinfo {year} {2019})},\ \Eprint {http://arxiv.org/abs/1906.04354} {arXiv:1906.04354 [astro-ph.HE]} \BibitemShut {NoStop}%
\bibitem [{\citenamefont {{Westernacher-Schneider}}\ \emph {et~al.}(2019)\citenamefont {{Westernacher-Schneider}} \emph {et~al.}}]{Westernacher-Schneider2019}%
  \BibitemOpen
  \bibfield  {author} {\bibinfo {author} {\bibfnamefont {J.~R.}\ \bibnamefont {{Westernacher-Schneider}}} \emph {et~al.},\ }\href {\doibase 10.1103/PhysRevD.100.123009} {\bibfield  {journal} {\bibinfo  {journal} {\prd}\ }\textbf {\bibinfo {volume} {100}},\ \bibinfo {eid} {123009} (\bibinfo {year} {2019})},\ \Eprint {http://arxiv.org/abs/1907.01138} {arXiv:1907.01138 [astro-ph.HE]} \BibitemShut {NoStop}%
\bibitem [{\citenamefont {{Westernacher-Schneider}}(2020)}]{Westernacher-Schneider2020}%
  \BibitemOpen
  \bibfield  {author} {\bibinfo {author} {\bibfnamefont {J.~R.}\ \bibnamefont {{Westernacher-Schneider}}},\ }\href {\doibase 10.1103/PhysRevD.101.083021} {\bibfield  {journal} {\bibinfo  {journal} {\prd}\ }\textbf {\bibinfo {volume} {101}},\ \bibinfo {eid} {083021} (\bibinfo {year} {2020})},\ \Eprint {http://arxiv.org/abs/2002.04468} {arXiv:2002.04468 [astro-ph.HE]} \BibitemShut {NoStop}%
\bibitem [{\citenamefont {{Sotani}}\ and\ \citenamefont {{Takiwaki}}(2020)}]{Sotani2020}%
  \BibitemOpen
  \bibfield  {author} {\bibinfo {author} {\bibfnamefont {H.}~\bibnamefont {{Sotani}}}\ and\ \bibinfo {author} {\bibfnamefont {T.}~\bibnamefont {{Takiwaki}}},\ }\href {\doibase 10.1093/mnras/staa2597} {\bibfield  {journal} {\bibinfo  {journal} {\mnras}\ }\textbf {\bibinfo {volume} {498}},\ \bibinfo {pages} {3503} (\bibinfo {year} {2020})},\ \Eprint {http://arxiv.org/abs/2008.00419} {arXiv:2008.00419 [astro-ph.HE]} \BibitemShut {NoStop}%
\bibitem [{\citenamefont {{Sotani}}\ \emph {et~al.}(2017)\citenamefont {{Sotani}}, \citenamefont {{Kuroda}}, \citenamefont {{Takiwaki}},\ and\ \citenamefont {{Kotake}}}]{Sotani2017}%
  \BibitemOpen
  \bibfield  {author} {\bibinfo {author} {\bibfnamefont {H.}~\bibnamefont {{Sotani}}}, \bibinfo {author} {\bibfnamefont {T.}~\bibnamefont {{Kuroda}}}, \bibinfo {author} {\bibfnamefont {T.}~\bibnamefont {{Takiwaki}}}, \ and\ \bibinfo {author} {\bibfnamefont {K.}~\bibnamefont {{Kotake}}},\ }\href {\doibase 10.1103/PhysRevD.96.063005} {\bibfield  {journal} {\bibinfo  {journal} {\prd}\ }\textbf {\bibinfo {volume} {96}},\ \bibinfo {eid} {063005} (\bibinfo {year} {2017})},\ \Eprint {http://arxiv.org/abs/1708.03738} {arXiv:1708.03738 [astro-ph.HE]} \BibitemShut {NoStop}%
\bibitem [{\citenamefont {{Torres-Forn{\'e}}}\ \emph {et~al.}(2019{\natexlab{b}})\citenamefont {{Torres-Forn{\'e}}}, \citenamefont {{Cerd{\'a}-Dur{\'a}n}}, \citenamefont {{Obergaulinger}}, \citenamefont {{M{\"u}ller}},\ and\ \citenamefont {{Font}}}]{Torres_et_al_letter}%
  \BibitemOpen
  \bibfield  {author} {\bibinfo {author} {\bibfnamefont {A.}~\bibnamefont {{Torres-Forn{\'e}}}}, \bibinfo {author} {\bibfnamefont {P.}~\bibnamefont {{Cerd{\'a}-Dur{\'a}n}}}, \bibinfo {author} {\bibfnamefont {M.}~\bibnamefont {{Obergaulinger}}}, \bibinfo {author} {\bibfnamefont {B.}~\bibnamefont {{M{\"u}ller}}}, \ and\ \bibinfo {author} {\bibfnamefont {J.~A.}\ \bibnamefont {{Font}}},\ }\href {\doibase 10.1103/PhysRevLett.123.051102} {\bibfield  {journal} {\bibinfo  {journal} {\prl}\ }\textbf {\bibinfo {volume} {123}},\ \bibinfo {eid} {051102} (\bibinfo {year} {2019}{\natexlab{b}})},\ \Eprint {http://arxiv.org/abs/1902.10048} {arXiv:1902.10048 [gr-qc]} \BibitemShut {NoStop}%
\bibitem [{\citenamefont {{Bizouard}}\ \emph {et~al.}(2021)\citenamefont {{Bizouard}}, \citenamefont {{Maturana-Russel}}, \citenamefont {{Torres-Forn{\'e}}}, \citenamefont {{Obergaulinger}}, \citenamefont {{Cerd{\'a}-Dur{\'a}n}}, \citenamefont {{Christensen}}, \citenamefont {{Font}},\ and\ \citenamefont {{Meyer}}}]{Bizouard2021}%
  \BibitemOpen
  \bibfield  {author} {\bibinfo {author} {\bibfnamefont {M.-A.}\ \bibnamefont {{Bizouard}}}, \bibinfo {author} {\bibfnamefont {P.}~\bibnamefont {{Maturana-Russel}}}, \bibinfo {author} {\bibfnamefont {A.}~\bibnamefont {{Torres-Forn{\'e}}}}, \bibinfo {author} {\bibfnamefont {M.}~\bibnamefont {{Obergaulinger}}}, \bibinfo {author} {\bibfnamefont {P.}~\bibnamefont {{Cerd{\'a}-Dur{\'a}n}}}, \bibinfo {author} {\bibfnamefont {N.}~\bibnamefont {{Christensen}}}, \bibinfo {author} {\bibfnamefont {J.~A.}\ \bibnamefont {{Font}}}, \ and\ \bibinfo {author} {\bibfnamefont {R.}~\bibnamefont {{Meyer}}},\ }\href {\doibase 10.1103/PhysRevD.103.063006} {\bibfield  {journal} {\bibinfo  {journal} {\prd}\ }\textbf {\bibinfo {volume} {103}},\ \bibinfo {eid} {063006} (\bibinfo {year} {2021})},\ \Eprint {http://arxiv.org/abs/2012.00846} {arXiv:2012.00846 [gr-qc]} \BibitemShut {NoStop}%
\bibitem [{\citenamefont {{Bruel}}\ \emph {et~al.}(2023)\citenamefont {{Bruel}}, \citenamefont {{Bizouard}}, \citenamefont {{Obergaulinger}}, \citenamefont {{Maturana-Russel}}, \citenamefont {{Torres-Forn{\'e}}}, \citenamefont {{Cerd{\'a}-Dur{\'a}n}}, \citenamefont {{Christensen}}, \citenamefont {{Font}},\ and\ \citenamefont {{Meyer}}}]{Bruel2023}%
  \BibitemOpen
  \bibfield  {author} {\bibinfo {author} {\bibfnamefont {T.}~\bibnamefont {{Bruel}}}, \bibinfo {author} {\bibfnamefont {M.-A.}\ \bibnamefont {{Bizouard}}}, \bibinfo {author} {\bibfnamefont {M.}~\bibnamefont {{Obergaulinger}}}, \bibinfo {author} {\bibfnamefont {P.}~\bibnamefont {{Maturana-Russel}}}, \bibinfo {author} {\bibfnamefont {A.}~\bibnamefont {{Torres-Forn{\'e}}}}, \bibinfo {author} {\bibfnamefont {P.}~\bibnamefont {{Cerd{\'a}-Dur{\'a}n}}}, \bibinfo {author} {\bibfnamefont {N.}~\bibnamefont {{Christensen}}}, \bibinfo {author} {\bibfnamefont {J.~A.}\ \bibnamefont {{Font}}}, \ and\ \bibinfo {author} {\bibfnamefont {R.}~\bibnamefont {{Meyer}}},\ }\href {\doibase 10.1103/PhysRevD.107.083029} {\bibfield  {journal} {\bibinfo  {journal} {\prd}\ }\textbf {\bibinfo {volume} {107}},\ \bibinfo {eid} {083029} (\bibinfo {year} {2023})},\ \Eprint {http://arxiv.org/abs/2301.10019} {arXiv:2301.10019 [astro-ph.HE]} \BibitemShut {NoStop}%
\bibitem [{\citenamefont {{Jakobus}}\ \emph {et~al.}(2023)\citenamefont {{Jakobus}}, \citenamefont {{M{\"u}ller}}, \citenamefont {{Heger}}, \citenamefont {{Zha}}, \citenamefont {{Powell}}, \citenamefont {{Motornenko}}, \citenamefont {{Steinheimer}},\ and\ \citenamefont {{St{\"o}cker}}}]{Jakobus2023}%
  \BibitemOpen
  \bibfield  {author} {\bibinfo {author} {\bibfnamefont {P.}~\bibnamefont {{Jakobus}}}, \bibinfo {author} {\bibfnamefont {B.}~\bibnamefont {{M{\"u}ller}}}, \bibinfo {author} {\bibfnamefont {A.}~\bibnamefont {{Heger}}}, \bibinfo {author} {\bibfnamefont {S.}~\bibnamefont {{Zha}}}, \bibinfo {author} {\bibfnamefont {J.}~\bibnamefont {{Powell}}}, \bibinfo {author} {\bibfnamefont {A.}~\bibnamefont {{Motornenko}}}, \bibinfo {author} {\bibfnamefont {J.}~\bibnamefont {{Steinheimer}}}, \ and\ \bibinfo {author} {\bibfnamefont {H.}~\bibnamefont {{St{\"o}cker}}},\ }\href {\doibase 10.1103/PhysRevLett.131.191201} {\bibfield  {journal} {\bibinfo  {journal} {\prl}\ }\textbf {\bibinfo {volume} {131}},\ \bibinfo {eid} {191201} (\bibinfo {year} {2023})},\ \Eprint {http://arxiv.org/abs/2301.06515} {arXiv:2301.06515 [astro-ph.HE]} \BibitemShut {NoStop}%
\bibitem [{\citenamefont {{Wolfe}}\ \emph {et~al.}(2023)\citenamefont {{Wolfe}}, \citenamefont {{Fr{\"o}hlich}}, \citenamefont {{Miller}}, \citenamefont {{Torres-Forn{\'e}}},\ and\ \citenamefont {{Cerd{\'a}-Dur{\'a}n}}}]{Wolfe2023}%
  \BibitemOpen
  \bibfield  {author} {\bibinfo {author} {\bibfnamefont {N.~E.}\ \bibnamefont {{Wolfe}}}, \bibinfo {author} {\bibfnamefont {C.}~\bibnamefont {{Fr{\"o}hlich}}}, \bibinfo {author} {\bibfnamefont {J.~M.}\ \bibnamefont {{Miller}}}, \bibinfo {author} {\bibfnamefont {A.}~\bibnamefont {{Torres-Forn{\'e}}}}, \ and\ \bibinfo {author} {\bibfnamefont {P.}~\bibnamefont {{Cerd{\'a}-Dur{\'a}n}}},\ }\href {\doibase 10.3847/1538-4357/ace693} {\bibfield  {journal} {\bibinfo  {journal} {\apj}\ }\textbf {\bibinfo {volume} {954}},\ \bibinfo {eid} {161} (\bibinfo {year} {2023})},\ \Eprint {http://arxiv.org/abs/2303.16962} {arXiv:2303.16962 [astro-ph.HE]} \BibitemShut {NoStop}%
\bibitem [{\citenamefont {{Jakobus}}\ \emph {et~al.}(2024)\citenamefont {{Jakobus}}, \citenamefont {{Mueller}},\ and\ \citenamefont {{Heger}}}]{Jakobus2024}%
  \BibitemOpen
  \bibfield  {author} {\bibinfo {author} {\bibfnamefont {P.}~\bibnamefont {{Jakobus}}}, \bibinfo {author} {\bibfnamefont {B.}~\bibnamefont {{Mueller}}}, \ and\ \bibinfo {author} {\bibfnamefont {A.}~\bibnamefont {{Heger}}},\ }\href {\doibase 10.48550/arXiv.2405.01449} {\bibfield  {journal} {\bibinfo  {journal} {arXiv e-prints}\ ,\ \bibinfo {eid} {arXiv:2405.01449}} (\bibinfo {year} {2024})},\ \Eprint {http://arxiv.org/abs/2405.01449} {arXiv:2405.01449 [astro-ph.HE]} \BibitemShut {NoStop}%
\bibitem [{\citenamefont {{Foglizzo}}(2002)}]{Foglizzo2002}%
  \BibitemOpen
  \bibfield  {author} {\bibinfo {author} {\bibfnamefont {T.}~\bibnamefont {{Foglizzo}}},\ }\href {\doibase 10.1051/0004-6361:20020912} {\bibfield  {journal} {\bibinfo  {journal} {\aap}\ }\textbf {\bibinfo {volume} {392}},\ \bibinfo {pages} {353} (\bibinfo {year} {2002})},\ \Eprint {http://arxiv.org/abs/astro-ph/0206274} {arXiv:astro-ph/0206274 [astro-ph]} \BibitemShut {NoStop}%
\bibitem [{\citenamefont {{Foglizzo}}\ \emph {et~al.}(2007)\citenamefont {{Foglizzo}}, \citenamefont {{Galletti}}, \citenamefont {{Scheck}},\ and\ \citenamefont {{Janka}}}]{Foglizzo2007}%
  \BibitemOpen
  \bibfield  {author} {\bibinfo {author} {\bibfnamefont {T.}~\bibnamefont {{Foglizzo}}}, \bibinfo {author} {\bibfnamefont {P.}~\bibnamefont {{Galletti}}}, \bibinfo {author} {\bibfnamefont {L.}~\bibnamefont {{Scheck}}}, \ and\ \bibinfo {author} {\bibfnamefont {H.~T.}\ \bibnamefont {{Janka}}},\ }\href {\doibase 10.1086/509612} {\bibfield  {journal} {\bibinfo  {journal} {\apj}\ }\textbf {\bibinfo {volume} {654}},\ \bibinfo {pages} {1006} (\bibinfo {year} {2007})},\ \Eprint {http://arxiv.org/abs/astro-ph/0606640} {arXiv:astro-ph/0606640 [astro-ph]} \BibitemShut {NoStop}%
\bibitem [{\citenamefont {Boyd}(2013)}]{BoydSpectral}%
  \BibitemOpen
  \bibfield  {author} {\bibinfo {author} {\bibfnamefont {J.}~\bibnamefont {Boyd}},\ }\href {https://books.google.es/books?id=b4TCAgAAQBAJ} {\emph {\bibinfo {title} {Chebyshev and Fourier Spectral Methods: Second Revised Edition}}},\ Dover Books on Mathematics\ (\bibinfo  {publisher} {Dover Publications},\ \bibinfo {year} {2013})\BibitemShut {NoStop}%
\bibitem [{\citenamefont {Trefethen}(2000)}]{ThefethenSpectral}%
  \BibitemOpen
  \bibfield  {author} {\bibinfo {author} {\bibfnamefont {L.~N.}\ \bibnamefont {Trefethen}},\ }\href@noop {} {\emph {\bibinfo {title} {Spectral methods in MATLAB}}}\ (\bibinfo  {publisher} {Society for Industrial and Applied Mathematics},\ \bibinfo {address} {USA},\ \bibinfo {year} {2000})\BibitemShut {NoStop}%
\bibitem [{\citenamefont {Voigt}\ \emph {et~al.}(1984)\citenamefont {Voigt}, \citenamefont {Gottlieb}, \citenamefont {Hussaini}, \citenamefont {for Computer Applications~in Science},\ and\ \citenamefont {Engineering}}]{VoigtSpectral}%
  \BibitemOpen
  \bibfield  {author} {\bibinfo {author} {\bibfnamefont {R.}~\bibnamefont {Voigt}}, \bibinfo {author} {\bibfnamefont {D.}~\bibnamefont {Gottlieb}}, \bibinfo {author} {\bibfnamefont {M.}~\bibnamefont {Hussaini}}, \bibinfo {author} {\bibfnamefont {I.}~\bibnamefont {for Computer Applications~in Science}}, \ and\ \bibinfo {author} {\bibnamefont {Engineering}},\ }\href {https://books.google.es/books?id=EV_TzMlpOiYC} {\emph {\bibinfo {title} {Spectral Methods for Partial Differential Equations}}},\ Proceedings in Applied Mathematics Series\ (\bibinfo  {publisher} {SIAM, Philadelphia},\ \bibinfo {year} {1984})\BibitemShut {NoStop}%
\bibitem [{\citenamefont {Dias}\ \emph {et~al.}(2016)\citenamefont {Dias}, \citenamefont {Santos},\ and\ \citenamefont {Way}}]{Dias:2015nua}%
  \BibitemOpen
  \bibfield  {author} {\bibinfo {author} {\bibfnamefont {O.~J.~C.}\ \bibnamefont {Dias}}, \bibinfo {author} {\bibfnamefont {J.~E.}\ \bibnamefont {Santos}}, \ and\ \bibinfo {author} {\bibfnamefont {B.}~\bibnamefont {Way}},\ }\href {\doibase 10.1088/0264-9381/33/13/133001} {\bibfield  {journal} {\bibinfo  {journal} {Class. Quant. Grav.}\ }\textbf {\bibinfo {volume} {33}},\ \bibinfo {pages} {133001} (\bibinfo {year} {2016})},\ \Eprint {http://arxiv.org/abs/1510.02804} {arXiv:1510.02804 [hep-th]} \BibitemShut {NoStop}%
\bibitem [{\citenamefont {Jansen}(2017)}]{Jansen:2017oag}%
  \BibitemOpen
  \bibfield  {author} {\bibinfo {author} {\bibfnamefont {A.}~\bibnamefont {Jansen}},\ }\href {\doibase 10.1140/epjp/i2017-11825-9} {\bibfield  {journal} {\bibinfo  {journal} {Eur. Phys. J. Plus}\ }\textbf {\bibinfo {volume} {132}},\ \bibinfo {pages} {546} (\bibinfo {year} {2017})},\ \Eprint {http://arxiv.org/abs/1709.09178} {arXiv:1709.09178 [gr-qc]} \BibitemShut {NoStop}%
\bibitem [{\citenamefont {Banyuls}\ \emph {et~al.}(1997)\citenamefont {Banyuls}, \citenamefont {Font}, \citenamefont {Ibanez}, \citenamefont {Marti},\ and\ \citenamefont {Miralles}}]{Banyuls:1997zz}%
  \BibitemOpen
  \bibfield  {author} {\bibinfo {author} {\bibfnamefont {F.}~\bibnamefont {Banyuls}}, \bibinfo {author} {\bibfnamefont {J.~A.}\ \bibnamefont {Font}}, \bibinfo {author} {\bibfnamefont {J.~M.~A.}\ \bibnamefont {Ibanez}}, \bibinfo {author} {\bibfnamefont {J.~M.~A.}\ \bibnamefont {Marti}}, \ and\ \bibinfo {author} {\bibfnamefont {J.~A.}\ \bibnamefont {Miralles}},\ }\href@noop {} {\bibfield  {journal} {\bibinfo  {journal} {Astrophys. J.}\ }\textbf {\bibinfo {volume} {476}},\ \bibinfo {pages} {221} (\bibinfo {year} {1997})}\BibitemShut {NoStop}%
\bibitem [{\citenamefont {{Font}}(2008)}]{2008Font_review}%
  \BibitemOpen
  \bibfield  {author} {\bibinfo {author} {\bibfnamefont {J.~A.}\ \bibnamefont {{Font}}},\ }\href {\doibase 10.12942/lrr-2008-7} {\bibfield  {journal} {\bibinfo  {journal} {Living Reviews in Relativity}\ }\textbf {\bibinfo {volume} {11}},\ \bibinfo {eid} {7} (\bibinfo {year} {2008})}\BibitemShut {NoStop}%
\bibitem [{\citenamefont {{Rezzolla}}\ and\ \citenamefont {{Zanotti}}(2013)}]{Rezzolla_book_GR_hydro}%
  \BibitemOpen
  \bibfield  {author} {\bibinfo {author} {\bibfnamefont {L.}~\bibnamefont {{Rezzolla}}}\ and\ \bibinfo {author} {\bibfnamefont {O.}~\bibnamefont {{Zanotti}}},\ }\href@noop {} {\emph {\bibinfo {title} {{Relativistic Hydrodynamics}}}}\ (\bibinfo  {publisher} {Oxford University Press},\ \bibinfo {year} {2013})\BibitemShut {NoStop}%
\bibitem [{\citenamefont {{Unno}}\ \emph {et~al.}(1989)\citenamefont {{Unno}}, \citenamefont {{Osaki}}, \citenamefont {{Ando}}, \citenamefont {{Saio}},\ and\ \citenamefont {{Shibahashi}}}]{1989nos..book.....U}%
  \BibitemOpen
  \bibfield  {author} {\bibinfo {author} {\bibfnamefont {W.}~\bibnamefont {{Unno}}}, \bibinfo {author} {\bibfnamefont {Y.}~\bibnamefont {{Osaki}}}, \bibinfo {author} {\bibfnamefont {H.}~\bibnamefont {{Ando}}}, \bibinfo {author} {\bibfnamefont {H.}~\bibnamefont {{Saio}}}, \ and\ \bibinfo {author} {\bibfnamefont {H.}~\bibnamefont {{Shibahashi}}},\ }\href@noop {} {\emph {\bibinfo {title} {{Nonradial oscillations of stars}}}}\ (\bibinfo  {publisher} {University of Tokyo Press, Tokyo},\ \bibinfo {year} {1989})\BibitemShut {NoStop}%
\bibitem [{\citenamefont {{Blanchet}}\ \emph {et~al.}(1990)\citenamefont {{Blanchet}}, \citenamefont {{Damour}},\ and\ \citenamefont {{Schaefer}}}]{Blanchet1990}%
  \BibitemOpen
  \bibfield  {author} {\bibinfo {author} {\bibfnamefont {L.}~\bibnamefont {{Blanchet}}}, \bibinfo {author} {\bibfnamefont {T.}~\bibnamefont {{Damour}}}, \ and\ \bibinfo {author} {\bibfnamefont {G.}~\bibnamefont {{Schaefer}}},\ }\href {\doibase 10.1093/mnras/242.3.289} {\bibfield  {journal} {\bibinfo  {journal} {\mnras}\ }\textbf {\bibinfo {volume} {242}},\ \bibinfo {pages} {289} (\bibinfo {year} {1990})}\BibitemShut {NoStop}%
\bibitem [{\citenamefont {{Courant}}\ and\ \citenamefont {{Friedrichs}}(1948)}]{1948Courant}%
  \BibitemOpen
  \bibfield  {author} {\bibinfo {author} {\bibfnamefont {R.}~\bibnamefont {{Courant}}}\ and\ \bibinfo {author} {\bibfnamefont {K.~O.}\ \bibnamefont {{Friedrichs}}},\ }\href@noop {} {\emph {\bibinfo {title} {{Supersonic flow and shock waves}}}}\ (\bibinfo  {publisher} {Springer New York, NY},\ \bibinfo {year} {1948})\BibitemShut {NoStop}%
\bibitem [{\citenamefont {{Marti}}\ and\ \citenamefont {{Muller}}(1994)}]{Marti_et_al_1994}%
  \BibitemOpen
  \bibfield  {author} {\bibinfo {author} {\bibfnamefont {J.~M.}\ \bibnamefont {{Marti}}}\ and\ \bibinfo {author} {\bibfnamefont {E.}~\bibnamefont {{Muller}}},\ }\href {\doibase 10.1017/S0022112094003344} {\bibfield  {journal} {\bibinfo  {journal} {Journal of Fluid Mechanics}\ }\textbf {\bibinfo {volume} {258}},\ \bibinfo {pages} {317} (\bibinfo {year} {1994})}\BibitemShut {NoStop}%
\bibitem [{\citenamefont {{Cox}}(1983)}]{1983_Cox}%
  \BibitemOpen
  \bibfield  {author} {\bibinfo {author} {\bibfnamefont {J.~P.}\ \bibnamefont {{Cox}}},\ }\href@noop {} {\emph {\bibinfo {title} {{Theory of stellar pulsations.}}}}\ (\bibinfo  {publisher} {Princeton Series in Astrophysics},\ \bibinfo {year} {1983})\BibitemShut {NoStop}%
\bibitem [{\citenamefont {{Guilet}}\ and\ \citenamefont {{Foglizzo}}(2010)}]{Guilet2010}%
  \BibitemOpen
  \bibfield  {author} {\bibinfo {author} {\bibfnamefont {J.}~\bibnamefont {{Guilet}}}\ and\ \bibinfo {author} {\bibfnamefont {T.}~\bibnamefont {{Foglizzo}}},\ }\href {\doibase 10.1088/0004-637X/711/1/99} {\bibfield  {journal} {\bibinfo  {journal} {\apj}\ }\textbf {\bibinfo {volume} {711}},\ \bibinfo {pages} {99} (\bibinfo {year} {2010})},\ \Eprint {http://arxiv.org/abs/0911.2795} {arXiv:0911.2795 [astro-ph.SR]} \BibitemShut {NoStop}%
\bibitem [{\citenamefont {{Foglizzo}}(2001)}]{Foglizzo2001}%
  \BibitemOpen
  \bibfield  {author} {\bibinfo {author} {\bibfnamefont {T.}~\bibnamefont {{Foglizzo}}},\ }\href {\doibase 10.1051/0004-6361:20000506} {\bibfield  {journal} {\bibinfo  {journal} {\aap}\ }\textbf {\bibinfo {volume} {368}},\ \bibinfo {pages} {311} (\bibinfo {year} {2001})},\ \Eprint {http://arxiv.org/abs/astro-ph/0101056} {arXiv:astro-ph/0101056 [astro-ph]} \BibitemShut {NoStop}%
\bibitem [{\citenamefont {{Nowak}}\ and\ \citenamefont {{Wagoner}}(1991)}]{Nowak1991}%
  \BibitemOpen
  \bibfield  {author} {\bibinfo {author} {\bibfnamefont {M.~A.}\ \bibnamefont {{Nowak}}}\ and\ \bibinfo {author} {\bibfnamefont {R.~V.}\ \bibnamefont {{Wagoner}}},\ }\href {\doibase 10.1086/170465} {\bibfield  {journal} {\bibinfo  {journal} {\apj}\ }\textbf {\bibinfo {volume} {378}},\ \bibinfo {pages} {656} (\bibinfo {year} {1991})}\BibitemShut {NoStop}%
\bibitem [{\citenamefont {{Kato}}\ \emph {et~al.}(2008)\citenamefont {{Kato}}, \citenamefont {{Fukue}},\ and\ \citenamefont {{Mineshige}}}]{Kato2008}%
  \BibitemOpen
  \bibfield  {author} {\bibinfo {author} {\bibfnamefont {S.}~\bibnamefont {{Kato}}}, \bibinfo {author} {\bibfnamefont {J.}~\bibnamefont {{Fukue}}}, \ and\ \bibinfo {author} {\bibfnamefont {S.}~\bibnamefont {{Mineshige}}},\ }\href@noop {} {\emph {\bibinfo {title} {{Black-Hole Accretion Disks --- Towards a New Paradigm ---}}}}\ (\bibinfo  {publisher} {Kyoto University Press, Kyoto},\ \bibinfo {year} {2008})\BibitemShut {NoStop}%
\bibitem [{\citenamefont {CAMPOS}\ and\ \citenamefont {KOBAYASHI}(2000)}]{CAMPOS_KOBAYASHI_2000}%
  \BibitemOpen
  \bibfield  {author} {\bibinfo {author} {\bibfnamefont {L.~M. B.~C.}\ \bibnamefont {CAMPOS}}\ and\ \bibinfo {author} {\bibfnamefont {M.~H.}\ \bibnamefont {KOBAYASHI}},\ }\href {\doibase 10.1017/S0022112000002068} {\bibfield  {journal} {\bibinfo  {journal} {Journal of Fluid Mechanics}\ }\textbf {\bibinfo {volume} {424}},\ \bibinfo {pages} {303–326} (\bibinfo {year} {2000})}\BibitemShut {NoStop}%
\bibitem [{\citenamefont {{Campos}}(1986)}]{Campos1986}%
  \BibitemOpen
  \bibfield  {author} {\bibinfo {author} {\bibfnamefont {L.~M.~B.~C.}\ \bibnamefont {{Campos}}},\ }\href {\doibase 10.1103/RevModPhys.58.117} {\bibfield  {journal} {\bibinfo  {journal} {Reviews of Modern Physics}\ }\textbf {\bibinfo {volume} {58}},\ \bibinfo {pages} {117} (\bibinfo {year} {1986})}\BibitemShut {NoStop}%
\bibitem [{\citenamefont {Kinsler}\ \emph {et~al.}(2000)\citenamefont {Kinsler}, \citenamefont {Frey}, \citenamefont {Coppens},\ and\ \citenamefont {Sanders}}]{Kinsler2000}%
  \BibitemOpen
  \bibfield  {author} {\bibinfo {author} {\bibfnamefont {L.~E.}\ \bibnamefont {Kinsler}}, \bibinfo {author} {\bibfnamefont {A.~R.}\ \bibnamefont {Frey}}, \bibinfo {author} {\bibfnamefont {A.~B.}\ \bibnamefont {Coppens}}, \ and\ \bibinfo {author} {\bibfnamefont {J.~V.}\ \bibnamefont {Sanders}},\ }\href@noop {} {\emph {\bibinfo {title} {Fundamentals of Acoustics}}}\ (\bibinfo  {publisher} {John Wiley \& Sons},\ \bibinfo {year} {2000})\BibitemShut {NoStop}%
\bibitem [{\citenamefont {{Pierce}}(2019)}]{2019_Pierce}%
  \BibitemOpen
  \bibfield  {author} {\bibinfo {author} {\bibfnamefont {A.~D.}\ \bibnamefont {{Pierce}}},\ }\href {\doibase 10.1007/978-3-030-11214-1} {\emph {\bibinfo {title} {{Acoustics: An Introduction to Its Physical Principles and Applications}}}}\ (\bibinfo  {publisher} {Springer International Publishing},\ \bibinfo {year} {2019})\BibitemShut {NoStop}%
\bibitem [{\citenamefont {Okamoto}(2006)}]{2006_Okamoto}%
  \BibitemOpen
  \bibfield  {author} {\bibinfo {author} {\bibfnamefont {K.}~\bibnamefont {Okamoto}},\ }\href {\doibase 10.1016/B978-0-12-525096-2.X5000-4} {\emph {\bibinfo {title} {Fundamentals of Optical Waveguides}}}\ (\bibinfo  {publisher} {Elsevier Inc.},\ \bibinfo {year} {2006})\BibitemShut {NoStop}%
\bibitem [{\citenamefont {{Balanis}}(2023)}]{2023_Balani}%
  \BibitemOpen
  \bibfield  {author} {\bibinfo {author} {\bibfnamefont {C.~A.}\ \bibnamefont {{Balanis}}},\ }\href {\doibase 10.1002/9781394180042.ch8} {\emph {\bibinfo {title} {Balanis' Advanced Engineering Electromagnetics}}}\ (\bibinfo  {publisher} {John Wiley \& Sons},\ \bibinfo {year} {2023})\BibitemShut {NoStop}%
\bibitem [{\citenamefont {{Bender}}\ and\ \citenamefont {{Orszag}}(1978)}]{1978_Bender_Orszag}%
  \BibitemOpen
  \bibfield  {author} {\bibinfo {author} {\bibfnamefont {C.~M.}\ \bibnamefont {{Bender}}}\ and\ \bibinfo {author} {\bibfnamefont {S.~A.}\ \bibnamefont {{Orszag}}},\ }\href@noop {} {\emph {\bibinfo {title} {{Advanced Mathematical Methods for Scientists and Engineers}}}}\ (\bibinfo  {publisher} {Springer},\ \bibinfo {year} {1978})\BibitemShut {NoStop}%
\end{thebibliography}%

\end{document}